\documentclass[twocolumn]{aastex631}

\graphicspath{{./}{figures/}}

\newcommand{\um }{$\mu $m}

\newcommand{\cmq}{cm{$^{-3}$}}
\newcommand{\cms}{cm{$^{-2}$}}

\newcommand{\kms}{km~s{$^{-1}$}}

\newcommand{\Msol}{M{$_{\odot}$}}

\newcommand{\Lsol}{L{$_{\odot}$}}
\newcommand{\Vlsr}{{V$_{LSR}$}}
\newcommand{\CO}{{$^{12}$CO}}

\newcommand{\Sii}{S~{\sc ii}}

\newcommand{\Ha}{H$\alpha$}

\newcommand{\Htwo}{H{$_2$}}

\newcommand{\vsini}{{\em v}\,sin{\em i}~}
\newcommand{\rsini}{{\em r}\,sin{\em i}~}

\newcommand{\tco}{{$^{13}$CO}}

\usepackage{amsmath}
\shorttitle{HP Tau/G2 as a Stellar Merger}
\shortauthors{Reipurth et al.}

\begin{document}

\title{\large\bf The Walkaway Star HP Tau/G2: Evidence for a Stellar Merger}

\correspondingauthor{Bo Reipurth}
\email{reipurth@hawaii.edu}

% This is my Orcid number
\author[0000-0001-8174-1932]{Bo Reipurth}
\affil{Institute for Astronomy, University of Hawaii at Manoa, 
640 N. Aohoku Place, Hilo, HI 96720, USA}
\affil{Planetary Science Institute, 1700 E Fort Lowell Rd Ste 106, Tucson, AZ 85719, USA  }

\author{J. Bally}
\affiliation{Center for Astrophysics and Space Astronomy, University of Colorado, Boulder, CO 80309, USA}

\author[0000-0002-8010-8454]{P. Friberg}
\affiliation{East Asian Observatory, 660 North Aohoku Place, Hilo, HI 96720, USA   }

\author[0000-0001-8603-8031]{D. M. Faes}
\affiliation{National Radio Astronomy Observatory, 1003 Lopezville Road, Socorro, NM 87801, USA}

\author[0000-0001-7124-4094]{C. Brice\~{n}o}
\affiliation{NSF's NOIRLab/Cerro Tololo Inter-American Observatory, Casilla 603, La Serena, Chile}

\author[0000-0002-8293-1428]{M. S. Connelley}
\affiliation{Institute for Astronomy, University of Hawaii at Manoa,
640 N. Aohoku Place, HI 96720}
\affiliation{Staff Astronomer at the Infrared Telescope Facility, which is operated by the University of Hawaii under contract NNH14CK55B with the National Aeronautics and Space Administration}

\author[0000-0002-8591-472X]{C. Flores}
\affiliation{Institute of Astronomy and Astrophysics, Academia Sinica, 11F of AS/NTU Astronomy-Mathematics Building, No.1, Sec. 4, Roosevelt Rd, Taipei 10617, Taiwan, R.O.C. }

\author[0000-0002-3656-6706]{A. M. Cody}
\affiliation{SETI Institute, 339 N Bernardo Ave, Suite 200, Mountain View, CA 94043}

\author[0000-0003-0504-3539]{H. Zinnecker}
\affiliation{Nucleo de Astroquimica y Astrofisica, 
Universidad Aut\'onoma de Chile, Av. Pedro de Valdivia 425, Santiago, Chile}


\begin{abstract}

HP~Tau/G2 is a luminous, short-period, fast-rotating G-type weak-line
T~Tauri star with a large radius, an oblate shape with
gravity-darkening, little circumstellar material, and centered in a
slowly expanding cloud cavity.  It is an X-ray source and a variable
nonthermal radio source. It forms, together with the late-type T~Tauri
star KPNO~15, a pair of oppositely directed walkaway stars launched
when a multiple system broke apart $\sim$5600~yr ago. Momentum
conservation indicates a mass of G2 of only $\sim$0.7~M$_\odot$, much
lower than $\sim$1.9~M$_\odot$ determined from evolutionary models.
G2 is virtually a twin of FK~Com, the prototype of a class of evolved
stars resulting from coalescence of W~UMa binaries. We suggest that G2
became a very close and highly eccentric binary during viscous
evolution in the protostellar stage and with KPNO~15 formed a triple
system, which again was part of a larger unstable group including the
binary G3 and the single G1.  Dynamical evolution led to multiple
bound ejections of KPNO~15 before it finally escaped after
$\sim$2~Myr. As a result the G2 binary recoiled and contracted 5600 yr
ago, became Darwin unstable and merged in a major outburst
$\sim$2000~yr ago.  The nearby compact triple system G1+G3 was also
disturbed, and broke up 4900 yr ago, forming another walkaway
pair. The G5 star HD~283572 has similar unusual properties, indicating
that G2 is not a pathological case. G2 is now fading towards a new
stable configuration.  YSO mergers may be rather common and could
explain some FUor eruptions.

\end{abstract}


\section{INTRODUCTION} \label{sec:intro}

Small reflection nebulae are ubiquitous in clouds associated with star
forming regions, for a catalog see \citet{magakian2003}. They represent
a relatively brief stage when the outflows and winds from recently
formed low mass stars break through the surface of the cloud, allowing
light to flood the surroundings until the stars have drifted away or
the cloud is destroyed. Most of these nebulae show very few changes on
a human timescale, but some show rapid variability as circumstellar
material creates a shadow play across the surrounding cloud surface
\citep[e.g.,][]{knoxshaw1916,hubble1916,dahm2017}. In at
least one case, the actual emergence of a reflection nebula from an
embedded infrared source has been witnessed \citep{reipurth1986}.

\begin{figure*}
\begin{center}
\includegraphics[angle=0,width=1.6\columnwidth]{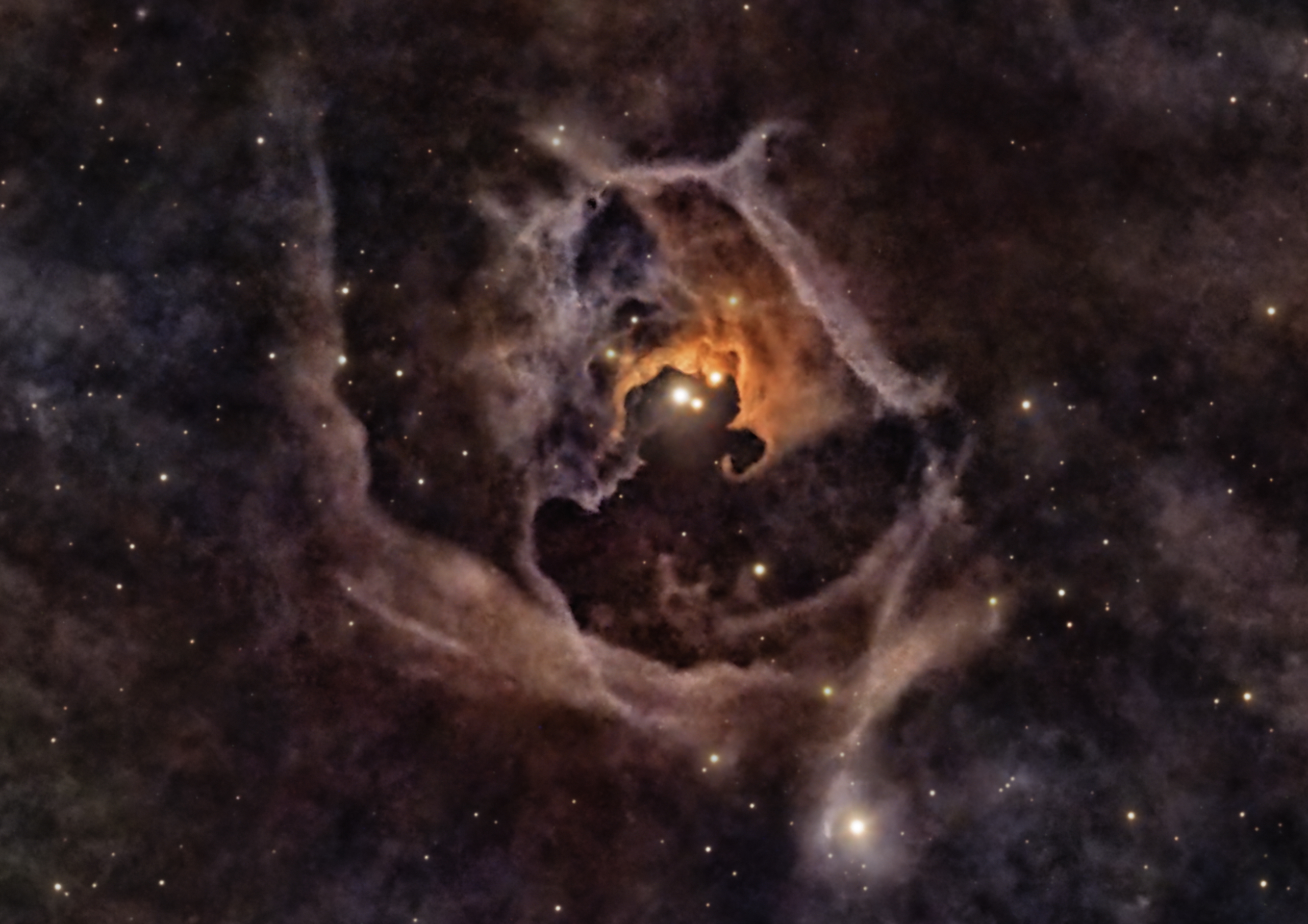} 
\caption{The large highly structured reflection nebula Magakian~77 surrounds the little group of young stars including HP~Tau. The figure is almost 8 arcmin wide. This is a 13hr 20min exposure with a 1m telescope through B, G, R filters plus luminance. 
Courtesy Alexandr Zaytsev and Mark Hanson.   
%[fig:zaytsev-hanson]  
\label{fig:zaytsev-hanson}} 
 \end{center}
 \end{figure*}

The large majority of reflection nebulae are diffuse. However, a few
are highly structured and display sharp edges. One such case,
Magakian~77 in the L1536 cloud in Taurus, surrounds a compact group of
stars known as the HP~Tau group, first noted by \citet{struve1962}
(Figure~\ref{fig:zaytsev-hanson}). \citet{cohen1979} found that
the group consists of five young stars, HP~Tau/G1, HP~Tau/G2, and
HP~Tau/G3 (hereafter G1, G2, and G3)\footnote{These systems are also known as CoKu~HP~Tau~G1/G2/G3} in addition to HP~Tau itself and
the young H$\alpha$ emission star Haro~6-28. Another 7 members have
subsequently been identified, see Appendix~A. The group, identified in
Figure~\ref{fig:2MASS+CC}, illuminates the large, sculpted reflection
nebula.  The brightest star near the center of the nebula is G2. Due to
their proximity, it is often assumed that HP Tau plus G1 and G2 form a
bound system, but we show here that it is not the case.

In this paper we study the structure and kinematics of the cloud
around and illuminated by the HP~Tau group.  We determine basic
properties of G2, and investigate the origin of the
unusual cloud structure in the context of the chaotic evolution of a
small multiple system. We argue that the properties of G2 and the
structure and kinematics of the surrounding gas can be understood if
G2 is a merger of two young low-mass stars, triggered by the breakup
of a multiple system about 5600 yr ago.

\vspace{0.3cm}

% \clearpage

% \begin{figure}
% \begin{center}
% \includegraphics[angle=0,width=0.99\columnwidth]{Fig1new.jpg} 
% \caption{The large highly structured reflection nebula Magakian~77 surrounds the little group of young stars including HP~Tau. The figure is almost 8 arcmin wide. This is an optical 13hr 20min exposure with a 1m telescope through B, G, R filters plus luminance. 
% Courtesy Alexandr Zaytsev and Mark Hanson.   
%[fig:zaytsev-hanson]  
% \label{fig:zaytsev-hanson}} 
% \end{center}
%  \end{figure}

\section{OBSERVATIONS} \label{sec:obs} 

We have observed G2 with a variety of telescopes and instruments to
better understand the nature of this peculiar star.

{\bf APO:}~~ 
Near-infrared observations were obtained with the NICFPS
camera on the Apache Point Observatory 3.5 meter telescope.  NICFPS
uses a 1024 $\times$ 1024 pixel Rockwell Hawaii 1-RG HgCdTe detector.
The pixel scale is 0.273\arcsec\ per pixel with a field of view
4.58\arcmin\ on a side.  Multiple images with 180 second exposures
were obtained in the 2.122 $\mu$m S(1) line of H$_2$ using a
narrow-band filter (FWHM= 0.4\% of the central wavelength).
% OMITTED:
% Separate sky frames were obtained using the same exposure time at a
% location 600\arcsec\ east.  A set of 5 dithered images were obtained
% both on-source and on the sky position.  A median-combined set of
% unregistered, mode-subtracted sky frames were used to form a master
% sky-frame that was subtracted from individual images.  Field stars
% were used to align the frames which were median-combined to produce
% the image.
Visual wavelength images were obtained using the ARCTIC CCD camera
which has an 8\arcmin\ field of view.  Images were obtained in the
broad-band SDSS g, r, and i filters and with 80\AA\ passband
narrow-band filters centered on the 6563\AA\ \Ha\ emission line and on
the $\lambda \lambda$ 6717 / 6731\AA\ [\Sii ] doublet.  Three
exposures were obtained in each filter with durations of 60 seconds
per exposure in the broad-band filters and 600 seconds in the
narrow-band filters.

{\bf SOAR:}~~ 
% OMITTED:
% We obtained optical and near-IR spectra of HP Tau, HP Tau/G1 and
% KPNO~Tau-15 during an available time slot on the engineering nights of
% UT2024-01-25 and 26.
The optical spectra were acquired the night of UT2024-01-26 with the
Goodman High Throughput Spectrograph \citep{clemens2004}, installed on the SOAR 4.1m telescope on
Cerro Pach\'on, Chile.
% OMITTED:
% The GHTS is a highly configurable imaging spectrograph that employs
% all-transmissive optics and Volume Phase Holographic Gratings, yieldis
% high throughput at low to moderate resolution spectroscopy over the
% 320-850 nm wavelength range.  We used the Goodman RED camera, which
% uses a deep-depletion CCD that provides extended sensitivity into the
% red end of the spectrum with minimal fringing.  The spatial scale of
% the Goodman detector is 0$\farcs15$/pixel; we used the instrument
% coupled with the Atmospheric Dispersion Corrector.
We used the 400 l/mm grating in its 400M1, and 400M2 $+$ GG 455
filter, preset modes, with 2 $\times$ 2 binning.  These two configurations
combined span the wavelength range $\sim 3600 \la \, \lambda \, \la
9000~$\AA. We used the $1\arcsec$ wide long slit, which results in a
FWHM resolution $=6.7$~{\AA } (equivalent to $\rm R=830$). We used
exposure times of 300s for all targets.  For HP Tau, we obtained $3
\times 300$s spectra in the 400M2 and $1\times 300$s spectrum in the
400M1 mode. For HP Tau/G1 and KPNO~Tau-15 we obtained a $1\times 300$s
spectrum in each 400M1 and 400M1 mode.
% OMITTED:
% The basic CCD reduction, up to and including the extraction of the 1-D
% spectra and wavelength calibration, was carried out using the Goodman
% Spectroscopic Pipeline (Torres-Robledo et al. 2020).  We did not
% perform flux calibration, since the main purpose of these spectra was
% to characterize each component: determining its spectral type,
% measuring line equivalent widths and determine its accretion status.
% We measured equivalent widths of spectral features using the {\sl
% splot} routine in IRAF.  The S/N ratio of the individual spectra was
% $\sim$30 at H$\alpha$ for the brighter HP Tau, sufficient for
% measuring equivalent widths down to $\sim 0.25$~{\AA }, at our
% spectral resolution of $\sim 7$~{\AA } FWHM.

The near-infrared spectra of HP Tau, HP Tau/G1, and KPNO~Tau-15 were
obtained on UT2024-01-25 using the TripleSpec 4.1 near-IR spectrograph
%(Schlawin2014)
at the SOAR telescope.
 We observed an A0 star at the same airmass as each of the three
targets, in order to perform the telluric correction.  The data were 
reduced with the Spextool IDL package \citep{Cushing2004}.

{\bf SpeX:}~~
Observations were carried out with the 3.2~m NASA Infrared Telescope
Facility (IRTF) on Maunakea, Hawaii with SpeX.  SpeX was used in the
short cross-dispersed mode, covering 0.7~\micron~ to 2.5~\micron.  We
used the 0\farcs5 wide slit, yielding a resolving power of R=1200.
Stars were nodded along the slit, with two exposures taken at each nod
position in the usual ABBA beam switch pattern.  An A0 telluric
standard star was observed at the same air mass as the target and
within a half-hour.  A thorium-argon lamp was observed for wavelength
calibration and a quartz lamp for flat fielding.  An arc/flat
calibration set was observed for each target/standard pair.  The data
were obtained on Sept 30, 2019, October 1, 2019, and Oct 15, 2019, and
were reduced using \emph{Spextool}. 
% \citep{2004PASP..116..362C}. 
JHKL aperture photometry was obtained on September 30, 2019.

{\bf iSHELL:}~~ 
We observed HP Tau G2 on October 23, 2020, using the
high-resolution near-infrared echelle spectrograph iSHELL on the IRTF
\citep{rayner2022}. The observations were performed in the K2 mode,
covering the range from 2.09 to 2.38 $\mu$m, using the 0.75 arcsecond
slit to achieve a spectral resolution of R $\sim$50,000.

{\bf JCMT:}~~
The J=3-2 \CO\ and $^{13}$CO at 345 and
330 GHz were observed with the 15-meter JCMT using the 16 channel HARP
receiver and the ACSIS correlator. The \CO\ observations cover an
850\arcsec\ (east to west) by 830\arcsec\ (north to south) field centered
on HP Tau. The bulk of the \CO\ data was obtained in the fall of
2019 with some additional data obtained in 2023. The $^{13}$CO 
observations cover an 800\arcsec\ (east to west) by 600\arcsec (north
to south) field. While some $^{13}$CO data was obtained in 2019 the
bulk was observed in 2023. The beam size is about 15\arcsec. The spectral
resolution of the \CO\ observations were 0.025 
%0.0246 
\kms\ while the spectral resolution of the $^{13}$CO observations
were 0.055 
%0.0554 
\kms. This intrinsic resolution has been smoothed as
specified in the maps and spectra presented. 

% C18O was observed together with \co\ - however has not been used in
% this paper due to the low signal to noise.\\ There was CO emission in
% the off position used for most of the 12CO observations. This has been
% corrected for by using frequency switch observations of the off
% positions.

\section{THE YOUNG STARS AROUND HP TAU} \label{sec:YSOs}

\begin{figure}
\begin{center}
\includegraphics[angle=0,width=0.77\columnwidth]{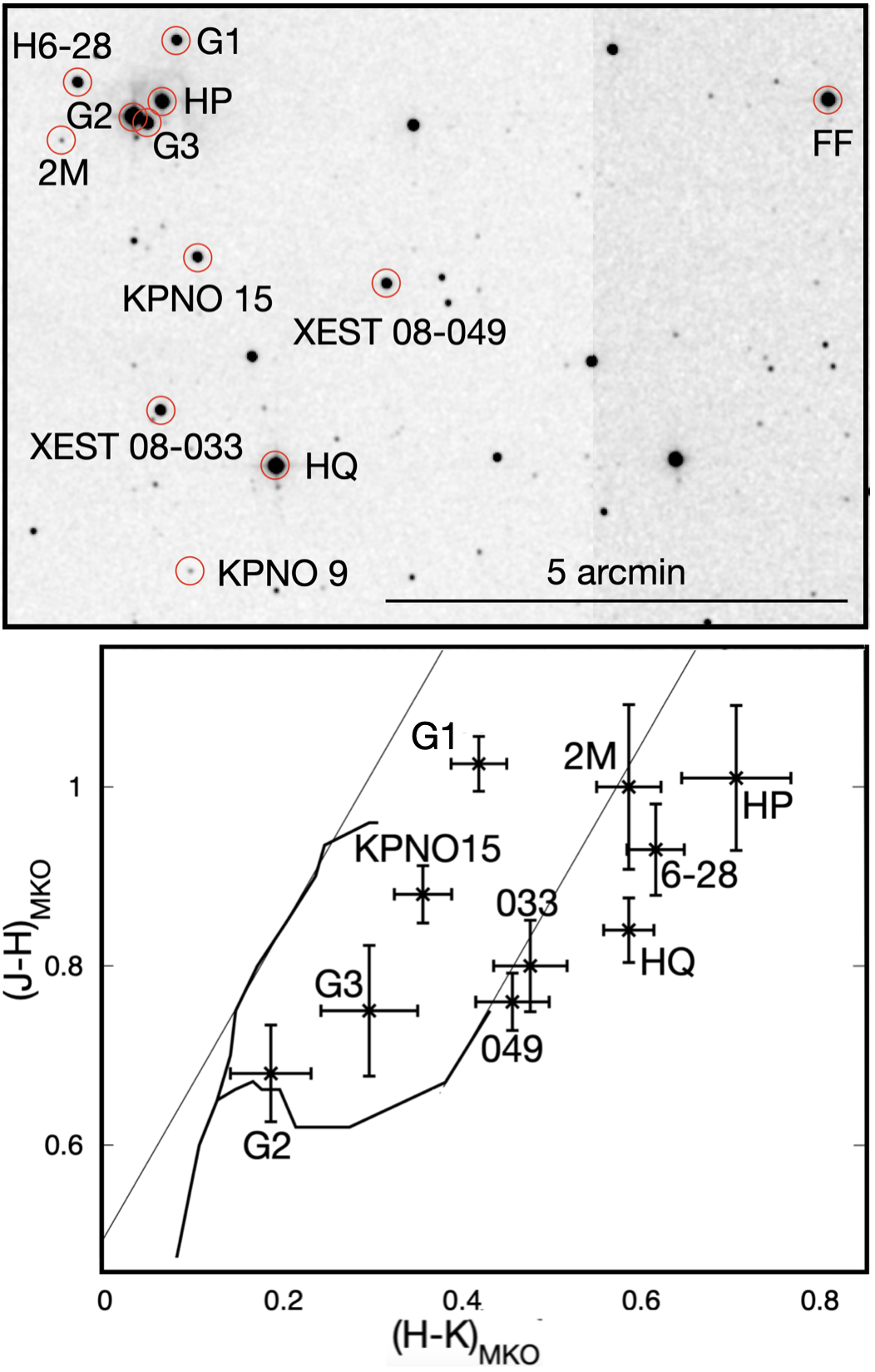}
\caption{ {\em (top)}  
The YSOs within the Magakian 77 reflection nebula are identified in
this 2MASS K-band image, and their properties are listed in
Table~1. H6-28 is Haro~6-28 and 2M is 2MASS J04355760+2253574. FF~Tau
is outside the field to the west. North is up and east is left.  {\em
(bottom)} A J-H vs H-K color-color diagram of ten of the sources in
the HP Tau multiple system. The dwarf and giant loci are from
 \citet{bessell1988} 
and the reddening lines from \citet{rieke1985}.  The
abbreviations refer to the names in Table~1.
%[fig:2MASS+CC]
\label{fig:2MASS+CC}} 
\end{center} 
\end{figure}

HP~Tau is a young H$\alpha$ emission star associated with the L1536
cloud, located in the southern part of the Taurus complex of star
forming clouds. Figure~\ref{fig:2MASS+CC}-top identifies the 12 known
young low-mass stars and brown dwarfs that are within 7.5 arcmin from
HP~Tau. This is one of the most compact aggregates in Taurus \citep{luhman2018}
and is part of a more scattered
distribution over $\sim$3$^\circ$ of more than 30 young stars that
share broadly common motions \citep{luhman2023}.
%The innermost part, consisting of HP Tau, G2, and G3, is often assumed
%to be bound due to their proximity.

% I have checked that the scale is correct

The properties of these 12 stars and brown dwarfs (some of which are
binaries) are listed in Table~1. G2, KPNO~Tau-15 (throughout the
following called KPNO~15), G1, and HP Tau are discussed in detail in
Section~\ref{sec:walkaway}. 
Additional information on the other 8 stars is given in Appendix~A. 
Most of the stars have near-infrared excesses or other evidence for circumstellar material.
%with some notable exceptions discussed in the following
(Figure~\ref{fig:2MASS+CC}-bottom).

%************************************
% SHORT VERSION OF TABLE 1
%************************************

%\begin{longrotatetable}
\begin{deluxetable*}{rlccrrrrrrl}
\tablecaption{Young Stars in the HP Tau Group} 
\tablecolumns{11}
%\tablenum{1}
\tablewidth{700pt}
\tablehead{
   \colhead{\#} &
   \colhead{Object} &
   \colhead{2MASS} &
%   \colhead{Sep.$^{a}$} &
   \colhead{SpT$^{a}$} &
   \colhead{G} &
   \colhead{J$^{b}$} &
   \colhead{H$^{b}$} &
   \colhead{K$^{b}$} &
%   \colhead{L'} &
%   \colhead{Dist.$^{d}$} &
   \colhead{pmra$^{c}$} &
   \colhead{pmdec$^{c}$} &
   \colhead{Notes}            \\
   \colhead{}         &
   \colhead{}         &
   \colhead{}         &
%   \colhead{[$''$]}   &
   \colhead{}         &
   \colhead{[mag]}    &
   \colhead{[mag]}    &
   \colhead{[mag]}    &
%   \colhead{[mag]}    &
   \colhead{[mag]}    &
%   \colhead{[pc]}     &
   \colhead{[mas/yr]} &
   \colhead{[mas/yr]} &
   \colhead{}
}
\startdata
\vspace{-0.17cm}
 1 & FF Tau         & J04352089+2254242  & K8       & 12.78 & 9.78 & 8.93 & 8.59  &  9.97 & -19.14    &  Binary 36 mas$^{d}$       \\
   &                &                    &          &       & 0.02 & 0.02 & 0.02  &  0.03 &   0.03    &                      \\
\vspace{-0.17cm}
 2 & XEST 08-049    & J04355286+2250585  & M4.25    & 14.72 &10.93 &10.17 & 9.71  & 10.55 & -16.93    &                      \\
   &                &                    &          &       & 0.03 & 0.01 & 0.04  &  0.06 &   0.04    &                      \\
\vspace{-0.17cm}
 3 & HQ Tau         & J04354733+2250216  & K2       & 11.53 & 8.85 & 8.01 & 7.42  & 10.87 & -18.98    &  Spec. Binary$^{e}$                    \\
   &                &                    &          &       & 0.03 & 0.02 & 0.02  &  0.24 &   0.18    &                      \\
\vspace{-0.17cm}
 4 & KPNO 15        & J04355109+2252401  & M2.75       & 14.73 &11.30 &10.42 &10.08  &  6.10 & -28.13    & Walkaway star         \\
   &                &                    &          &       & 0.03 & 0.01 & 0.03  &  0.04 &   0.03    &                      \\
\vspace{-0.17cm}
 5 & KPNO 9         & J04355143+2249119  & M8.5V    & 20.57 &15.48 &14.66 &14.19  & 11.12 & -15.04    & Brown dwarf          \\
   &                &                    &          &       & 0.04 & 0.04 & 0.05  &  1.59 &   1.10    &                      \\
\vspace{-0.17cm}
 6 & XEST 08-033    & J04354203+2252226  & M4.75    & 15.31 &11.25 &10.45 & 9.97  & 10.02 & -16.89    &                      \\
   &                &                    &          &       & 0.05 & 0.01 & 0.04  &  0.07 &   0.05    &                      \\
\vspace{-0.17cm}
 7 & HP Tau/G1      & J04355209+2255039  & M4.5     & 15.47 &11.31 &10.23 & 9.81  &  7.23 &  -7.65    & Walkaway star         \\
   &                &                    &          &       & 0.02 & 0.02 & 0.02  &  0.08 &   0.05    &                      \\
\vspace{-0.17cm}
 8 & HP Tau         & J04355277+2254231  & K4.0     & 13.03 & 9.36 & 8.35 & 7.64  &  9.04 & -13.85    & Binary 17 mas$^{f}$     \\
   &                &                    &          &       & 0.08 & 0.01 & 0.06  &  0.06 &   0.05    &                      \\
\vspace{-0.17cm}
 9 & HP Tau/G3      & J04355349+2254089  & M0.6     & 13.32 & 9.80 & 9.05 & 8.75  & 11.67 & -18.76    & Binary 31 mas$^{g}$        \\
   &                &                    &          &       & 0.07 & 0.02 & 0.05  &  0.05 &   0.03    &                      \\
\vspace{-0.17cm}
10 & HP Tau/G2      & J04355415+2254134  & G2       & 10.51 & 8.25 & 7.57 & 7.38  & 13.46 & -11.40    & Walkaway star         \\
   &                &                    &          &       & 0.05 & 0.02 & 0.04  &  0.07 &   0.05    &                      \\
\vspace{-0.17cm}
11 & Haro 6-28      & J04355684+2254360  & M3.1     & 14.95 &11.08 &10.15 & 9.53  & 10.56 & -17.13    & Binary 0.66''$^{h}$        \\
   &                &                    &          &       & 0.05 & 0.01 & 0.03  &  0.36 &   0.24    &                      \\
\vspace{-0.17cm}
12 &                & J04355760+2253574  & M5       & 19.91 &15.06 &14.06 &13.47  &       &           &                      \\
   &                &                    &          &       & 0.09 & 0.02 & 0.03  &       &           &                        
\enddata      
\tablecomments{
%(a): Separation in arcsec from HP~Tau. 
(a): \#1,3,8,9,10,11  \citet{herczeg2014} %Herczeg \& Hillenbrand 2014; 
\#2,6,7  \citet{luhman2009}       %Luhman et al. 2009, 
\#4 \citet{luhman2003}   %Luhman et al. 2003; 
\#5 \citet{briceno2002}            %Brice\~no et al. 2002; 
\#12  \citet{luhman2017}           %Luhman et al. 2017. 
(b): The JHK photometry is in the MKO system. The values for FF Tau, KPNO 9, and G1 are transformed from 2MASS using the transformations of 
\citet{connelley2007}        %Connelley et al. (2007). 
(c): Proper motions from Gaia DR3.
(d): \citet{simon1987}            %Simon et al. 1987.
(e): \citet{herczeg2014}  suggest that HQ~Tau is a spectroscopic binary, which is confirmed by Gaia's very large RUWE value.
(f): \citet{richichi1994}             %Richichi et al. 1994.
(g): \citet{rizzuto2020}              %Rizzuto et al. 2020.
(h): \citet{leinert1993}   }                              %Leinert et al. 1993.
\end{deluxetable*}
%\end{longrotatetable}

\begin{deluxetable*}{lrccccccr}
\tablecaption{Proper Motions within Rest Frame$^a$  and Radial and Space Velocities} 
\tablecolumns{7}
%\tablenum{1}
\tablewidth{700pt}
\tablehead{
   \colhead{Object}    &
   \colhead{Dist.}     &
   \colhead{pmra}      &
   \colhead{pmdec}     &
   \colhead{Tang.Vel.}      &      
   \colhead{PA}        &
   \colhead{Rad.Vel.$^b$}  &
   \colhead{Spa.Vel.$^c$}   &
   \colhead{RUWE$^d$}       \\  
   \colhead{}          &
   \colhead{[pc]}      &
   \colhead{[mas/yr]}  &
   \colhead{[mas/yr]}  &
   \colhead{[km/sec]}  &
   \colhead{[deg.]}  &
   \colhead{[km/sec]}    &
   \colhead{[km/sec]}  &
   \colhead{}
}
\startdata
%\vspace{-0.17cm}
FF Tau         & 161.0/0.8 & -0.55 & -1.03 & 0.9 & $\sim$208 & 16.9 & 0.9 & 1.48  \\
%               &   0.8 &       &       &     &              & 0.01 &  &   \\
%\vspace{-0.17cm}
XEST 08-049    & 162.3/1.2 & +0.03 & +1.18 & 0.9 & $\sim$89  & 17.7 & 1.3  & 1.24  \\
%               &   1.2 &       &       &     &              & 0.01 &  &   \\
%\vspace{-0.17cm}
HQ Tau         & 161.3/6.3 & +0.35 & -0.87 & 0.7 & $\sim$158 & 16.8 & 0.7 & 13.39  \\
%               &   6.3 &       &       &     &              & 0.01 &  &   \\
%\vspace{-0.17cm}
KPNO 15        & 160.5/0.8 & -4.42 &-10.02 & 8.4 & $\sim$204 & 12.9 & 9.3  & 1.10  \\
%               &   0.8 &       &       &     &              & 0.01 &  &   \\
%\vspace{-0.17cm}
KPNO 9         & 155.8/27 & +0.60 & +3.07 & 2.4 & $\sim$11   & .. &  & 1.20  \\
%               &  27.0 &       &       &     &              &  &  &   \\
%\vspace{-0.17cm}
XEST 08-033    & 164.2/1.6 & -0.50 & +1.22 & 1.0 & $\sim$337 & 16.8 & 1.0 & 1.06  \\
%               &   1.6 &       &       &     &              & 0.01 &  &   \\
%\vspace{-0.17cm}
HP Tau/G1      & 160.8/1.5 & -3.29 &+10.46 & 8.5 & $\sim$343 & 1.4 & 17.5 & 1.18  \\
%               &   1.5 &       &       &     &              & 6.95 &  &   \\
%\vspace{-0.17cm}
HP Tau         & 171.2/1.5 & -1.48 & +4.26 & 3.5 &  $\sim$341 & ..  &  & 2.77 \\
%               &   1.5 &       &       &     &               &  &  &  \\
%\vspace{-0.17cm}
HP Tau/G3      & 159.5/1.0 & +1.15 & -0.65 & 1.0 & $\sim$119 & .. &  &  2.43 \\
%               &   1.0 &       &       &     &              &  &  &   \\
%\vspace{-0.17cm}
HP Tau/G2      & 167.2/1.7 & +2.94 & +6.71 & 5.7 & $\sim$24  & 16.6 & 5.7 &  3.85 \\
%               &   1.7 &       &       &     &              & 1.0 &  &   \\
%\vspace{-0.17cm}
Haro 6-28      & 149.0/6.5 & -0.99 & +0.98 & 1.1 &  $\sim$315 & 16.6 & 1.1  & 10.04 \\
%               &   6.5 &       &       &     &               & 0.01  &  &  \\
%\vspace{-0.17cm}
% 2MASS         &       &   &  &   \\
%               &       &   &  &                                   
\enddata      
\tablecomments{(a): The rest frame is 10.52 mas/yr, -18.11 mas/yr}
(b): The radial velocities listed are heliocentric and are from APOGEE2 \citep{jonsson2020}, except for G2 which is from \citet{nguyen2012}. The uncertainty for G1 is larger, about $\pm$7~\kms. 
(c): Formal values of the space velocities are given in the rest frame of the cloud with uncertainties of a few \kms except for G1 (see note {\em b}). The heliocentric velocity of the cloud is 16.7~\kms.    
(d): Gaia Renormalized Unit Weight Error; a large value indicates the presence of a very close companion
\end{deluxetable*}
%\end{longrotatetable}

Table~1 lists the Gaia DR3 proper motions for confirmed members of the HP
Tau group.  We have selected the six stars that have distance
uncertainties of less than $\pm$1.6~pc: HP~Tau, G1, G3,
KPNO~15, XEST 08-049, and FF~Tau.  These data suggest a distance to
the HP~Tau group of 162.6$\pm$0.6~pc, which is consistent within the
errors to the VLBA distance for G2 of 161.2$\pm$0.9~pc by \citet{torres2009}
%Torres et al. (2009) 
and the VLBI distance for G2 of 162.7$\pm$0.8~pc by \citet{galli2018}.
%Galli et al. (2018). 
The weighted mean of these three distance estimates is
162.2$\pm$0.4~pc, which we adopt for the HP Tau group. 
% For a discussion of distances to various parts of the Taurus complex,
% see Galli et al. (2019).

A number of studies have explored the star formation history of
the Taurus young star population. \citet{krolikowski2021}
estimated the age of the young stars associated with the
L1536 cloud, which includes the HP~Tau group, as
2.01$^{+0.32}_{-0.29}$ Myr.  \citet{kerr2021}
%Kerr et al. (2021) 
determined
an age of 3.3$\pm$0.9 Myr, 
and finally \citet{luhman2023} suggested 2.1$^{+4.0}_{-1.4}$ Myr. 
 We here adopt the more precise age estimate of 2.0$\pm$0.3~Myr for the group.

% If we take a weighted mean of these independent age estimates we get
% an age for the group of 2.3$\pm$0.3~Myr.
 
\citet{rizzuto2020} used non-redundant aperture masking
interferometry with NIRC2 on the Keck-II telescope and presented
orbits and dynamical masses for the two binaries G3 (P$\sim$27~yr) and
FF~Tau (P$\sim$15~yr).   
They estimated ages for the primary components
of $\sim$3~Myr, but the secondaries have ages half of that.

%============================\\
%Definition of walkaway and walkaway stars:\\

%Blaauw 1956 (PASP 68,495) finds that among O-B0 stars 25% have high
%%space velocities ($>$30 {\kms}), this drops to 16$\%$ for B0.5 stars and 1$\%$%
%for stars B1V to B5V. Blaauw 1961 coined the term walkaway
%star. Eldridge+(2011) suggested to also include slower ejectees and
%suggested a lower limit of 5 {\kms}. de Mink+(2014) suggested to call
%these slower objects between 5 and 30 {\kms} for walkaway stars, see
%e.g. Schoettler+2020 for this use.\\
%==========================

\section{GAIA PROPER MOTIONS: WALKAWAY STARS} \label{sec:walkaway}

%\subsection{Nomenclature} \label{subsec:nomenclature}

%Blaauw (1961) 
\citet{blaauw1961} identified a number of high-velocity OB stars, which he
assumed had been ejected from their regions of birth. For those with a
velocity larger than 30~{\kms} he coined the term runaway stars.
Since then it has been estimated that roughly 10-30$\%$ of O~stars and
about 2-10$\%$ of B~stars have such high velocities, 
dependent on different definitions of runaway stars \citep[e.g.,][]{renzo2019}. It was predicted
that young low-mass runaways would also be discovered 
{\citep{sterzik1995,schoettler2019}
%(Sterzik \& Durisen 1995, Schoettler et al. 2019), 
but only recently have the first of this population been identified, 
\citep[e.g.,][]{mcbride2019,farias2020,platais2020}.
%McBride \& Kounkel 2019, Farias et al. 2020, Platais et al. 2020)

Eldridge et al. (2011) analyzed the distribution of runaway velocities
for massive stars and proposed to include also slower ejectees with a
lower limit of 5~{\kms}.  It has been suggested to use the term
walkaway stars for velocities between 5 and 30~{\kms}
\citep{demink2014}.

%\subsection{Proper Motions of the HP~Tau Stellar Group} \label{subsec:hptaugroup}

%Luhman (2018) 
\citet{luhman2018} studied the motion of the Taurus stars with Gaia
DR2 data, and made the remarkable discovery that three stars in the HP
Tau group, KPNO~15, G2, and G1 (the latter labeled in the Luhman paper
as 2MASS J04355209+2255039) are high proper motion stars, and he
suggested that G1 and KPNO~15 were once part of a multiple system that
broke up about 7,200~yr ago. These conclusions were based on Gaia~DR2
data, and we have revisited the proper motions using the newer Gaia
DR3 data. The improvement from DR2 to DR3 is significant. But an
uncertainty for interpretation comes from transforming the observed
proper motions to the peculiar velocity frame of the little compact
group of YSOs around HP~Tau. The reference frame obviously does not
affect the relative motions of the stars, but the {\em relative} sizes
of the vectors depend on the reference frame, and thus will impact 
stellar masses derived from momentum conservation.
\citet{luhman2023} obtained the mean motion of many stars in the whole L1536 
complex, finding a group velocity of 10.3, -17.0 mas/yr in $\alpha$ and
$\delta$. In Appendix~B we discuss this further.  
% In a more recent paper, Luhman (2023) has used Gaia DR3 data to study
% the motions of stars in the Taurus clouds, and he derives a group
% motion of 10.3 mas, -17.0 mas for the entire population of the L1536
% complex, and we note that this motion is very close to the motion we
% have derived.
%Kerr et al. (2021) have also used Gaia DR3 to measure a group
%motion for the L1536 YSO population of 10.7 mas, -17.4 mas.

We have chosen to use only the YSOs immediately around HP~Tau within a
radius of 7.5 arcmin. These are the 12 stars listed in Table~1. Of the
11 members with Gaia measurements, four stars - G1, G2, KPNO~15, and
HP~Tau - have higher proper motions and are known or suspected
walkaway stars and as such are not included in the determination of
the rest frame. 
% Four stars (XEST 08-033, XEST 08-049, G3, and FF~Tau) have high
% precision proper motions, with the remaining three having a bit larger
% errors.
We calculate a weighted mean proper motion of the remaining 
7 stars of 10.52, -18.11 mas/yr. Table~2 lists the motions of the
individual members of the HP~Tau group when corrected for this group
motion.

% The new Gaia DR3 data show that G1 and KPNO~15 indeed were in each
% others vicinity about 7000~yr ago, as noted by Luhman. The problem
% with a pairing of G1 and KPNO~15 is that in the rest frame of the
% cloud there is an angle of $\sim$40$^\circ$ between their two lines of travel,
% much more than can be accounted for by the uncertainties in the proper
% motions. Momentum conservation requires the two vectors to lie
% precisely along the same line.

\begin{figure}
\begin{center}
\includegraphics[angle=0,width=0.7\columnwidth]{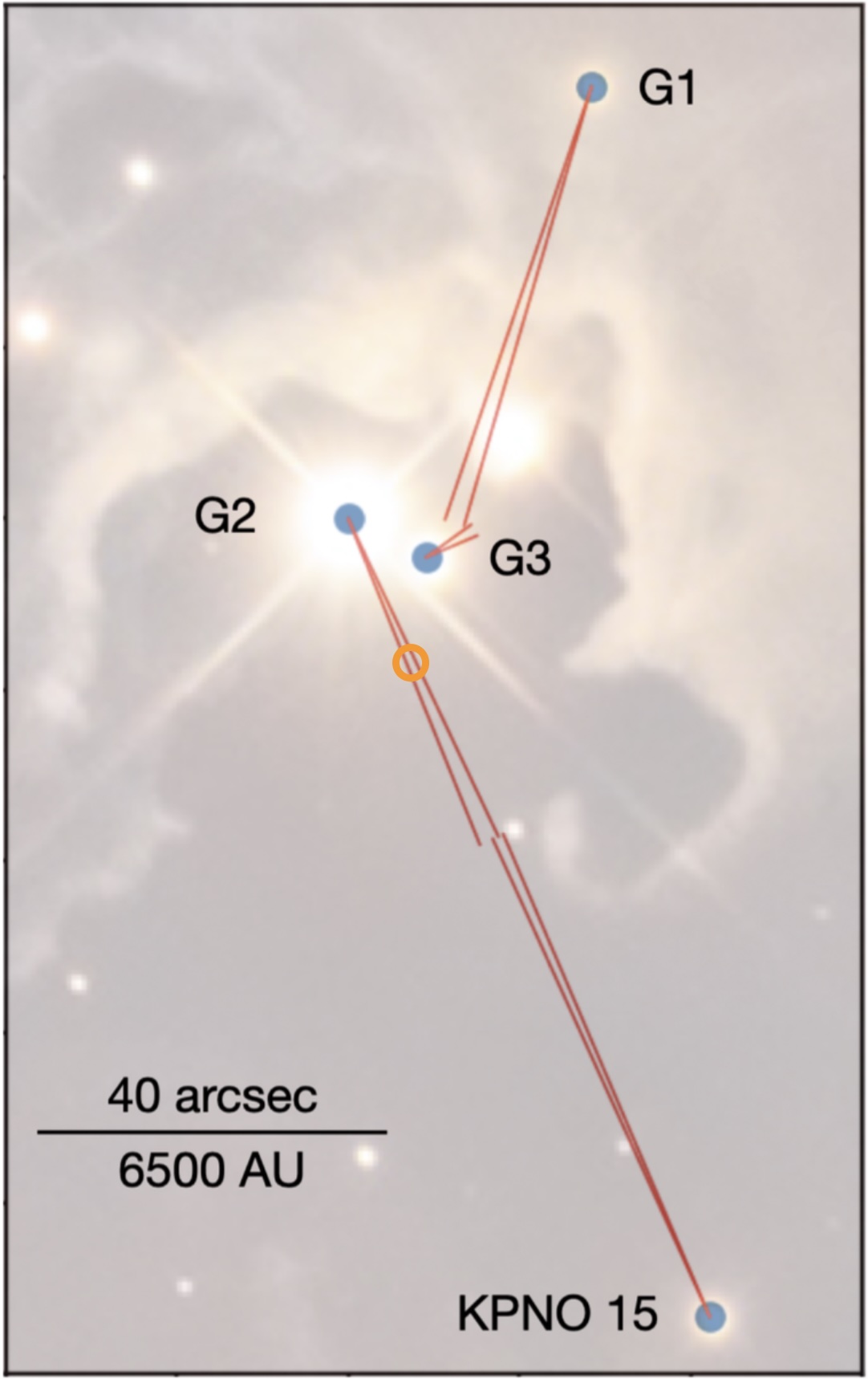} 
\caption{The relative proper motions of the stars G2, KPNO~15, 
G1, and G3 in a reference system based on 7 stars from the surrounding
group. G2 and KPNO~15 were within one arcsecond of each other about
5600 yr ago.  G1 is another walkaway star,
possibly related to the binary G3.  The orange circle marks the approximate
center of the inner rim of the cavity and the presumed site of the merger, 
suggesting it happened roughly 2000~yr ago.
\label{fig:propermotions}}
\end{center}
\end{figure}

%\subsection{Scenario 1} 

We note that in the past G2 and KPNO~15 were once close to each
other. In the reference frame we determined, the vectors of G2 and
KPNO~15 are pointing in opposite directions to within 0.6$^{\circ}$,
as must be the case if one recoiled from the other, see
Figure~\ref{fig:propermotions}.  With the current (DR3) accuracy of
the Gaia astrometry the two stars were within 1.1\arcsec\ of each
other $\sim$5600~yr ago (at the 2000-position 4:35:52.2 +22:53:36).
If they broke up it implies that one of the two is a binary.

% This is correct, but not important:
% It should be noted that the 0.6$^{\circ}$ is not dependent on the
% choice of rest frame, but results from a direct subtraction of the two
% stars.

The tangential velocity of G2 (5.65~{\kms}) 
% THIS IS THE FINAL CORRECT VALUE
is less than
that of KPNO~15 (8.44~{\kms}) 
% THIS IS ALSO THE FINAL CORRECT VALUE
by a factor of $\sim$0.67 so from momentum conservation it follows
that in this reference frame G2 must be more massive than KPNO~15 by a
factor of $\sim$1.5, which is consistent with G2 having an earlier
spectral type than KPNO~15.

\citet{torres2009} measured a heliocentric radial velocity for G2 of
+17.7$\pm$1.0~ {\kms}.  \citet{nguyen2012} found +16.6$\pm$1.0~{\kms}, and 
\citet{rivera2015} determined 16.6$\pm$1.7~{\kms}.  Gaia DR3 lists a velocity 
of 23.0~{\kms}, but without an error estimate. The weighted mean of
the Torres, Nguyen, and Rivera values is
17.0$\pm$0.7~{\kms}. \citet{jonsson2020} included KPNO~15 in their
catalog of radial velocities from the APOGEE2 DR16 survey and found a
radial velocity of 12.86~{\kms}.  The heliocentric radial velocity
of the cloud from CO data is 16.7~{\kms} (see Section~\ref{sec:mm}) 
and if we allow an error not exceeding 0.7~{\kms} for each of these
three numbers, we find that - 
within the resulting range of possible angles -  
 for an assumed angle  to the
plane of the sky of $\sim$16$^\circ$, the two stars would move precisely in opposite directions,
with G2 being redshifted and KPNO~15 blueshifted.

This scenario does not explain the large velocity of the walkaway star G1.
The star has a tangential velocity of 8.5~{\kms} and Gaia
DR3 lists a heliocentric radial velocity of 1.4$\pm$6.9~{\kms}. Given the
heliocentric velocity of the gas of 16.7~{\kms} it follows that G1 has
a space velocity of about 17.5$\pm$7~{\kms}.

To explain this peculiar motion we would expect to find a counterpart
moving in the opposite direction. However, we have searched the Gaia
DR3 catalog for a walkaway star moving precisely away from G1, but
within errors no plausible candidate was found.

Within errors, however, G1 is moving straight away from G3. Since it
is the only young star in the path of G1, it is a possible candidate
for pairing with G1, see Figure~\ref{fig:propermotions}.
The problem is that the proper motions of G3 are
at a $\sim$45 degree angle to the motion of G1. But G3 is a 31~mas binary
\citep{rizzuto2020}, which could easily affect the Gaia proper motions. 
The angular separation of G1 and G3 is 58 arcsec, and at the relative
speeds of G1 and G3 this implies a travel time of roughly 4700 yr. So
if G1 was once bound to G3, then they broke apart less than 1000 yr
after the breakup of G2 and KPNO~15.

The unsatisfactory aspect of this scenario is that it requires two
nearby but apparently independent triple systems to each break apart
within a few thousand years, which is statistically unlikely, unless
there is some connection between the two systems. We have instead considered
whether the three stars could have originated from the breakup of a single
multiple system. However, even with a generous interpretation of the
Gaia errors, there is no solution where the three stars were at the same
point at the same time and maintained momentum conservation.

% If one accepts a 20\% error on the G2 vector, a 10\% error on KPNO~15,
% and a 5\% error on G1, then there is a solution, but when one applies
% realistic estimates of the masses for the three stars, then momentum
% is not conserved.

In Section~\ref{subsec:G2merger} we discuss a possible dynamical
history of the small aggregate of young stars surrounding G2.

\section{THE WALKAWAY STARS} \label{sec:walkaways}

We here discuss what is known about each of the three stars with anomalous velocities. 

\subsection{The Walkaway Star G2} \label{subsec:G2}

G2 is generally classified as a G2 star 
\citep[e.g.,][]{hartigan1994, herczeg2014}, but sometimes also as G0  
\citep[e.g.,][]{cohen1979, luhman2010}.
% The energy distribution peaks at longer wavelengths, around 8500~\AA,
% due to extinction.
A low-resolution spectrum from LAMOST shows a G-type spectrum, but peaking
around 8000~\AA, indicating extinction. Indeed 
%Herczeg \& Hillenbrand (2014) 
\citet{herczeg2014} determine an A$_V$ = 2.55$\pm$0.2.
%We have compared the near-infrared
%spectrum of G2 with a similar spectrum of the G2~IV star HD~126868
%(Figure~\ref{fig:ir-spectrum-G2-mosaic}). Assuming that HD~126868 has
%no extinction and that G2 has no veiling we find that G2 is obscured
%by an A$_V$ = 2.0$\pm$0.4, which is close to the value of 
G2 is not a known H$\alpha$ emission star, but a spectrum from George
Herbig's archives reveals a very wide $\sim$1200~\kms absorption line
overlaid with a double-peaked emission
(Figure~\ref{fig:herbig-spectrum}).  Given the absence of detectable
circumstellar material (see below), the absorption component is
unlikely to be due to accretion. From comparison with the star FK~Com
(see Section~{7.3.2}) 
% and Appendix~E 
it is well explained as the
result of a very strong and broad H$\alpha$ line ($\sim$1200~\kms )
overlaid with a double-peaked emission line whose blue and red
components vary in anti-phase, leading to the central absorption to
shift back and forth.

% THIS IS NOT AN INVERSE P CYGNI, THE ABSORPTION IS REDSHIFTED _RELATIVE_ TO THE CENTER OF HALPHA at a velocity of 80~\kms, perhaps spectrum not fully wavelength calibrated.

A 2.0 - 2.4~$\mu$m SpeX spectrum of the star shows no evidence for a
near-infrared excess that could indicate the presence of a circumstellar
disk. Figure~\ref{fig:ir-spectrum-G2-mosaic} shows a comparison with
the G2IV standard star HD~126868 to which an extinction A$_V$=2.0 has
been applied.

\begin{figure}
 \begin{center}
\includegraphics[angle=0,width=0.45\columnwidth]{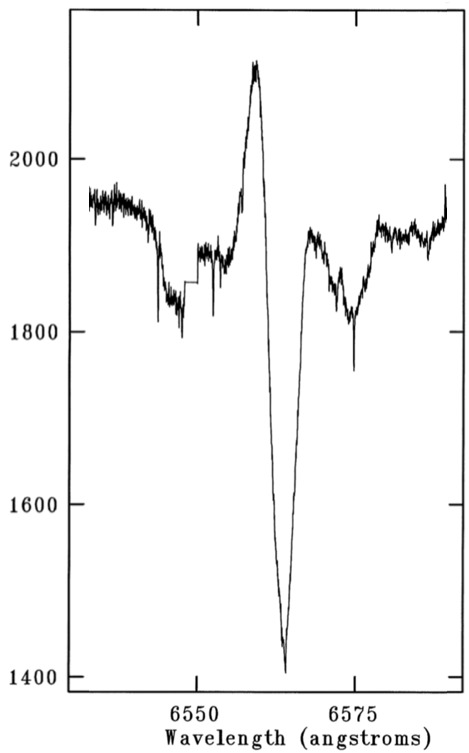}
%\vspace{0.5cm} 
\caption{HIRES spectrum taken on Dec 21, 2005 by George Herbig showing
the H$\alpha$ emission line profile of G2 overlaid a very broad absorption line with a width of $\sim$1200~\kms . The central absorption is likely formed by 
the superposition of two blue and red-shifted emission peaks that vary in anti-phase, causing the central dip to move back and forth, see Section~\ref{subsec:G2}.
%[fig:herbig-spectrum]
\label{fig:herbig-spectrum}}
\end{center}
\end{figure}

\begin{figure} 
\begin{center}
\includegraphics[angle=0,width=0.77\columnwidth]{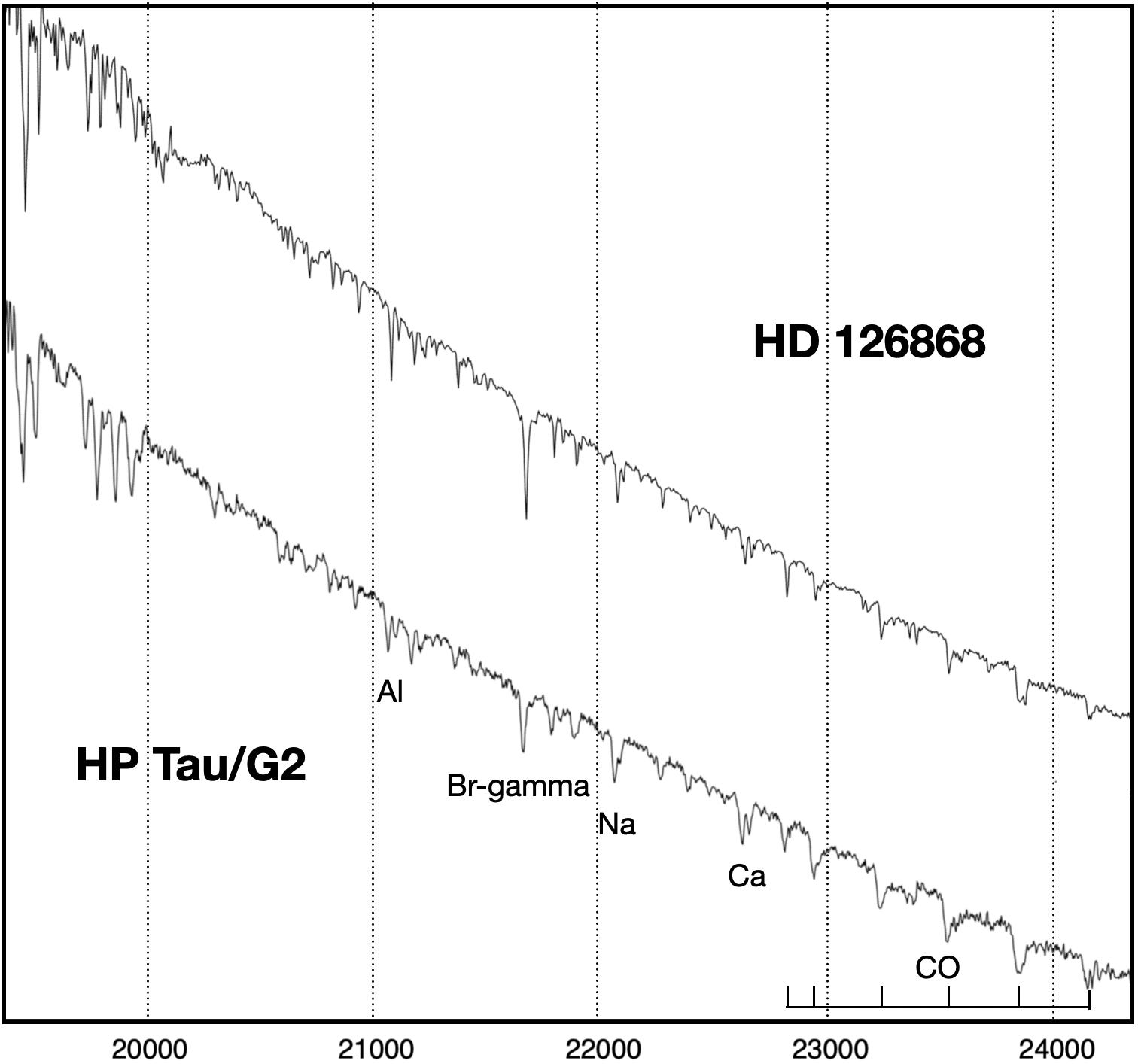}
\caption{Comparison of the 2.0 - 2.4 $\mu$m SpeX spectrum of G2 and the G2IV
standard star HD126868 with an extinction of A$_V$=2.0 applied. The spectra
confirm G2's spectral type as G2 with very minor differences apart
from the faster rotation of G2. The abscissa is wavelength in Angstroms.
%and the ordinate is flux in ergs s$^{-1}$cm$^{-2}$A$^{-1}$.
% [fig:ir-spectrum-G2-mosaic]  
\label{fig:ir-spectrum-G2-mosaic}} 
\end{center}
\end{figure}

\begin{figure}
\begin{center}
\includegraphics[angle=0,width=0.77\columnwidth]{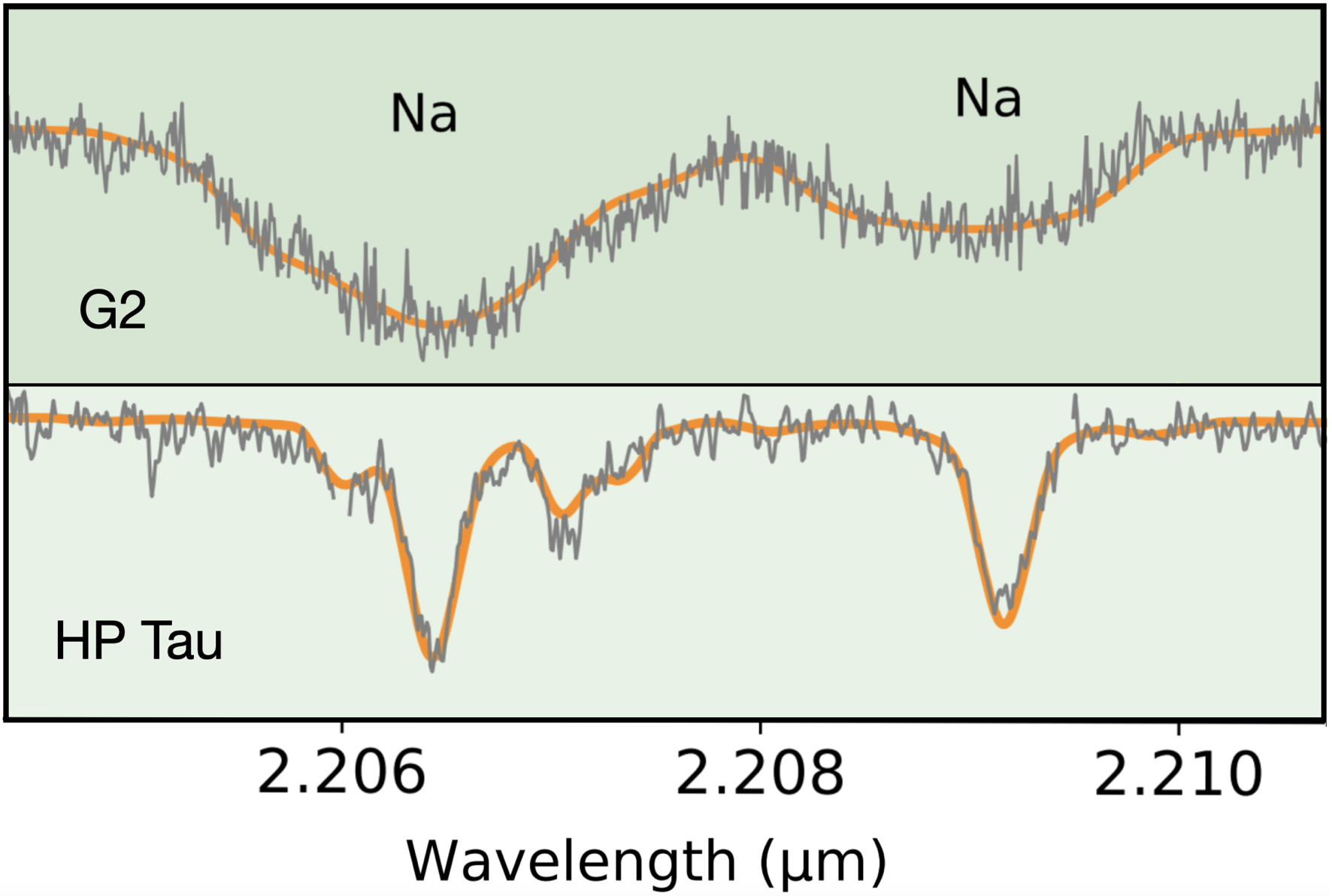}
 \caption{Part of an iSHELL spectrum of HP~Tau/G2 and HP~Tau with \vsini of $\sim$130~\kms\ and 17~\kms , respectively.
The rotational line broadening is evident. The spectra have been shifted to
the same velocity. 
%[fig:iSHELL]
\label{fig:iSHELL}} 
\end{center}
\end{figure}
%Figure A7

Our near-infrared photometry also shows that G2 has no near-infrared
excess (Figure~\ref{fig:2MASS+CC}) and the energy distribution of G2
based on WISE photometry
% seen in Figure~\ref{fig:G2-energydistribution}
shows a clean Planck curve at least out to about 22~$\mu$m. 
A small excess is seen with Spitzer at 70 and 160 $\mu$m,
but at those wavelengths the point spread function is so large that it includes the
T~Tauri stars G3 and HP~Tau which are only 10$''$ and 21$''$
away, respectively.
%and a small excess only thereafter, indicating that G2 has
%circumstellar material in the form of a transition disk (van der Marel 2023).
%Garufi et al. (2024) 
\citet{garufi2024} puts an upper limit on any disk of G2 of at most
1~M$_\earth$ of dust. In Section~\ref{sec:discussion} we discuss a
mechanism that would remove any remnant disk material.}

% {\color{red} while L. Stapper et al. (in prep.) find
% 1.2~M$_\earth$. We conclude that G2 no longer has any significant
% circumstellar disk. 

% {\color{green} CHECK IF GARUFI USED ALMA ACA}

We have obtained a high resolution spectrum in the wavelength range
from 2.107 to 2.304~$\mu$m with iSHELL on the IRTF 3m telescope, part
of which is seen in Figure~\ref{fig:iSHELL}, and as is obvious from
the line profiles, G2 is a fast rotator. We measure a \vsini of
$\sim$130~{\kms}, (see Section~\ref{subsec:Teff}), which confirms the
extreme rotational velocity of $\sim$100$\pm$20~{\kms} measured by
%Hartmann et al. (1986) 
\citet{hartmann1986} and 127$\pm$4~{\kms} by 
%Nguyen et al. (2012)  
\citet{nguyen2012}. 
This makes G2 one of the faster known rotating low-mass young stars,
\citep[e.g.,][]{malo2014}.

G2 is an X-ray source: XEST 08-051 \citep{guedel2007}, and
\citet{bieging1984} 
found it to be a point-like non-thermal radio source,
with significant variability at 15 and 5~GHz 
% (Cohen \& Bieging 1986, O'Neal et al. 1990).
\citep{cohen1986, oneal1990}. 
This is unusual, since radio emission
from low-mass young stars is generally found to be thermal 
% (e.g., Anglada 1996) 
\citep[e.g.,][]{anglada1996}
and indicative of partially ionized outflows \citep{reynolds1986}.
%(Reynolds 1986). 
A few objects are known to emit synchrotron radiation, 
%(e.g., Rodriguez et al. 1989, Ray et al. 1997, Purser et al. 2018), 
\citep[e.g.,][]{rodriguez1989, ray1997, purser2018},
which is ascribed to diffusive shock acceleration of electrons to
mildly relativistic velocities in a highly supersonic flow of
turbulent magnetized plasma where it interacts with the ambient medium
%(Henriksen et al. 1991, Mohan et al. 2022).
\citep[e.g.,][]{henriksen1991, mohan2022}.

\begin{figure} 
\begin{center}
\includegraphics[angle=0,width=0.6\columnwidth]{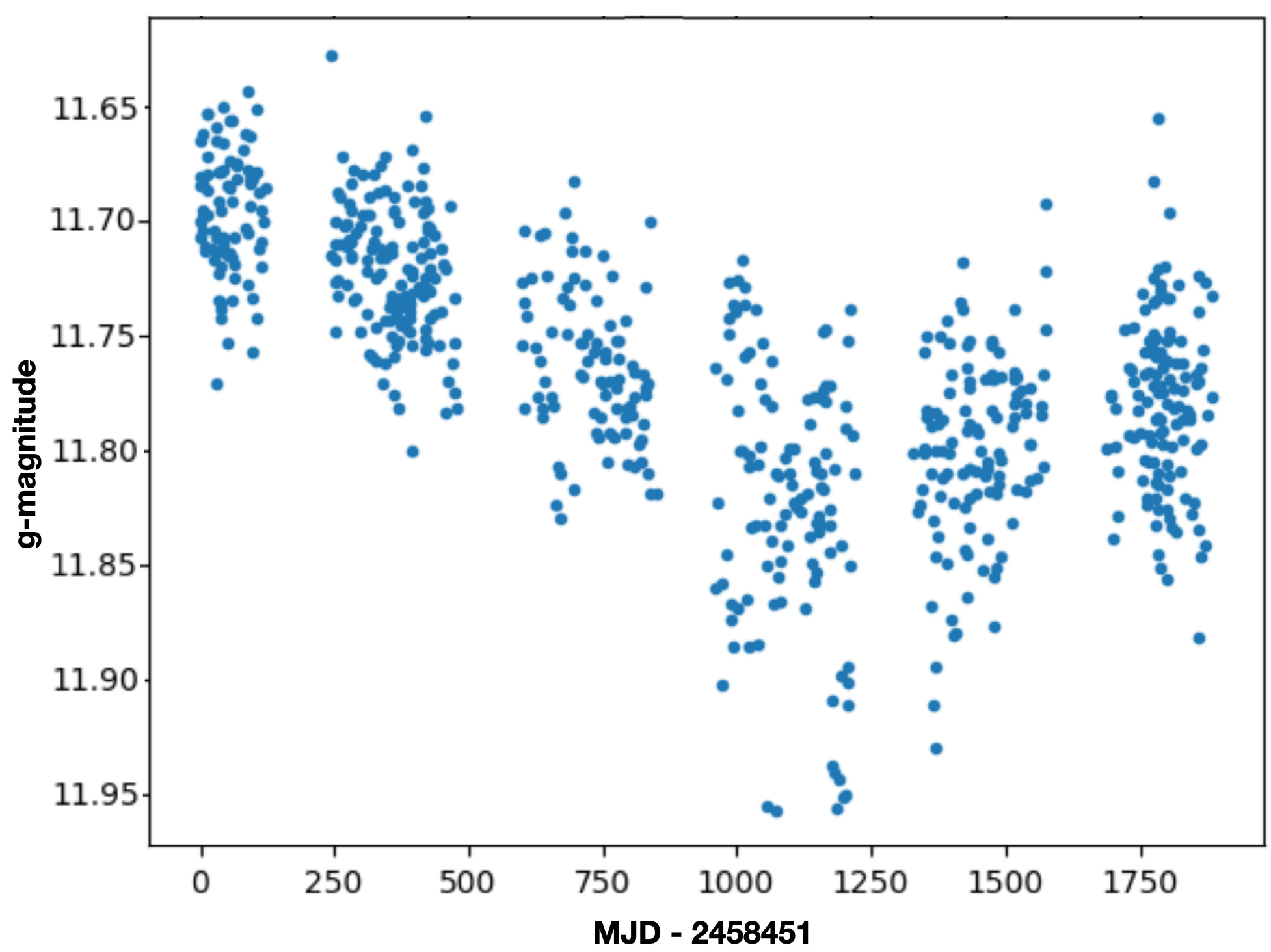}
\caption{The {\em g} light curve of HP~Tau~G2 from Nov 28, 2018 to
Jan 16, 2024 as observed by the ASAS-SN survey (Shappee et al. 2014,
 Kochanek et al. 2017).  
%[fig:ASASSN] 
\label{fig:ASASSN}}
\end{center}
\end{figure}

G2 is a variable star named V1025~Tau. The ASAS-SN {\em g}-light curve
from Nov 28, 2018 to Jan 16, 2024 in Figure~\ref{fig:ASASSN} shows a
low-amplitude irregular variability within a range of 0.5 mag. based
on {\em g}-photometry from multiple sites. There appears to be a long
term undulation in the light curve, and {\em if} it is periodic then
the period would be at least 6.5 years.
%\footnote{However, we note that
%if the long-term variability describes a 6.5 year activity cycle, then
%the known relation between cycle period and rotation period 
%\citep[e.g.,][]{mittag2023} that - for main sequence stars - shows longer
%activity cycles corresponding to longer rotation periods, is violated
%by G2, which rotates at least ten times faster than expected.}  
This might then represent the activity cycle of the star.  The ASAS-SN
point spread function is about 15~arcsec, and as previously noted G3 is
10~arcsec away, hence any variability of G3 could in principle affect
the photometry. However, G3 is more than 3~1/2 magnitudes fainter in
V, so the impact would be minimal.

% ** OBS: check paper by Gaidos on ASASSN and ATLAS photometry

%G2:  26 days from April 16, 2017 to May 12, 2017.\\
%No further TESS observations have been done. 

%Figure A5a

\begin{figure} 
\begin{center}
\includegraphics[angle=0,width=0.9\columnwidth]{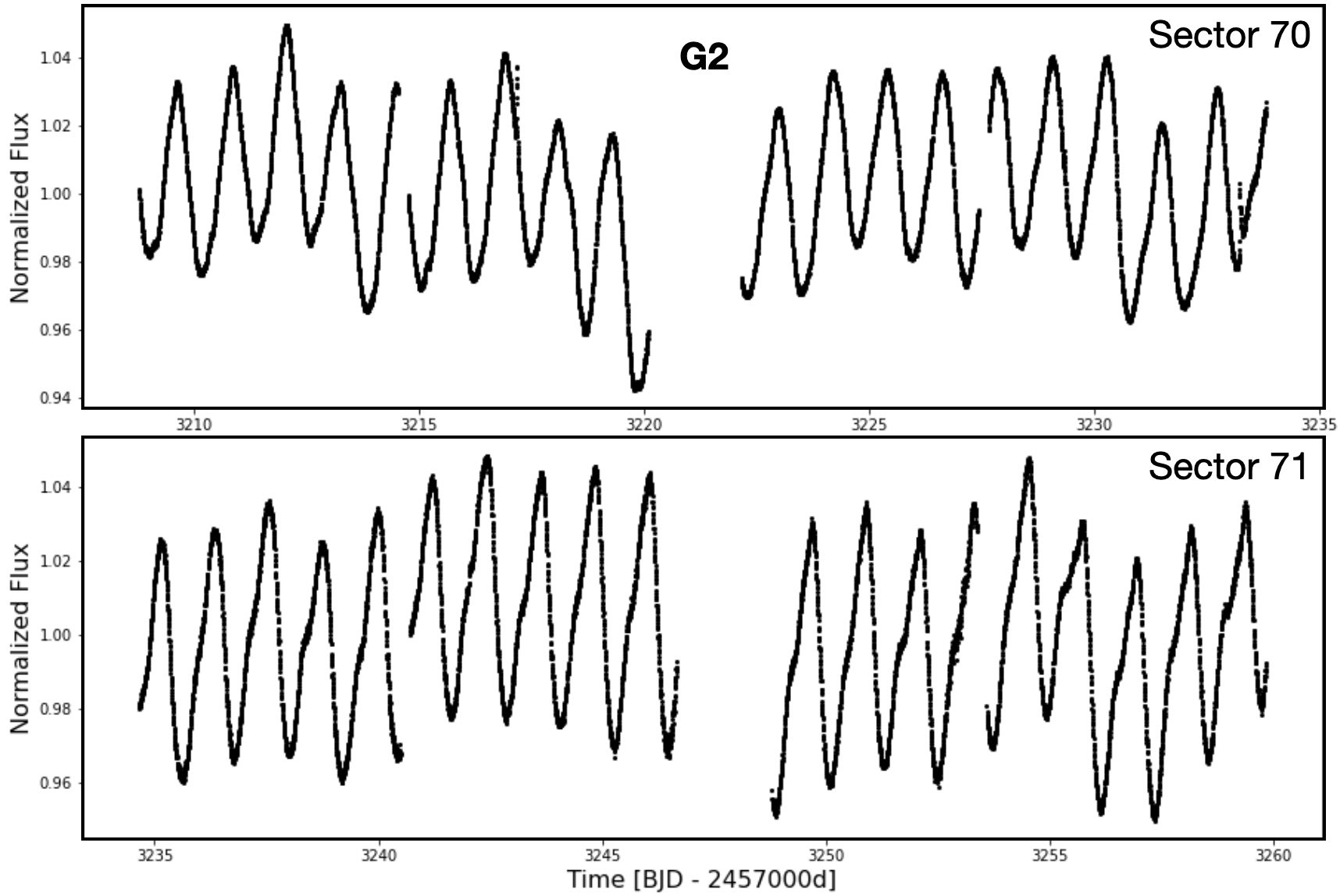}
\caption{G2 was observed by TESS in Sectors 70 and 71 in October 2023. The observations confirm the well established period of 1.2 days. The variable asymmetries indicate the presence of large star spots. A couple of flares are detected in Sector 70.  
%[fig:TESS-G2] 
\label{fig:TESS-G2}}
\end{center}
\end{figure}
%Figure A5a

% \begin{figure}
% \begin{center}
% \includegraphics[angle=0,width=0.99\columnwidth]{Fig7new.jpg} 
% \caption{The light curve of HP~Tau~G2 as observed by the Kepler K2 mission. The upper panel shows the entire light curve observed between March 9 and June 1, 2017, and the lower panel shows 26 days covering the transition phase when spot group~A (blue dots) shrinks and the absorption from spot group~B (yellow dots) becomes dominant. 
% %[fig:K2-G2]
% \label{fig:K2-G2}} 
% \end{center}
% \end{figure}

G2 has a well determined and stable rotational period of 1.2d which
was first measured by \citet{vrba1989} and later confirmed by
\citet{bouvier1995}, \citet{rebull2004}, and \citet{guedel2007}.
A light curve is shown in Figure~\ref{fig:TESS-G2}, which shows subtle
changes ascribed to star spots. \citet{rebull2020} and \citet{cody2022} 
have used K2 data to measure two well-determined simultaneous periods, 
P1=1.1978d and 1.222d, with P2/P1=1.02, which we discuss in 
Appendix~C.

G2 was observed in a high-contrast polarization imaging study by
% Garufi et al. (2024) 
\citet{garufi2024} and it was the only star in their sample of 41
YSOs in Taurus for which a disk was not observed. 
%Duvert et al. (2000)
\citet{duvert2000} observed it with the IRAM Plateau de Bure interferometer in the
1.3~$\mu$m continuum, and detected no emission in a 3\arcsec\ beam
with a 3$\sigma$ detection limit of $\sim$2.5~mJy. Unfortunately, at
the time of writing G2 has not been observed with ALMA. So there is no
direct evidence for any circumstellar material.

\begin{figure}
\begin{center} 
\includegraphics[angle=0,width=0.99\columnwidth]{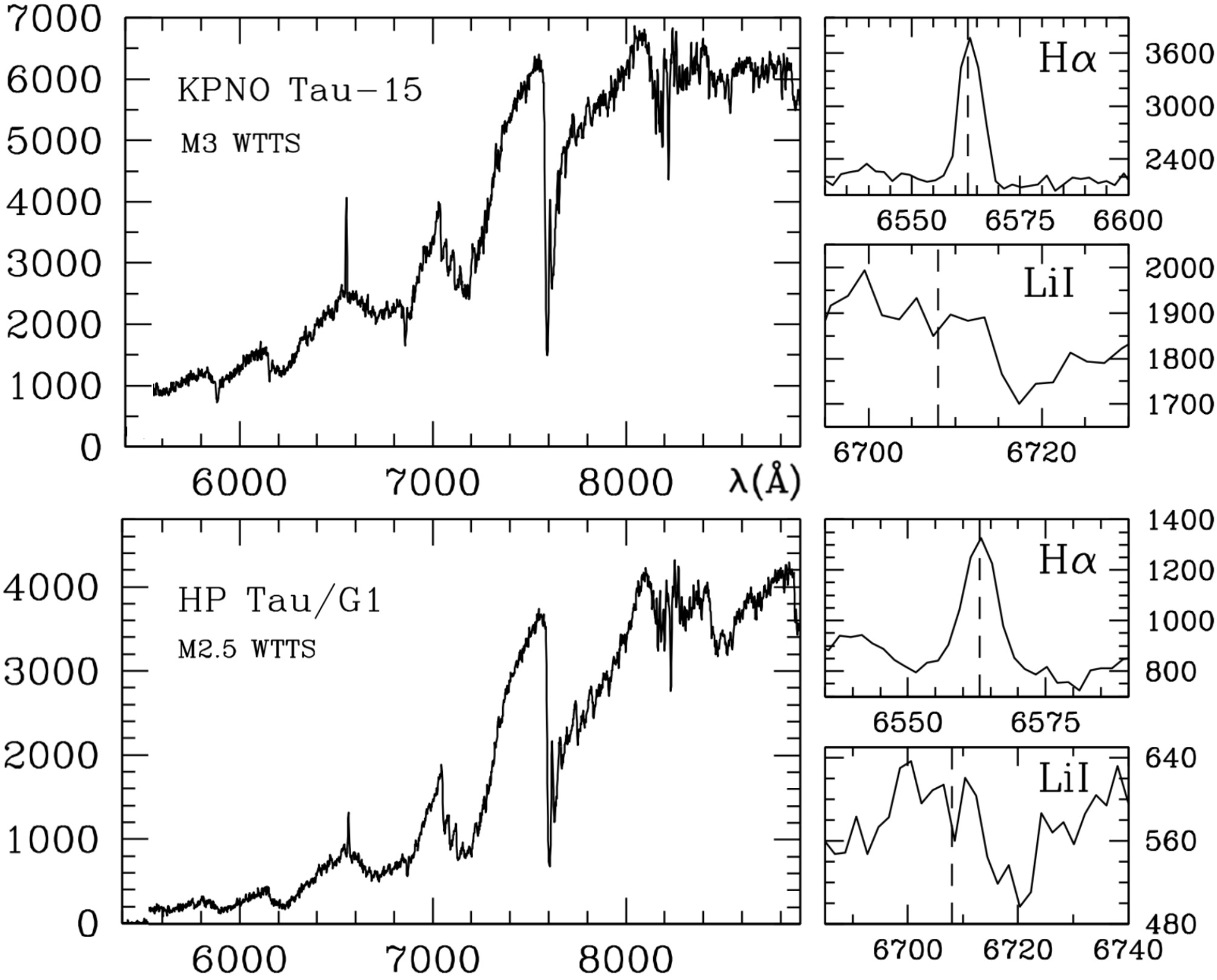} 
\caption{Low-resolution optical spectra of KPNO~15 and G1 show that both stars are weakline T~Tauri stars with lithium in their spectra. The axes are counts against wavelength.
% [fig:opt-spectra]  
\label{fig:opt-spectra}} 
\end{center}
\end{figure}

\begin{figure}
\begin{center} 
\includegraphics[angle=0,width=0.6\columnwidth]{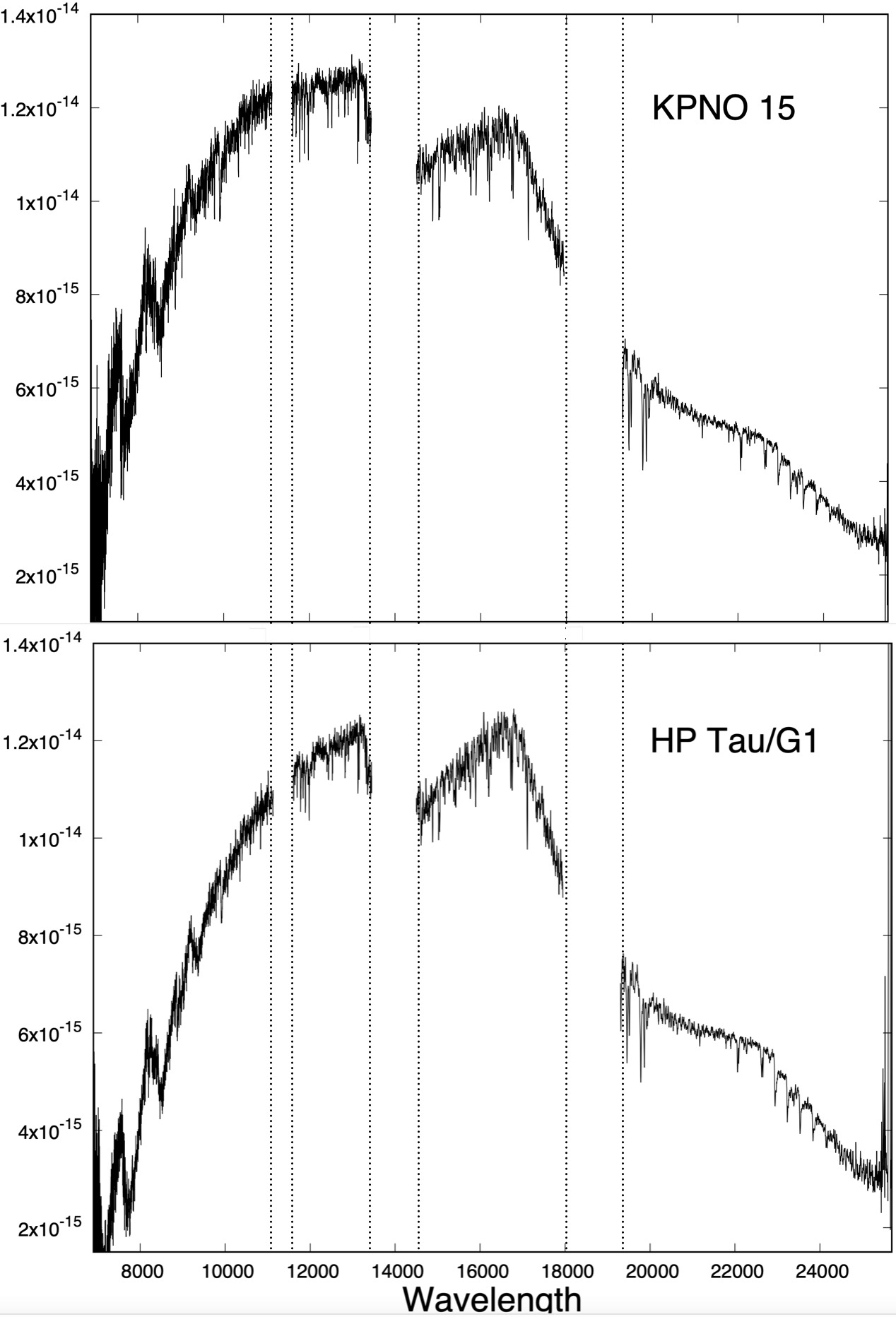} 
\caption{Low-resolution near-infrared spectra of  KPNO~15 and G1 which show that they have very late spectral types and thus very low masses.  
% [fig:ir-spectra]  
\label{fig:ir-spectra}} 
\end{center}
\end{figure}

%\subsection{{\color{red} The Search for a Companion to G2} \label{subsec:G2binary}}

\citet{richichi1994} used lunar occultations to search for a companion
to G2, and put an upper separation limit (perpendicular to the lunar
horizon) of about 1.6 mas (about 0.2~AU) on any equally bright
companion.  \citet{kraus2011} used non-redundant aperture
masking at the Keck telescope, and the observations are sensitive to
equal brightness companions at separations less than $\sim$10 mas
($\sim$2~AU) and can detect companions 5-6 magnitudes fainter at
separations larger than $\sim$40 mas ($\sim$8~AU). The observations
were able to detect companions out to a distance of 0.36~arcsec, but
none were found. Kraus et al. additionally used AO observations and
also found no companion in the range $\sim$0.3~-~2 arcsec and flux
ratios $\leq$2.5.

We have examined archival HST images of G2 and find a very faint star
2.5~arcsec ESE of G2, which we confirmed on unpublished
archival images from SPHERE (PI: Ginski). Nothing is known about it
and while in principle it could be a companion to G2, it is more
likely to be a background star because G2 is located in a cavity in
the cloud where other background stars can be seen, see
Figure~\ref{fig:duchene}. Future proper motion measurements will
clarify this.

% These observations still leave room for G2 to be a close spectroscopic
% binary or a contact binary.

Gaia provides two parameters that in most cases can be indicative of
the presence of a close companion, these are the renormalized unit
weight error ({\tt RUWE}), which evaluates whether a given star is a
clean point source, and {\tt ipd\_frac\_multi\_peak} which is the
fraction of observations for which the algorithm identified a double
peak. High values of {\tt RUWE} indicate probable binarity, with some
authors using a limit of 1.2, others 1.4, and some 2.0. For G2
the {\tt RUWE} is 3.84. But {\tt ipd\_frac\_multi\_peak} is 0.  There
are several secondary indicators of binarity, but only one of those
points to G2 being a binary \citep{cifuentes2025}.
% (Cifuentes et al. 2025). 

Circumstellar disks can increase the value of {\tt RUWE},
\citep[e.g.,][]{fitton2022}, but G2 has very little or no 
circumstellar material. We therefore in the following assume that G2
may have a very close, dim companion, so far only detected by Gaia.

% However, by itself a faint close companion does not explain the odd
% characteristics of G2. 

\subsection{The Walkaway Star KPNO~15} \label{subsec:KPNO15}

\begin{figure}
 \begin{center} 
 \includegraphics[angle=0,width=0.90\columnwidth]{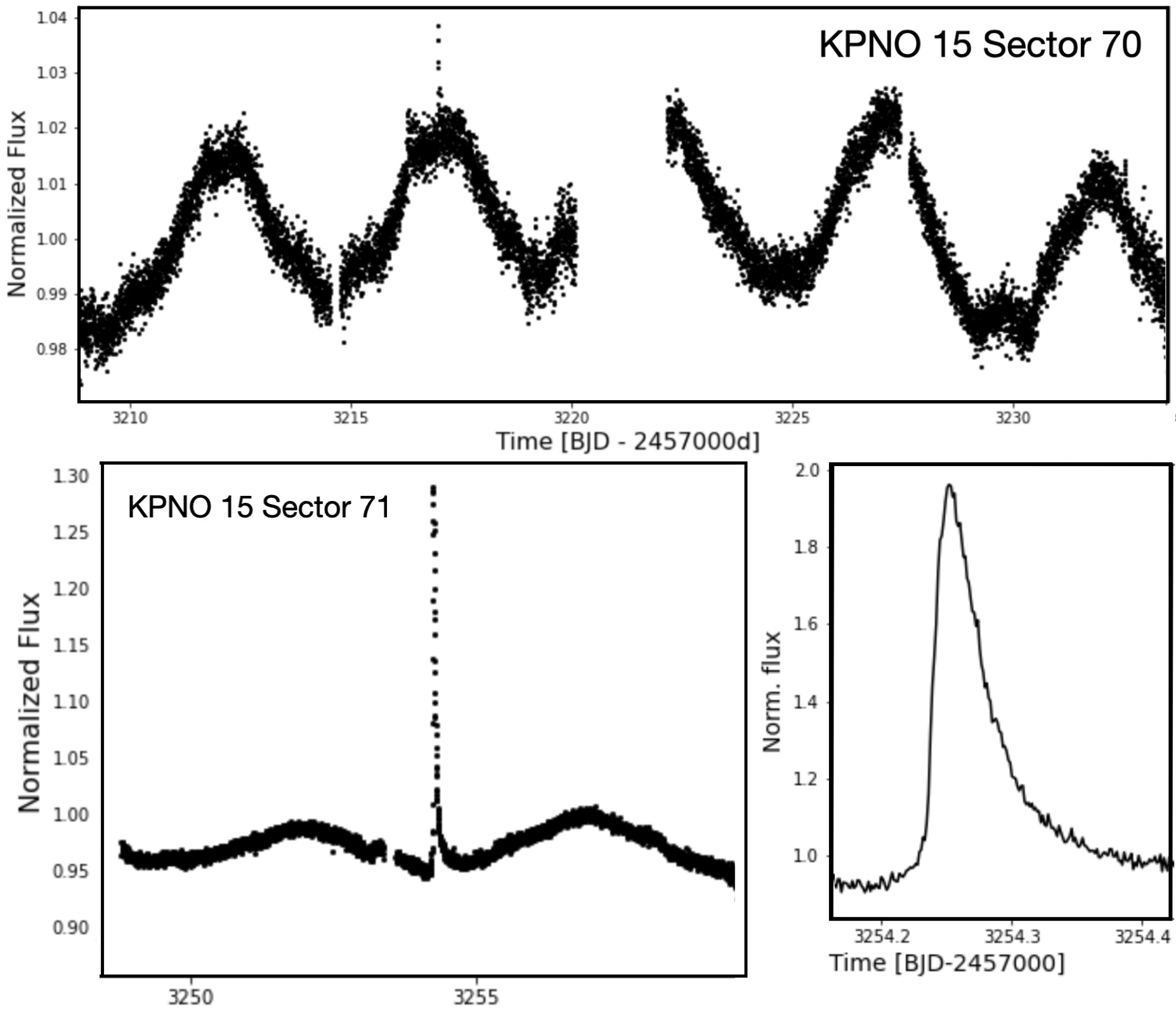} 
\caption{The TESS light curve of KPNO 15 as observed in Sector 70 and
71.  KPNO~15 is a well-behaved rotationally modulated source with a
period of $\sim$5 days. A major flare doubling the brightness of
KPNO~15 was observed in Sector~71. 
%[fig:TESS-KPNO15]
\label{fig:TESS-KPNO15}}
 \end{center}
 \end{figure}

KPNO~15 was identified as a weakline T~Tauri star by
\citet{luhman2003}, who classified it as M2.75 (the full identification in SIMBAD is [BLH2002] KPNO-Tau 15). We have obtained low-resolution
optical spectra of KPNO~15 (Figure~\ref{fig:opt-spectra}-top), and by
comparing with spectra of already known M-type TTS from the extensive
sample of Brice\~no et al. (2019) we determine a spectral type of M3 for
KPNO~15, with an overall uncertainty of 1 subclass, in agreement with
Luhman.  Gaia indicates an effective temperature of 3716~K, which
corresponds to M1 in the spectral-type to T{\em eff} conversions of
%Luhman (2003) and Herczeg \& Hillenbrand (2014).  
\citet{luhman2003} and \citet{herczeg2014}.
We also obtained a near-infrared spectrum which displays metal
absorption lines and CO bands and a triangular H-band continuum due to
water absorption, consistent with the optical late spectral type
(Figure~\ref{fig:ir-spectra}-top). Comparison with a second
near-infrared spectrum taken 4 years later shows weak emission at
Br$\gamma$, indicating occasional accretion. KPNO~15 is an X-ray
source cataloged as XEST 08-043 \citep{guedel2007}.
% G\"udel et al. 2007).  
Our JHK-photometry in Table~1 shows that KPNO~15 does not have a
near-infrared excess (Figure~\ref{fig:2MASS+CC}-bottom). KPNO~15 was
observed by TESS in Sectors 70 and 71, and we created a light curve
from full frame image files (30 minute cadence) using the
\texttt{lightkurve} package \citep{lightkurve2018} and a custom
2$\times$2 pixel aperture. The resulting light curve was corrected for
scattered light systematics using the \texttt{lightkurve} linear
regression corrector. We observe modest variability with an amplitude
of about 0.05~mag (Figure~\ref{fig:TESS-KPNO15}). Running a Fourier
transform periodogram we estimate a period of $\sim$5.0$\pm$0.7
days. We note that no evidence was found for a second period that
could have indicated the presence of a companion.  A small flare was
seen in Sector 70, but a major flare was observed in Sector~71,
doubling the brightness of KPNO~15 and with a duration of over
6~hours. For a young M3 dwarf
% {\color{red} [ask Ward for an age of $\sim$2-3 Myr]}, 
a flare that doubles the brightness of the star in the TESS
bandpass would have an energy of about 10$^{35.5}$ erg, 
which corresponds to a 'superflare' 
%(Vasilyev et al. 2024) 
\citep{vasilyev2024} and a
characteristic waiting-time between events of 1000 days (Ward Howard,
pers. comm.). For a typical TESS sector of 27d duration, this suggests
that one flare of this strength will be seen once every 37 sectors,
i.e., it is very rare. 
%{\color{red} see statistics in Zhang+2024-flares.pdf}

% TESS comment from Ward Howard: "For an M3 dwarf with an age of ~10
% Myr, a flare that doubles the brightness of the star in the TESS
% bandpass would have an energy of about 10$^{35.5}$ erg and a
% characteristic waiting-time between events of 1000 days." And Ann
% Marie says: "So basically, in a TESS sector worth of data (27d
% duration), you'd see such flares for roughly one in every 37 similar M
% dwarfs."

%Figure A5b

\subsection{The Walkaway Star G1} \label{subsec:G1}

G1, also known as 2MASS J04355209+2255039, was identified as a young
H$\alpha$ emission star by 
% Cohen \& Kuhi (1979), 
\citet{cohen1979}, and as an X-ray source 
% XEST 08-047 
by \citet{guedel2007}.
%G\"udel et al. (2007).
% Cody et al. (2022) found photometric variability with a very fast
% period of 0.85 days and a normalized flux amplitude of 0.06.
%Luhman (2018) 
\citet{luhman2018} found it to have a large proper motion, and suggested
that it is a walkaway star. We have obtained a low-resolution optical
spectrum of G1 (see Figure~\ref{fig:opt-spectra}-bottom). G1 is a
weakline T~Tauri star which exhibits a clear Li I 6708{\AA }
absorption line, with W(Li I)$\sim 0.3${\AA }. Our low-resolution
near-infrared spectrum shows a spectral type slightly later than
KPNO~15, and we concur with the spectral type of M4.5 assigned by
\citet{luhman2009}. Comparison with a second near-infrared spectrum
taken 4~years later shows very weak
Brackett$\gamma$ emission, indicative of some weak accretion
(Figure~\ref{fig:ir-spectra}-bottom). Our near-infrared photometry
indicates a modest near-infrared excess
(Figure~\ref{fig:2MASS+CC}-bottom).

% THIS HAS BEEN REPLACED WITH A REFERENCE TO CODY ET AL 2022:
% G1 was observed by TESS in Sector 70 and 71. The
% star is faint and the data noisy, but the periodogram shows a clear
% periodicity with a period of 0.85$\pm$0.02d, based on the cleaner data
% in Sector 70, see Figure~\ref{fig:TESS-G1}.
% and Cody et al. (2022). 
% When fitting a sinusoid to the light curve the semi-amplitude is 0.004
% - 0.005 magnitudes but with significant photometric noise causing
% scatter. 

\citet{cody2022} observed G1 with K2 and found it to be a
low-amplitude periodic variable with a period of 0.853 days. Clearly
G1 is a very fast rotator.

%\begin{figure} 
%\begin{center}
%\includegraphics[angle=0,width=0.75\columnwidth]{Fig11.jpg}
%\caption{A 5-day segment of the TESS light curve of G1 as observed 
%in Sector 70. The data are noisy so the variability is subtle, but
%reveals a well-determined period of 0.85$\pm$0.02 days, with a
%semi-amplitude of a sinusoidal fit of only $\sim$0.005 mag.
%%[fig:TESS-G1]
%\label{fig:TESS-G1}} 
%\end{center}
%\end{figure}

\section{PHYSICAL PARAMETERS OF G2}\label{sec:physpara} 

In the preceeding sections we have presented all current information,
from our own data as well as from the literature, on the three walkaway
stars. In the following we use the available observations in an
attempt to understand the nature of the peculiar star G2.

% INCLINATION ANGLE IS ANGLE TO LINE-OF-SIGHT

\subsection{Age and Mass of G2}  \label{subsec:age-massG2}
 
%%% AGE:

\citet{rizzuto2020} used the effective temperature and luminosity of
G2 as determined by \citet{herczeg2014} to compare with non-magnetic
evolutionary models, indicating a high age of $\sim$5~Myr, which is
significantly older than what the authors determined for two young
binaries (G3 and FF~Tau) in the same little HP~Tau group ($\sim$2.0
Myr). \citet{garufi2024} suggest an age in the range 3.6 - 5.0 Myr.

\citet{mullan2020} followed up on this puzzling result and
considered both non-magnetic and magnetic models. For the former they
found an age of G2 of 4.5~Myr and a mass of 1.9~M$_\odot$. For the
latter they found that a magnetic field gives an even larger age and
mass.

It is highly unlikely that G2, as part of the little HP~Tau group,
should have an age significantly higher than the other members of the
group for which \citet{krolikowski2021} determined an age of
2.0$\pm$0.3~Myr (see Section~\ref{sec:YSOs}), and we show in
Section~\ref{sec:discussion} that the peculiarities of G2 cause the
very high ages. In the following we adopt 2~Myr as the age of G2.

%%% MASS:

\citet{torres2009} used their accurate VLBA distance measurement for
G2 to refine its position in the H-R diagram and compare it with four
different stellar evolutionary models. The models agree that G2 has a
mass between 1.7 and 1.9~M$_\odot$. \citet{garufi2024} suggest a mass
of 2~M$_\odot$. However,  as will be
discussed in Section~\ref{sec:discussion}, it is declining from a major
outburst, which makes it appear hotter and more luminous.  The fact
that G2 was ejected when a triple system including KPNO~15 broke up
places some useful constraints on the mass of G2. As discussed in
Section~\ref{sec:walkaway} conservation of momentum considerations
imply that the ratio of masses of G2 to KPNO~15 is about 1.5. All we
know about KPNO~15 is that it has an $\sim$M3 spectral
type.  
Evolutionary models from \citet{baraffe2015} then suggest a mass for
KPNO~15 of $\sim$0.36~M$_\odot$. 
\citet{luhman2025} has established a relation between spectral type
and mass for young low mass stars based on dynamical masses determined
from binary orbits or rotation of circumstellar disks, and using that
it follows that KPNO~15 has a probable mass of
$\sim$0.45$\pm$0.2~M$_\odot$.  Relying more on the empirical relation
we then adopt a mass for G2 of $\sim$0.7~M$_{\odot}$, while keeping
the uncertainties in mind. One or the other of the components of a
disrupted triple system must be a binary. If G2 is an unresolved
binary with identical components, then they would be mid-M-dwarfs,
which is inconsistent with the optical spectral type. If instead
KPNO~15 is an unresolved twin binary with a combined mass of
$\sim$0.9~M$_\odot$ then G2 could, at least in principle, have a mass
as high as $\sim$1.3~M$_{\odot}$. However, it is usually the lowest
mass member of a multiple system that is ejected, and hence it is most
probable that KPNO~15 is single, which we assume in the following.

% Finally, both stars might be binaries from a quadruple system that
% broke up, but this is statistically unlikely and we do not consider it
% further here.

% As we will suggest in Section~\ref{sec:discussion}, G2 could be the merger of
% two low-mass stars. The mass ratio derived from proper motions then
% implies a mass for G2 much lower than what the evolutionary models
% discussed above suggest.

\subsection{Radius, Inclination, and Shape of G2} \label{subsec:size-shape}

For G2 we have both a rotational velocity \vsini $\sim$ 130~{\kms} and
a rotation period of 1.2 days, and that yields \rsini =
3.1~R$_\odot$. This is a lower limit to R$_*$, primarily because of
the unknown inclination angle of the star, but also because we do not
know the latitude of the spot groups, and G2 is likely to rotate
differentially.

\citet{herbst1986} and \citet{weaver1987} used the Stefan-Boltzmann law with
observed values for T{\em eff} and L$_*$ to determine R$_*$  for G2 and they
find an inclination angle to the line of sight of 67$^\circ$. 
%Bouvier et al. (1995) 
\citet{bouvier1995} similarly suggest an inclination angle of
50$^{\circ}$. None of these studies list uncertainties, but they must
be considerable, especially for a peculiar star as G2, as discussed in
Section~\ref{subsec:Teff}. We here assume an inclination of
60$^\circ$, consistent with the above rough estimates, but mainly
chosen because at that angle the probability to be higher or lower is
equal. For this value the \vsini $\sim$ 130 {\kms} corresponds to a
true equatorial rotational velocity of $\sim$150~{\kms} and a true
radius R$_e$ of 3.6~R$_\odot$. 
This is unusually large, for comparison
only three stars of the $\sim$400 T~Tauri stars studied by
\citet{rebull2002} have comparable or larger radii.

\begin{figure}
\begin{center}
\includegraphics[angle=0,width=0.8\columnwidth]{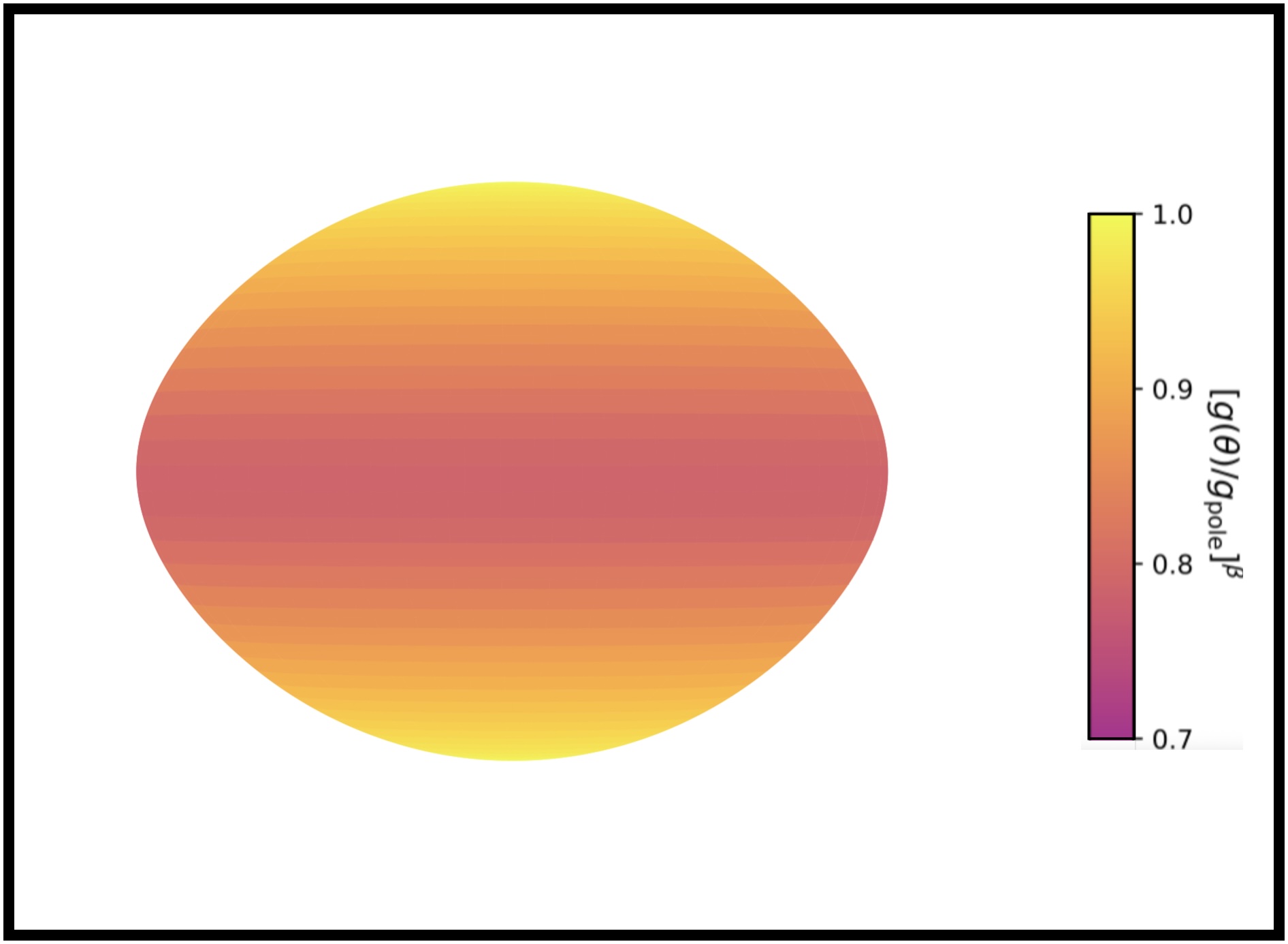} 
\caption{ 
A figure illustrating the oblate shape of G2 with gravity darkening. 
The flattening (R$_{eq}$/R$_{pol}$) is $\sim$1.3, assuming that the mass is
0.7~M$_\odot$, \rsini = 3.1~R$_\odot$, \vsini = 130~\kms, and {\em i} =
60$^\circ$. The gravity darkening exponent is $\beta$ = 0.16, as derived by 
\citet{espinosa2011}. 
The mass and inclination are the most uncertain parameters. For these parameters log~$g_{pole}$ = 3.40 and log~$g_{eq}$ = 2.75. This can be compared to the (uncertain) measurement of $g$$\sim$3.6 (see Section~\ref{subsec:Teff}). 
\label{fig:oblate}}
\end{center}
%\end{figure}

\vspace{0.5cm}

%\begin{figure}
\begin{center}
\includegraphics[angle=0,width=0.90\columnwidth]{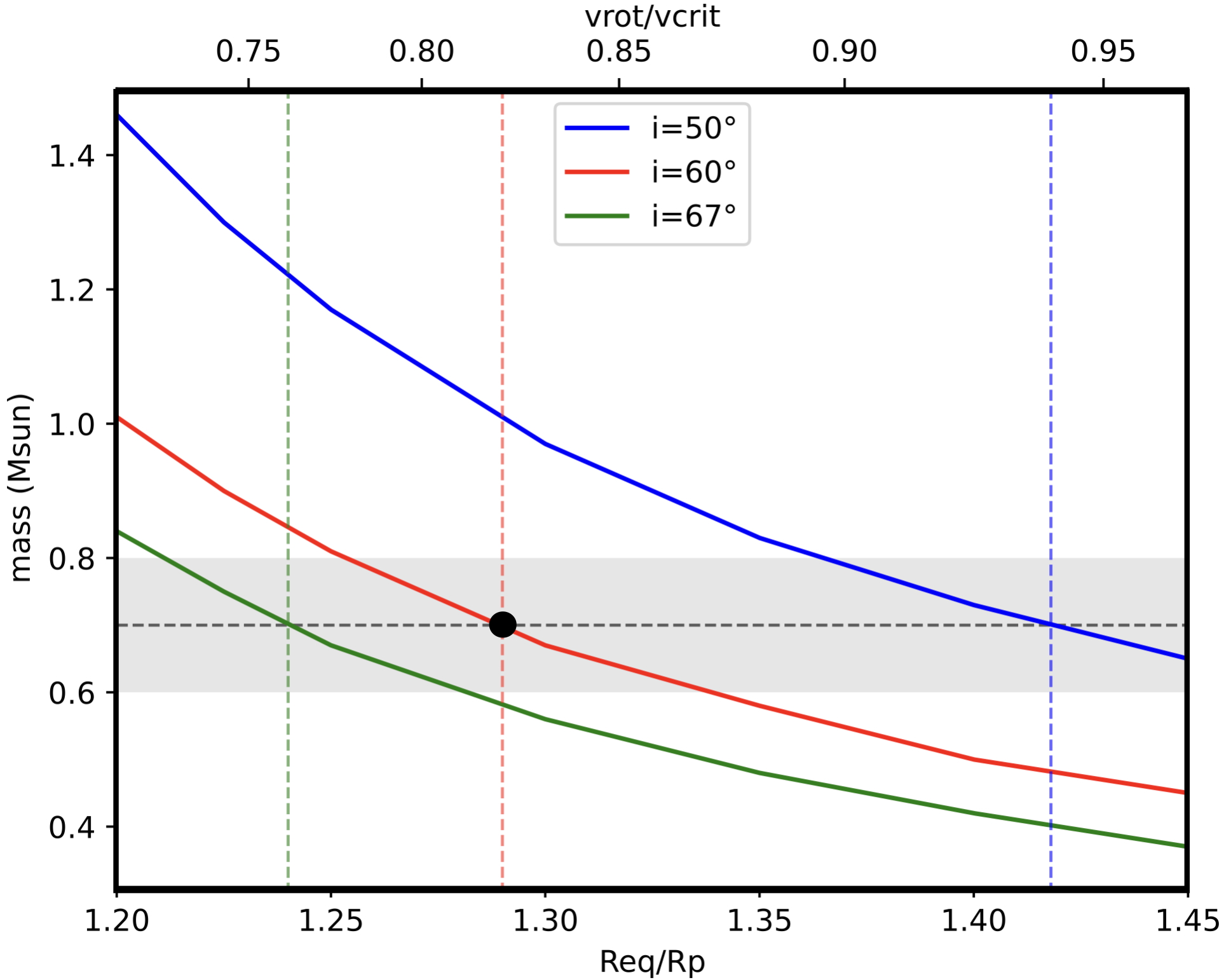} 
\caption{  
Application of the rigid rotator model to the star HP Tau/G2, showing
stellar mass ($M$) as a function of the rotation rate. The model
incorporates observational constraints: projected rotational velocity
$v\sin i \approx$ 130 \,km/s, and projected equatorial radius $R\sin i
\approx 3.1 R_\odot$. The plotted curves correspond to stellar
inclinations ($i$): $50^\circ$ (blue), $60^\circ$ (red), and
$67^\circ$ (green). Within an assumed mass range of 0.6--0.8
$M_\odot$, the star is identified as a fast rotator, with its rotation
speed constrained between 0.76 and 0.94 of the break-up velocity,
depending on the assumed inclination.
\label{fig:curves}}
\end{center}
\end{figure}

The breakup speed of a star is the Keplerian velocity at the stellar
surface (GM/R$_e$)$^{1/2}$, where M is the stellar mass, G the
gravitational constant, and R$_e$ is the stellar radius at the equator
\citep[e.g.,][]{maeder2009}. 
The Roche model for rapidly rotating stars assumes rigid
rotation and mass concentrated at the center of the star, and is valid
for flattenings (equatorial to polar radius) up to about 1.5.  With an
equatorial velocity {\em v} of 150 {\kms} and an equatorial radius
{\em r} of 3.6~R$_\odot$ G2 would break up at a rotational velocity of
$\sim$180~{\kms} for a mass of $\sim$0.7~M$_\odot$ suggested by
momentum conservation of the two walkaway stars (see the arguments for
choosing this mass in Section~\ref{subsec:age-massG2}). If so, G2
would be rotating at roughly 80$\%$ of its breakup velocity, and
consequently it would be oblate, with a flattening of roughly 1.3
(Figure~\ref{fig:oblate}). This indicates that G2 would appear
significantly flattened, almost as much as the extreme case of the fast
rotating B3V star Achernar ($\sim$1.35) \citep{domiciano2014}. If G2
has a companion then the companion mass would have to be subtracted,
implying that G2 would be less massive and therefore even more oblate. This estimate of the
flattening of G2 also depends on the uncertain value of the inclination,
see Figure~\ref{fig:curves} for the range of inclinations suggested by
observations.
% and on the assumption that KPNO~15 is a single star with a mass of $\sim$
One consequence of the oblate shape of G2 is that gravity darkening
takes place around the equator regions, which is discussed in
Section~\ref{subsec:Teff}.

% G2 would have to have a mass less than $\sim$0.XXXX~M$_\odot$ to break
% up, so this is a lower limit to its mass.  } %end red

\subsection{Effective Temperature of G2} \label{subsec:Teff}

From our high resolution K-band spectrum of G2 it is in principle
possible to derive or constrain the temperature, \vsini, gravity,
magnetic field, and veiling, see \citet{flores2019, flores2020,
flores2022}.  Using synthetic models, we fit eight wavelength regions
containing more than 20 atomic and molecular lines, including Ti,
Al, Na, Ca, Fe, Sc, and CO lines,

We derive a K-band temperature of 4200~K, a \vsini of $\sim$130
{\kms}, and an IR veiling of 0.4. We also derive a gravity log(g)
$\sim$3.6 and a magnetic field strength of $<$ 1~kG, but due to
the extreme width of the lines, we consider the values of
gravity and magnetic field to be unreliable. We are confident,
however, in the derived temperature values, which are primarily based
on the line depth ratios of the more than 20 lines analyzed in the
stellar spectrum. We have previously tested the temperature accuracy
against selected dwarf, giant, and sub-giant sources with precisely
derived temperatures, obtaining median temperature differences of
$\sim$50 K and standard deviations of $\sim$100 K 
\citep[see Appendix of][]{flores2019}.

\citet{flores2022} studied the effect of star spots on the
determination of stellar temperature, and found major differences
between those derived from optical and infrared spectra. While the
K-band temperature we derive for G2 is about 4200~K, the optical
temperature is around 5700~K \citep{herczeg2014}.
%Herczeg \& Hillenbrand 2014). 
This is the largest discrepancy between optical and infrared
temperatures found for any of the young stars studied in 
\citet{flores2022}. We have used a simple two-component model with a hot
component,
%(T\_hot), 
a cold component, 
%(T\_cold), 
and a geometrical filling factor 
%(f) 
to determine which combinations of these three parameters can
reproduce the observed optical (0.5~$\mu$m) temperature of $\sim$5700~K 
\citep{herczeg2014} and K-band T$_{K-band}$ temperature of
4200~K. In this two-component model, we assume that: {\em (a)} the
cold and hot components are represented by Planck functions B(T); {\em
(b)} the filling factor is a geometrical filling factor; and {\em (c)}
at each wavelength, the observed temperature is the weighted average
of the hot and cold temperatures, see equations C1 to C3 in
\citet{flores2020}. We find that there is a large family of solutions 
for the specific cold and hot temperature components that satisfy both
spectroscopic observations, with assumed temperature uncertainties of
150~K. However, a minimum filling factor of 70\% is required, with
even larger filling factors also possible, to satisfy these
conditions. This demonstrates that if the surface of G2 is indeed
mainly characterized by two temperatures, then surface spots covering
a significant fraction of the stellar area are unavoidable.

In principle we could instead determine T$_{eff}$ from the luminosity
and radius of the star. But this assumes that the star is spherical,
and we have shown that G2 must have an oblate shape. As such it is
affected by gravity darkening, which occurs because the centrifugal
acceleration at the equator offsets the star's
gravity. \citet{vonzeipel1924} showed that the effective temperature
for fast rotating radiative stars are a function of latitude, with
T$_{eff}$ $\propto$ g$^{\beta}$, where g is the local gravity, and
$\beta$ is the gravity-darkening exponent. Modifications to $\beta$
are required for convective stars such as G2
\citep[e.g.,][]{lucy1967,claret2012}.

It follows that both spots and gravity darkening affect G2, and that
determinations of effective temperature for a star like G2 should be
treated with caution.

\section{DISCUSSION} \label{sec:discussion}

In the preceeding pages we identified G2 as a highly unusual star, and
showed that G2 and KPNO~15 are walkaway stars moving apart following a
dynamical dissociation of a young triple system about 5600 yr ago.
In the following we discuss the dynamics of such small
multiple systems. 

% centered in and illuminating the walls of a cavity in the surrounding
% gas.  G2 and KPNO~15 form a a pair of walkaway stars moving apart from
% each other, which suggests that the event forming the cavity is
% related to the breakup of what must originally have been a multiple
% system.

\subsection{The Dynamical Evolution of Multiple Systems} \label{subsec:triples}

Non-hierarchical multiple systems are inherently unstable, and their
decay is a stochastic process, which occurs on a time scale of
$\sim$100 crossing times \citep[e.g.,][]{anosova1986}, for a recent
review on triple systems see \citet{perets2026}. This most often
happens in the embedded protostellar phase
\citep[e.g.,][]{reipurth2010}.  The end result, after a period of
cgaotica motions, is either a multiple system with a hierarchical
structure or the system breaks apart.

% A non-hierarchical quadruple system, for example, would after a period
% of chaotic motions develop into a hierarchical configuration, turning
% into either a 2+2 (a 2-tier hierarchy) or a 2+1+1 system (a 3-tier
% hierarchy) or break up, see
% \citet{tokovinin2021}.  A 3-tier quadruple has a central binary with
% two single stars bound at different distances. Such systems can form
% when an initially non-hierarchical quadruple first ejects one star,
% leading the remaining stars to form a more compact triple system,
% which then subsequently ejects a second star, and as a result the
% remaining binary contracts even more. 

As a result, the remaining system becomes increasingly
eccentric, statistically following a thermal distribution, and hence
the ultimate bound binary often possesses a very high eccentricity.  During
protostellar evolution when the stars are surrounded by circumstellar
and ambient material, periastron passages can lead to dissipative disk
interactions that eventually cause the binary orbit to shrink.  Close
binaries (less than $\sim$10~AU) are not formed in the collapse of a
cloud core, so such dynamical evolution explains the existence of
spectroscopic binaries among young stars.  In some more extreme cases,
the end result can be a merger of the two components
\citep[e.g.,][]{rawiraswattana2012,toonen2022}.
\citet{shariat2023}  estimate that at least 30\% of solar-type stars were
formed in hierarchical triples.

%%%%%%%%%%%%%%%%%%%%%%%%%%%%%%%%%%%%%%%%%%%%%%%%%%%%

\begin{figure}
\begin{center}
\includegraphics[angle=0,width=0.99\columnwidth]{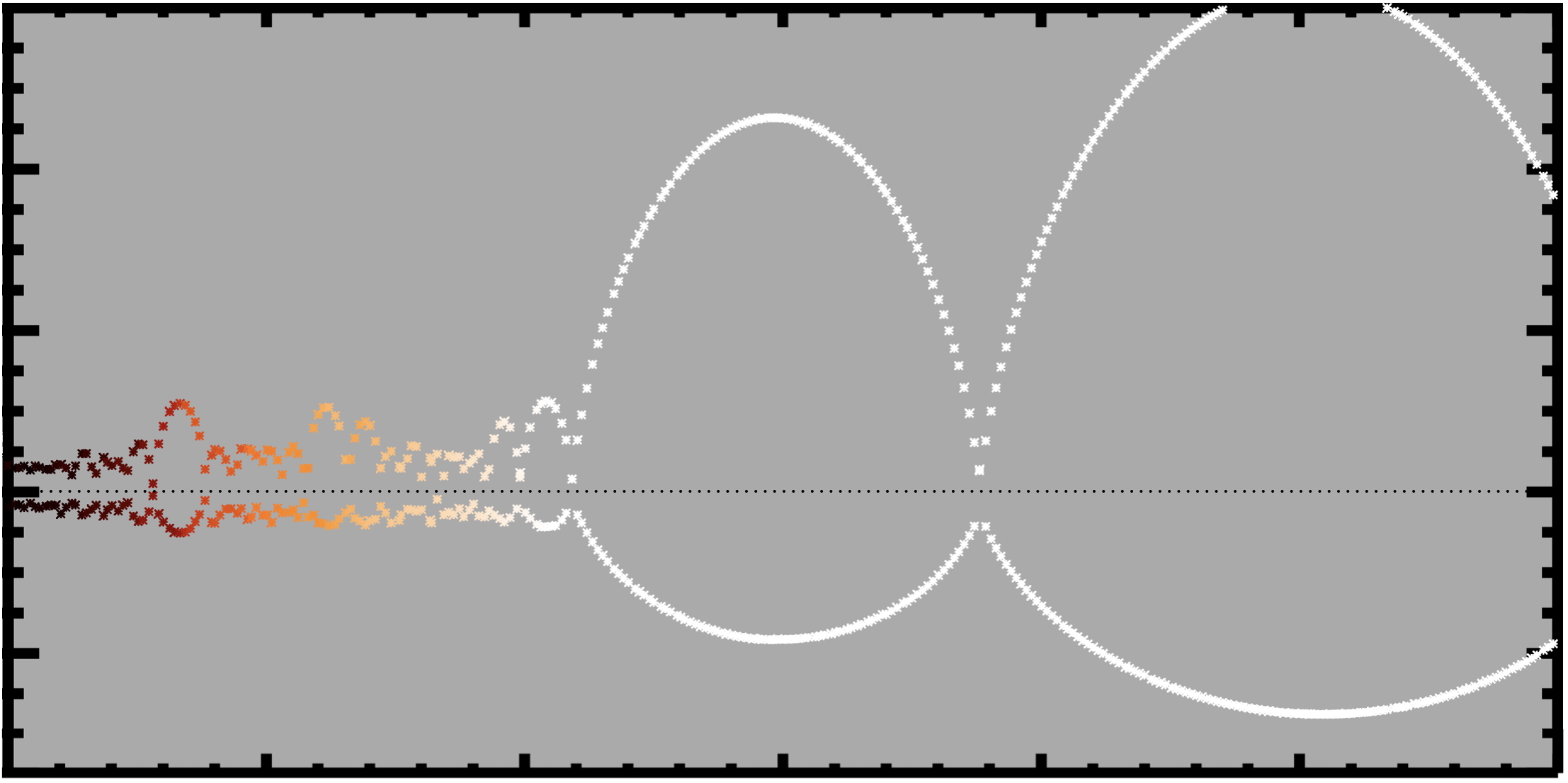} 
\caption{ 
Non-hierarchical triple systems are unstable,
and will eject one component typically after about 100
crossing times \citep{valtonen1991}. However, while the ejected body in some cases can escape, it is more common that it is ejected into a bound orbit. This
may be a stable orbit, or it may return to trigger another episode of
interactions which leads it to be ejected again. This can numerous
times before either a stable triple configuration is achieved or the
third body escapes. An example of a simulation, shown here for the
first 300,000 yr, reveals numerous ejections that all return the third
body before it finally escapes (outside the figure) after about
1~million yr \citep{reipurth2010}. The colors indicate extinction from
a slowly disappearing cloud core. As the core loses mass, the gravity
well weakens, facilitating larger excursions.  The dashed line
indicates the center of mass of the system. 
\label{fig:orphan}}
\end{center}
\end{figure}

Non-hierarchical triple systems are inherently unstable, and one
component, often the least massive, can be ejected again and again into a
bound orbit before either being ejected into an escape or achieving a
stable hierarchical configuration.  The ratio of inner to outer
separation in the triple system that is required to achieve stability
is a complex function of the orbital characteristics, but is roughly
of the order of 10. If stability is not achieved, the three components
will interact again. Numerical simulations show that this can go on
for a long time, up to ten million years, with the third body being
repeatedly ejected into bound but unstable orbits. Some ejections can
last from a few thousand years to in some cases more than a million
years before the third body returns to the remaining binary
\citep{reipurth2010,reipurth2012}. As a result, protostars are more
often found with a distant companion than more evolved young stars
\citep{connelley2008a,connelley2008b,tobin2022}. 

The disruption of a multiple system of N bodies will eventually lead
to the existence of at least one stable binary. Depending on the
masses of the components, the energies involved in the disruptions,
and the amount of circumstellar material that can alter the binary
orbit, the resulting binary can have a large range of semimajor
axes. In the ultimate case the two components can merge.

%   Figure~XXXX is a sketch of a such a possible evolution.

% \begin{figure*}
%  \begin{center}
% \includegraphics[angle=0,width=1.99\columnwidth]{sketch.jpg} 
% \caption{
%  A sketch of a possible dynamical evolution of a quadruple
% system that could lead to the currently observed configuration. (1)
% The components of a non-hierarchical quadruple system is moving
% chaotically; (2) one star (KPNO~15) is ejected, shrinking the size of
% the remaining triple system and gently moving its center of mass in
% the opposite direction; (3) the non-hierarchical triple system evolves
% into a hierarchical configuration, in the process tightening the
% newborn binary and making it highly eccentric; (4) dissipative
% periastron passages in the eccentric binary system leads to a merger,
% leaving the third star as a close (and unresolved) companion only
% detected by Gaia. It is evidently impossible today to know what
% actually occurred in this system, but this is at least one possible
% evolution.
% [fig:sketch]
% \label{fig:sketch}}
% \end{center}
% \end{figure*}

% We examine the characteristics of G2 identified in
% the previous pages, and relate them to the expectations of a merger.

\subsection{Coalescence}  \label{subsec:mergers}

Two stars can coalesce either via collisions or via mergers. The
former may occur in environments with extreme stellar densities and
are unpredictable, whereas the latter may occur in binary systems
which in various ways are losing angular momentum and spiral in. 
For a comprehensive review, see \citet{schneider2025}.

\begin{figure} 
\begin{center}
\includegraphics[angle=0,width=0.75\columnwidth]{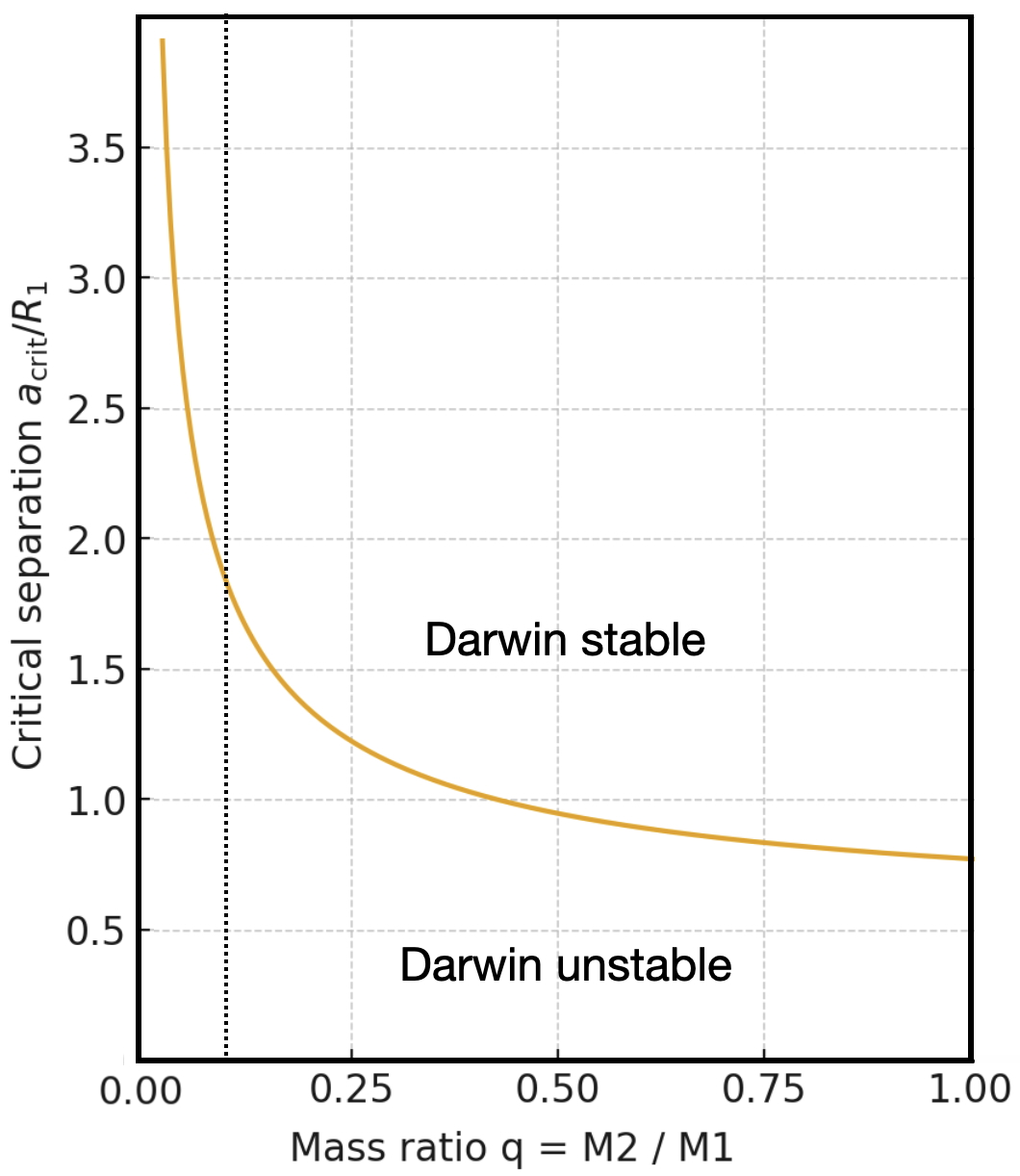} 
\caption{
The limit between Darwin stable and Darwin unstable binaries is a
steep function of the mass ratio. Very small mass ratios $<$ 0.1 are
far more likely to lead to a merger than binaries with higher mass ratios. 
R$_1$ is the radius of the primary and a$_{crit}$ is the separation at which
the binary becomes Darwin unstable.
%[fig:contreras]
\label{fig:Darwin-stability}} 
\end{center} 
\end{figure}

When binary systems evolve they can in some cases reach the point
where the Darwin instability is triggered
\citep{darwin1879, hut1981, stepien2011,
nandez2014}. In a binary that has become Darwin-unstable, tides try to
spin up the primary star at the expense of orbital angular momentum,
but the spin of the orbit changes faster than the spin of the primary,
so synchronization is never achieved. More specifically the Darwin
instability limit is crossed when the orbital angular momentum $J_{orb}$
is less than three times the spin angular momenta of the two stars,
i.e. $J_{orb} < 3(J_{spin1}+J_{spin2})$.  As orbital angular momentum
continues to be removed, the orbit shrinks, and the primary spins
up. Eventually the stars may reach contact and the stars will merge on
an orbital time scale.  This can happen for a variety of reasons, for
example if stellar evolution leads one component to expand, or if the
Kozai-Lidov mechanism shrinks the binary orbit. 
\citet{henneco2024} identify five  pathways that can lead to contact
and mergers.  Mergers are more likely to happen if the secondary is
much smaller than the primary (Figure~\ref{fig:Darwin-stability}).

Coalescence of two stars is not uncommon at late evolutionary stages
\citep[e.g.,][]{kochanek2014}.  Blue straggler stars are regularly found
in globular and open clusters, and are in many cases well interpreted
as the result of collisions between two cluster members
\citep[e.g.,][]{hills1976}.  
% In a series of papers, \citet{leigh2012,leigh2015} and
\citet{leigh2017,leigh2018} have studied collision probabilities in
small N-body systems and demonstrate that the probability rises as
N$^2$.
\citet{perets2009} and \citet{naoz2014} have proposed that blue stragglers may instead
originate in hierarchical triple systems in which the inner binaries
evolve into close binaries or contact binaries by a combination of
Kozai-Lidov cycles leading to increasing eccentricity and then tidal
friction which will lead to orbital circularisation
\citep[e.g.,][]{eggleton2001}. Mass transfer in such
close binaries could then lead to merger events. New 
simulations of three-body dynamics show that this may be a common
phenomenon \citep{shariat2025}.

The modelling of stellar mergers has a long history 
\citep[e.g.,][]{seidl1972,benz1987,soker2006}. 
Recent three-dimensional magnetohydrodynamic simulations of massive
stars
% (1.5~M$_\odot$ or more) 
have resulted in a detailed understanding of
the merging process \citep{schneider2019}.
During a thermal relaxation phase
after the coalescence, the merger product reaches critical surface
rotation and begins to shed mass. The merger product can only lose
angular momentum through mass loss (from stellar winds or from
shedding mass when the star rotates critically) and from magnetic
braking. 
% However, the models show that the ensuing spin-down of the star is not
% because of angular momentum loss but is caused by internal
% restructuring of the star.
%The time-scale of the ensuing spin-down is of the order of 50~Myr.
Eventually the internal structure of the merger product adjusts and
subsequently it evolves as a normal single star of the combined mass.
%\citep{schneider2020}.

An important characteristic of coalesced stars is their strong
magnetic fields, which result because turbulent mixing greatly
amplifies magnetic field energies
\citep{ferrario2009, wickramasinghe2014, schneider2020}.

Young stars are subject to the same processes as evolved stars, but
in addition the association with molecular clouds and the presence of
massive circumstellar and circumbinary disks can play additional
important roles for the dynamical evolution of newborn binaries and
multiples. Binaries closer than roughly 10~AU cannot be formed by
fragmentation, but both mass accretion \citep[e.g.,][]{bate2002} and
dynamical friction \citep[e.g.,][]{stahler2010} 
% as well as magnetic fields \citep[e.g.,][]{lee2019}
play important and efficient roles in the orbital evolution of newborn
multiple systems. \citet{lee2019} have carried out 3D
magnetohydrodynamic simulations of turbulent star forming clouds, and
find that such evolution can occur on time scales as short as
$\sim$0.1~Myr or less. For an up-to-date review of the literature on
newborn multiples, see
\citet{offner2023}. In short, conditions are commonly available that
can facilitate the merger of young stars.

%Young binaries with abundant ambient gas have an efficient way to
%become unstable because of dynamical friction \citep{stahler2010}. If
%accretion from circumstellar disks spin up one or both components
%while keeping the orbit roughly fixed, then the total spin angular
%momentum increases while the orbital angular momentum does not, moving
%the system towards the Darwin threshold.

%Additionally, circumbinary
%disks can in some cases lead to orbital decay
%\citep{bate2002, heath2020}, although this may require significant
%disk masses \citep{valli2024}.  Also, mass loss through the outer
%Lagrange point L2 is an additional mechanism to shrink a binary
%\citep{pejcha2017}.} The end result can be that the secondary spirals
%in and is violently engulfed by the primary, resulting in a 
%fast-rotating single star with an envelope inflated from the energy
%liberated during the merger. If the primary is a solar-type star, the
%merger with a low-mass object may result in a luminosity
%$\sim$10$^6$~L$_\odot$ \citep{soker2006}.

%%%%%%%%%%%%%%%%%%%%%%%%%%%%%%%%%%%%%%%%%%%%%%%

Observationally, numerous studies exist, 
%particularly of the blue stragglers in globular and open clusters. 
and especially two events have
provided important empirical information on mergers: the eruption of
V838~Mon in 2002, and the eruption of V1309~Sco in 2008, see
Appendix~D.

%\begin{longrotatetable}
\begin{deluxetable*}{l|cc|cc|cc}
\tablecaption{Comparison of HP Tau/G2 with HD 283572 and FK Com \label{tab:comparison}}
\tablewidth{0pt}
\tablehead{
\colhead{Parameter} & \colhead{HP Tau/G2} & \colhead{Ref.$^a$} & \colhead{HD 283572} & \colhead{Ref.$^a$} & \colhead{FK Com} & \colhead{Ref.$^a$}
}
%\decimalcolnumbers
\startdata
SpT              & G2                   & H14           & G5IVe                     & W87        & G4III & S09 \\
H$\alpha$ emiss. &      weak            & C79           &    no                     & S94        & weak &  R81, W93 \\
Lithium          &        yes           &  {***}        &    yes                    & W87,P93   & no & \\
%Type             & WTTS                 & C79           & WTTS                      & W87        & FK Com & B81 \\
Distance         & 162.2/0.9 pc         & T09           & 128.5/0.6 pc              & T07        & 224.7/1.5 pc & G22  \\
%Cloud            &  L1536               &               & B214                      & S94        & none    & \\
%V-mag            & 11.2                 &               & 9.1                       &            & 8.2 & \\
Low-ampl. var.   &   yes (V1025 Tau)    & V89           & yes (V987 Tau)            & G08        & yes (FK Com) & S23\\
Period           & 1.2 days             & V89           & 1.5 days                  & W87,G08   & 2.4 days & J93 \\
%Differential Rot.&       yes            & R20a          & yes                       & S11        &  ??    & \\
\vsini           & 127$\pm$4 {\kms}     & N12           & 79 - 136 {\kms}           & N12,W87   & 159$\pm$4 & R90\\
\rsini           &  3.1~R$_\odot$       & R26           & 2.39$\pm$0.03 R$_{\odot}$ & S98        & 7.5 R$_\odot$ & J93   \\
Inclination      & 50-67${^\circ}$  & H86,W87,B95   &  $\sim$35$^\circ$($\substack{+15 \\ -5}$) & S98 & $\sim$60${^\circ}$    & K99  \\
Mass             & 1.9$\pm$0.2~M$_\odot$$^b$ & T09,M20      & 1.8$\pm$0.2~M$_\odot$$^b$     & S98        &     & \\
L$_{bol}$$^c$    &   10.3 L$_{\odot}$   & R26           & 6.4~L$_\odot$             & R26        & 23~L$_\odot$ & R26 \\
Age              &   3.6-5.0 Myr$^d$    & R20,G24       & 2.5 - 9 Myr              & L24        & 5-10~Gyr & G86\\
Flares           &       yes            & R26           &  yes                     & F98        & yes      & O99 \\
Radio Cont.      &     non-thermal      & B84           & thermal                   & O90        & non-thermal      & R91 \\       
X-rays           &       yes            & G07           &  yes                      & W87,S94,F98&  yes    &  D08   \\
Gaia RUWE        &       3.85           & G22           &  0.93                     & G22        & 1.16    &  G22 \\
Large star spots &       yes            & R26           &   yes                     & J94,S98    & yes & K07  \\
Walkaway/Runaway &       yes            & L18,R26       &   no                       & R26         & yes & T11  \\
%Radial velocity  &               &          &        14.2$\pm$1.0                 & R15  &   & 
%Binary           & none detected        & T09, K11      & none detected             & T07        &     & \\
%Bipolar Outflow  &       yes            & D00           &   ??                      &            & ?? &  \\
\enddata
\tablecomments{(a):   
B81:~\citet{bopp-rucinski1981};
%Bopp \& Rucinski 1981; 
B84:~\citet{bieging1984};
%Bieging et al. 1984; 
B95:~\citet{bouvier1995};
%Bouvier et al. 1995; 
C79:~\citet{cohen1979};
%Cohen \& Kuhi 1979; 
D00:~\citet{duvert2000};
%Duvert et al. 2000; 
D08:~\citet{drake2008};
%Drake et al. 2008;
F98:~\citet{favata1998};
%Favata et al. 1998; 
G86:~\citet{guinan1986};
%Guinan \& Robinson 1986; 
G07:~\citet{guedel2007};
%G\"udel et al. 2007; 
G08:~\citet{grankin2008};
%Grankin et al. 2008; 
G22:~Gaia DR3; 
G24:~\citet{garufi2024};
%Garufi et al. 2024; 
H14:~\citet{herczeg2014};
%Herczeg \& Hillenbrand 2014; 
H86:~\citet{herbst1986};
%Herbst 1986; 
J93:~\citet{jetsu1993};
%Jetsu et al. 1993; 
J94:~\citet{joncour1994};
%Joncour et al. 1994; 
\citet{kenyon1995};
%K95:~Kenyon \& Hartmann 1995;
K99:~\citet{korhonen1999},
%Korhonen et al. 1999; 
K00:~\citet{korhonen2000};
%Korhonen et al. 2000; 
K07:~\citet{korhonen2007}
%Korhonen et al. 2007; 
K11:~\citet{kraus2011};
%Kraus et al. 2011; 
L18:~\citet{luhman2018};
%Luhman 2018; 
L24:~\citet{lovell2024};
%Lovell et al. 2024; 
M20:~\citet{mullan2020};
%Mullan \& McDonald 2020; 
N12:~\citet{nguyen2012};
%Nguyen et al. 2012; 
O90:~\citet{oneal1990};
%O'Neal et al. 1990; 
O99:~\citet{oliveira1999}
%Oliveira et al. 1999;
P93:~\citet{patterer1993}
%Patterer et al. 1993; 
R81:~\citet{ramsey1981};
%Ramsey et al. 1981; 
R15:~\citet{rivera2015};
%Rivera et al. 2015;
R20:~\citet{rizzuto2020};
%Rizzuto et al. 2020;
R26:~this paper; 
R90:~\citet{rucinski1990};
%Rucinski 1990; 
R91:~\citet{rucinski1991};
%Rucinski 1991;  
S94:~\citet{strom1994};
%Strom \& Strom 1994; 
S98:~\citet{strassmeier1998};
%Strassmeier \& Rice 1998; 
S09:~\citet{strassmeier2009};
%Strassmeier 2009; 
S11:~\citet{siwak2011};
%Siwak et al. 2011; 
S23:~\citet{savanov2023};
%Savanov et al. 2023; 
T07:~\citet{torres2007};
%Torres et al. 2007; 
T09:~\citet{torres2009};
%Torres et al. 2009; 
T11:~\citet{tetzlaff2011};
%Tetzlaff et al. 2011;
V89:~\citet{vrba1989};
%Vrba et al. 1989; 
W87:~\citet{walter1987};
%Walter et al. 1987; 
W93:~\citet{welty1993};
%Welty et al. 1993; 
W96:~\citet{wolk1996}.\\
(b): Masses from evolutionary tracks assuming normal stars. We argue that these masses are wrong, and that the mass of G2 is much smaller, $\sim$0.7~M$_\odot$, based on conservation of momentum considerations.\\
(c): Luminosities were calculated by integrating under the
optical/infrared photometry. For G2 an A$_V$ = 2.55 was adopted
(HH14), for HD 283572 an A$_V$ = 0.75 was used (SS94), and for FK~Com
no extinction was assumed.\\
(d) Age determined from evolutionary models assuming a normal star. In reality G2 is likely to have the same age as the surrounidng group, $\sim$2.0$\pm$0.3~Myr, see Section~\ref{sec:YSOs}.
}
\end{deluxetable*}
%\end{longrotatetable}

%Near-IR excess   &        no           & K95           &    no                     & W87,S94    &      & \\
%Mid/Far-IR excess &       yes          & R25           &    weak                   & W96        &       & \\

% 35$\substack{+15 \\ -5$

\subsection{What can we learn from late stellar evolution?} 
\label{subsec:late-evolution}

It is tempting to ask what similarities there may be between mergers
among pre-main sequence stars and among evolved stars.

Binaries that have evolved past the main sequence are involved in a
plethora of high energy stellar phenomena divided into many classes of
stars and displaying an almost bewildering variety of spectral
signatures and types of variability. Young stars and evolved stars are
different in very many ways, so in the following we will merely
examine certain stars in late stellar evolution that may act as a
broad guide to understand the peculiarities of G2.

\subsubsection{W~UMa stars and the formation of contact binaries}

W~UMa stars are low-mass eclipsing contact binaries that touch each
other at the L1 Lagrangian point and share a common envelope. They
generally have spectral types from late A to mid K and have stellar
activity which likely derives from a coupling of convective motions
with fast rotation that drives a dynamo. They have periodic light
curves, typically with periods shorter than 1 day, but there may be a
tail to 1.3-1.5 days. Most contact binaries are observed to be part of
triple or higher order multiple systems \citep[e.g.,][]{tokovinin2006,rucinski2007}. 
% Observations of W~UMa binaries are widely interpreted using the Lucy
% (1968a,b) model, but a more elaborate model of Stepie\'n (2009) is
% supported by new observations (Rucinski 2024).

% W~UMa stars are believed to have evolved from initially more detached
% binaries to a common envelope stage, and are often found within triple
% systems that through Kozai-Lidov cycles after several Gyr reach
% contact.

%%%%%%%%%%%%%%%%%%%%%%%%%%%%%%%%%

\subsubsection{FK~Com stars} \label{FKCom}

Some W~UMa binaries will merge during their common envelope evolution
and become single fast-rotating stars \citep[e.g.,][]{henneco2024}. It is
generally assumed that such newly single stars emerge as FK~Com
stars. This is a small group of peculiar stars identified by
\citet{bopp-rucinski1981} and \citet{bopp-stencel1981}. Not many of
these stars are known, possibly indicating that they are in a
relatively shortlived evolutionary phase.  The prototype FK Com is a
high-probability runaway star with a velocity of $\sim$33
\kms \citep{tetzlaff2011},
a spectral type of G4III \citep{strassmeier2009},
and a spectrum showing a weak and variable H$\alpha$ emission line
\citep[e.g.,][]{ramsey1981,welty1993}, see Appendix~E.  
Ca~II H and K lines are in emission, indicating
chromospheric activity \citep[e.g.,][]{vida2015}.
%(e.g., K. Vida et al. 2015). 
The star rotates very rapidly, with a \vsini around 160~{\kms} 
\citep[e.g.,][]{rucinski1990}
and its rotational period is 2.4 days
\citep{jetsu1993}, which implies a radius \rsini of 7.5~R$_\odot$.
\citet{huenemoerder1993} argue that
the star is rotating near its breakup velocity.
\citet{welty1994a} interpret a series of H$\alpha$ profiles (width$\sim$1000~\kms ) as evidence for
nonradial pulsations which they postulate is evidence for a recent
merger. On longer timescales the star shows irregular low-amplitude
variability \citep[e.g.,][]{panov2007}.  The variability is ascribed
to giant star spots, which have been observed with Doppler imaging
that shows spots preferentially located at high latitudes
\citep{korhonen2000,korhonen2007}. 
These are an indication of strong magnetic fields \citep{korhonen2009}.
FK Com is an active X-ray source 
\citep[e.g.,][]{welty1994b,drake2008}
and has been studied at ultraviolet wavelengths by \citet{ayres2016}.
Non-thermal radiation in the 3.6~cm radio continuum has
also been detected \citep{rucinski1991}.
\citet{guinan1986} conclude
that it is a member of the old disk population, with an inferred age
around 5-10~$\times$10$^9$~yr.

%==============================\\
% * In contrast to the W~UMa stars, FK~Com
% shows no radial velocity variations (McCarthy \& Ramsey 1984). 
% * H$\alpha$ absorption line with a central emission core superposed: Ramsey et al. (1981), Walter \& Basri (1982),  Welty et al. (1993)
% * It rotates very rapidly, vsini about $\sim$160~{\kms} (e.g., Rucinski 1990).
% * Light curve: Panov \& Dimitrov (2007), 
% * FK Com shows high chromospheric activity ...Halpha and Ca H&K: Vida+2015
% * Radial vel. variations: McCarthy \& Ramsey (1984)
% * 3.6 cm detection: Rucinski (1991)
% * Magnetic field: Korhonen et al. (2009)
% * Spots and Doppler imaging: Korhonen et al. (2000,2007)
% * Rotational PERIOD: 2.4 days Jetsu et al. (1993)
% * Age of FK Com: 5-10 x 10^9 yr Guinan \& Robinson 1986
% * X-rays: Drake et al. (2008)
% * UV: Ayres et al. (2016)\\
% =====================\\

Table~3 gives a detailed comparison between G2 and FK~Com
itself. Remarkably, we find that they have almost identical properties.
At first glance it is puzzling that  two stars can be so similar
considering their very different evolutionary stages.

Once two stars have merged, they lose their identity and for a while
the new star is instead dominated by the aftermath of the merger as it
evolves on a thermal time scale towards an undisturbed star with the
characteristics of a more massive star. Hence, a merger of either two
young stars or of two evolved stars will for a while look if not
identical then very similar.

A star newly formed in a merger will at least initially be rotating
very fast, but the star will eventually (and sometimes even rather
quickly) spin down and become a slow rotator.  This can be due to
internal structure changes that lead to a re-distribution of angular
momentum, or from mass loss (magnetized winds, outflows, or equatorial
decretion disks) that will carry a significant amount of angular
momentum away \citep{ryu2025}.

% ... chemical differences

We suggest that the merger of two young stars will result in a bloated
star like the FK Com stars, and that G2 may be
such a case.

Neither G2 nor FK~Com are similar to the post-eruption appearance of
the evolved star V1309~Sco (see Appendix~D), which immediately after
the eruption had an F-type spectrum that turned into a red giant
spectrum within a month.  However, V1309~Sco erupted less than 20
years ago, while the G2 merger most likely 
happened several thousand years ago. During that time-span, if a cool
dusty envelope was ejected in the eruption, it would have dispersed.

%%%%%%%%%%%%%%%%%%%%%%%%%%%%%%%%%%%%%%%%%%%%%%%%%
%%%%%%%%%%%%%%%%%%%%%%%%%%%%%%%%%%%%%%%%%%%%%%%%%

% \begin{figure}
% \begin{center}
% \includegraphics[angle=0,width=0.60\columnwidth]{FigX.jpg} 
% \caption{{\color{red} Proper motions of the stars G2, KPNO~15, G1, and
% HP Tau in a reference system based on 7 stars from the surrounding group. 
%  The red star indicates where G2 and KPNO~15 were within
% two arcseconds of each other 5600~yr ago at the position (2000)
% 4:35:52.2 +22:53:36, and marks where a disintegration of the multiple
% system took place, ejecting KPNO~15 and G2.  The red dotted lines
% indicate past motion of the stars. HP~Tau drifted past where G2 was a
% few thousand years ago, but it is unclear if it had anything to do
% with the evolution of the system. The orange circles mark 
% the approximate center of the semi-circular cloud cavity surrounding G2
% and is likely close to the place where the merger and ensuing explosion took place.
% Background image courtesy Adam Block.}
% \label{fig:propermotions-2}}
% \end{center}
% \end{figure}

\subsection{G2 as a Merger}  \label{subsec:G2merger}

% We propose that the observations of G2 and its surrounding cavity is
% best understood if it is the result of a recent merger of two low-mass
% young stars in a close binary system, which was part of a higher-order
% multiple system.

If G2 is a merger, an immediate question is why this happened now,
$\sim$2~Myr after the group formed, when there is very little
dissipative material left in the system. Such a viscous environment
would be essential to make two components spiral in and merge.

It is evidently not possible now to reconstruct the precise dynamical
evolution of a chaotic system that led to the current state of G2 and
the other walkaway stars in the group.  But in the following we hazard
a guess to what {\em might} have taken place.

The group of 12 stars and brown dwarfs around HP~Tau is one of the
most compact and rich groupings in Taurus
\citep{joncour2018}, yet is about 2~Myr old \citep{krolikowski2021}.
If we assume that the individual stars in the central group have space
velocities of $\sim$1~{\kms}, and if we take the angular separation
between G2 and FF Tau as the size of the multiple system in the plane
of the sky (radius 7.3 arcmin at 162 pc, see
Figure~\ref{fig:2MASS+CC}), then it should take of the order of
400,000 years to spread beyond this volume. But the age of the system
is about 5 times larger. This suggests that the group may have
remained tenously bound until more recently, probably because the
cloud core that gave rise to the stars survived for a long time,

% or that the aggregate remained bound, albeit only tenuously. Over time
% some members started to drift slowly away, like the more distant stars
% FF Tau and HQ Tau (both of these are subarcsecond binaries, so their
% proper motions are compromised and cannot be used to determine their
% origin).

A high order multiple system of at least 6 bodies then remained,
consisting of G2 (at the time a binary, and possibly a
compact triple if the Gaia companion is real), G3 (a binary), and the
two late M-dwarfs KPNO15 and G1. In order to stay bound together, even
loosely, this multiple system had to be hierarchical, consisting of
two separate groups (the G2 binary plus KPNO~15, and the G3 binary
plus G1) tenuously bound together.

As noted in Section~{\ref{subsec:triples}} such systems are generally
born unstable and typically become hierarchical by ejecting a
component (either bound or unbound) already within about 100 crossing
times, which in most cases occurs during the protostellar stage.  But
the resulting hierarchical configuration is not necessarily stable, so
upon return, an ejected body may be sent away again and again
(Figure~{\ref{fig:orphan}}). This process can sometimes go on for a
very long time, numerical simulations show that up to 10~million years
or even more can pass before the ejected star, after numerous
excursions, is finally either escaping or parked in a stable eccentric
orbit \citep{reipurth2010,reipurth2015,toonen2022}.  KPNO~15 and G1
must have undergone such extensive 'yo-yo' episodes over the past
$\sim$2~million years.

\begin{figure}
 \begin{center}
 \includegraphics[angle=0,width=0.70\columnwidth]{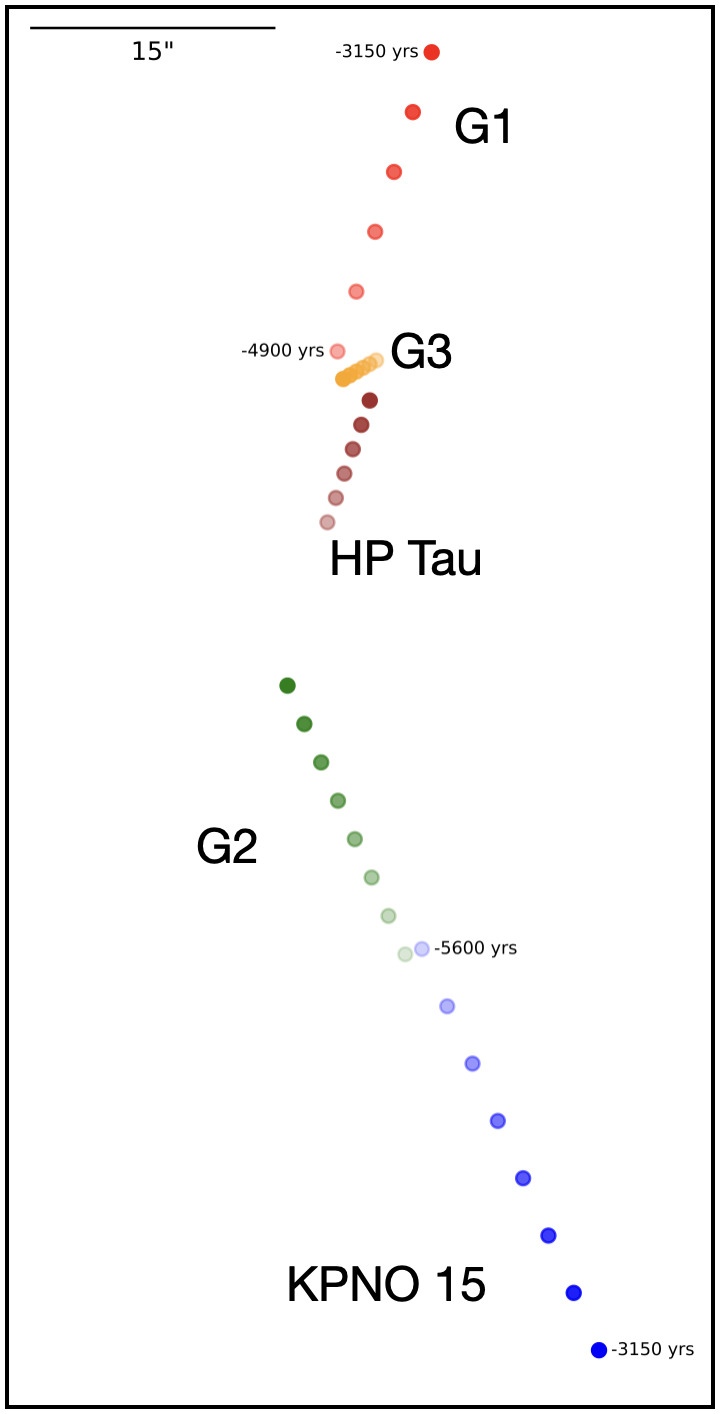}
\caption{The motions of G1, G2, G3, KPNO~15 and HP~Tau are shown in
this figure in the time interval from 5600 to 3150 yr ago in steps of
350 years. The nominal time of breakup for the G2/KPNO~15 pair is
5606~yr and 4869~yr for G1/G3. Note that the position and proper
motion of G3 is affected by its close (31 mas) companion. The K4 star
HP~Tau is slowly crossing the stellar group and possibly disturbing it. 
It has an unusually high velocity, although not high enough to be labeled 
a walkaway star.
\label{fig:paths}} 
\end{center} 
\end{figure}

At some point, maybe 6000 years ago, this fragile little aggregate
somehow became disturbed, perhaps by the nearby more massive
binary HP~Tau.  Whatever the disturbance was, the G2+KPNO15
and the G3+G1 multiple systems each broke up roughly at the same time
($\sim$5600~yr and $\sim$4900~yr, respectively)
forming two slowly separating walkaway pairs 
(Fig~\ref{fig:paths}).

%{\color{red} 

%} %end red

In the case of G2, the
ejection of KPNO~15 into an escape was the final step in the cascade
that had bound the components of the G2 binary ever tighter in a
highly eccentric orbit, and rather than resulting in a tight
spectroscopic binary as G3, in this case the binary 
became Darwin unstable and merged.

A few details of the process can be extracted.  If KPNO~15 has a mass
of $\sim$0.4--0.5~M$_\odot$ then the G2 precursors (plus the putative 
Gaia companion) would have a total mass of roughly $\sim$0.7~M$_\odot$,
indicating that the components would have been very low mass stars or
brown dwarfs. That KPNO~15 is a walkaway star and not a runaway star
indicates that the dynamical interaction triggering the ejection was
not a particularly close one, with separations not much closer than
$\sim$10 AU.

The above broadly outlined scenario is of course very speculative, but
the general dynamical process is well understood, and in fact is
probably an integral part of how parts of dense clusters dissolve.

It is possible that more
accurate data from future Gaia releases may provide a fuller
understanding of a possible relation between these three walkaway
stars.

% In Appendix~F we discuss an
% unsuccessful attempt to connect the three walkaway stars to the
% single breakup of a higher-order multiple system.

%%%%%%%%%%%%%%%%%%%%%%%%%%%%%%%%%%%%%%%%%%%%%%%%%%%%%%%%%%%%%%%%%%

% The curious fact that the more massive star HP~Tau and G1 both were
% together at one point (seen along the line of sight) and what role
% this may have played when they passed the path of G2 slightly before
% G2 arrived remains unexplained in the above scenario, and might
% indicate that it is either wrong or incomplete.

%%%%%%%%%%%%%%%%%%%%%% END OF NEW TEXT %%%%%%%%%%%%%%%%%%%%

%%%%%%%%%%%%%%%%%%%%%%% OLD TEXT %%%%%%%%%%%%%%%%%%%%%%%%%%
% In this scenario G2 and KPNO~15 belonged to a multiple system of (at
% least) 4 stars: a very close binary that eventually would merge to
% become G2, a faint companion detected only by Gaia, KPNO~15, and
% possibly also G1.  The specific details of this evolution are
% obviously no longer observable, but the following is one possible
% pathway. We know that KPNO~15 at some point was ejected, and  
% It is possible that G1 was also ejected from the multiple
% system, which would have further contracted the system, although the
% details are obscure.  
% Proper motions from Gaia~DR4 will clearly put significant constraints
% on the evolution that led to the present state of the
% system. Meanwhile we can only speculate, and in Appendix~C we outline
% one possible scenario that possibly could explain the observations.

\begin{figure}
 \begin{center}
\includegraphics[angle=0,width=0.80\columnwidth]{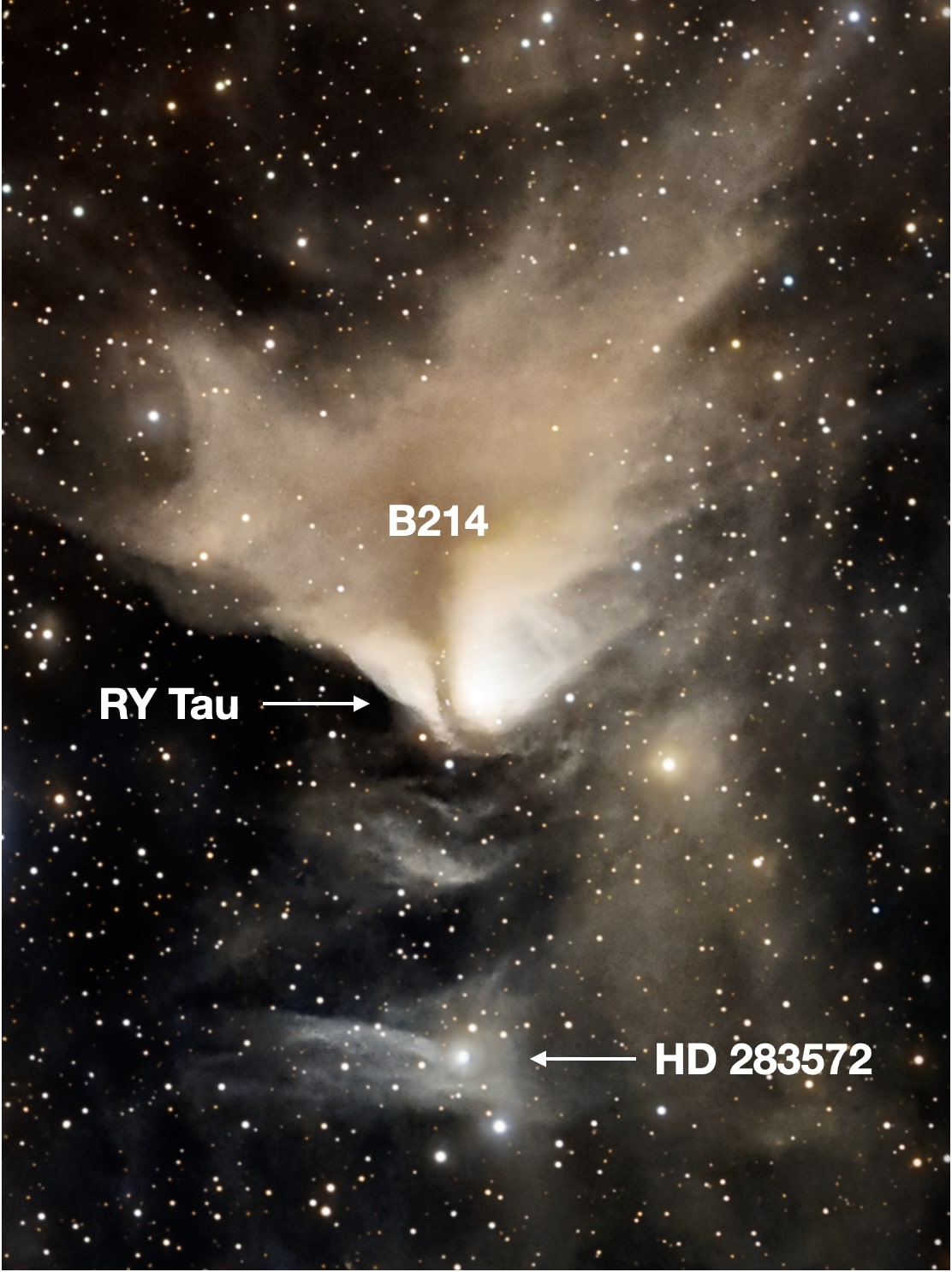} 
\caption{HD 283572 is located near the little cloud B214, close to
RY~Tau which illuminates the LBN~785 = vdB27 reflection nebula. Note
that the well defined cometary shape of the B214 cloud points straight
towards HD~283572.  No other such cometary clouds are seen in this
region of Taurus. Courtesy C\'edric Mauro and Christophe Marsaud from
Team Astrofleet.
%[fig:HD283572]
\label{fig:HD283572}}
\end{center}
\end{figure}

%} %end blue

\subsection{Are there other cases like G2?} \label{sec:other}

G2 is a highly unusual star, and we believe that it is best understood
as the result of a merger. If so, could similar characteristics in
other young stars suggest recent mergers? We are aware of only one
star that shares almost all important features with G2, and it is HD
283572 (V987~Tau) near the small Barnard~214 cloud in Taurus.  It is
part of an aggregate of $\sim$35 YSOs near L1495E \citep{strom1994}.
%(Strom \& Strom 1994). 
%Torres et al. (2007) 
\citet{torres2007} used VLBA to determine a distance of
128.5~pc and Gaia DR3 gives 127.0~pc.  HD~283572 was initially
recognized as a young X-ray source by \citet{feigelson1981}
%Feigelson \& Kriss (1981) 
and \citet{walter1981}.
% Walter \& Kuhi (1981).  
Subsequently \citet{walter1987}
%Walter et al. (1987)
carried out a detailed study and classified HD~283572 as a G5IV star
with lithium and a fast rotational period P of 1.54 days
\citep{strassmeier1998}, which we confirmed with TESS data.
It has no
infrared excess out to 60~$\mu$m \citep{wolk1996,stassun2001}.
%(S.J. Wolk \& F.M. Walter 1996, K.G. Stassun et al. 2001). 
It is a weak-line TTS with variable H$\alpha$ profile and
signs of chromospheric activity \citep{fernandez1998}.
%(Fernandez \& Miranda 1998).  
The star is a fast rotator with a \vsini of 79.5$\pm$3.0 {\kms} 
\citep{johnskrull1996},
and confirmed by \citet{donati1997} [77~{\kms}] and 
\citet{strassmeier1998} [78$\pm$1~{\kms}].  An extreme millimeter
flare was detected by \citet{lovell2024}.

\begin{figure} \begin{center}
\includegraphics[angle=0,width=0.55\columnwidth]{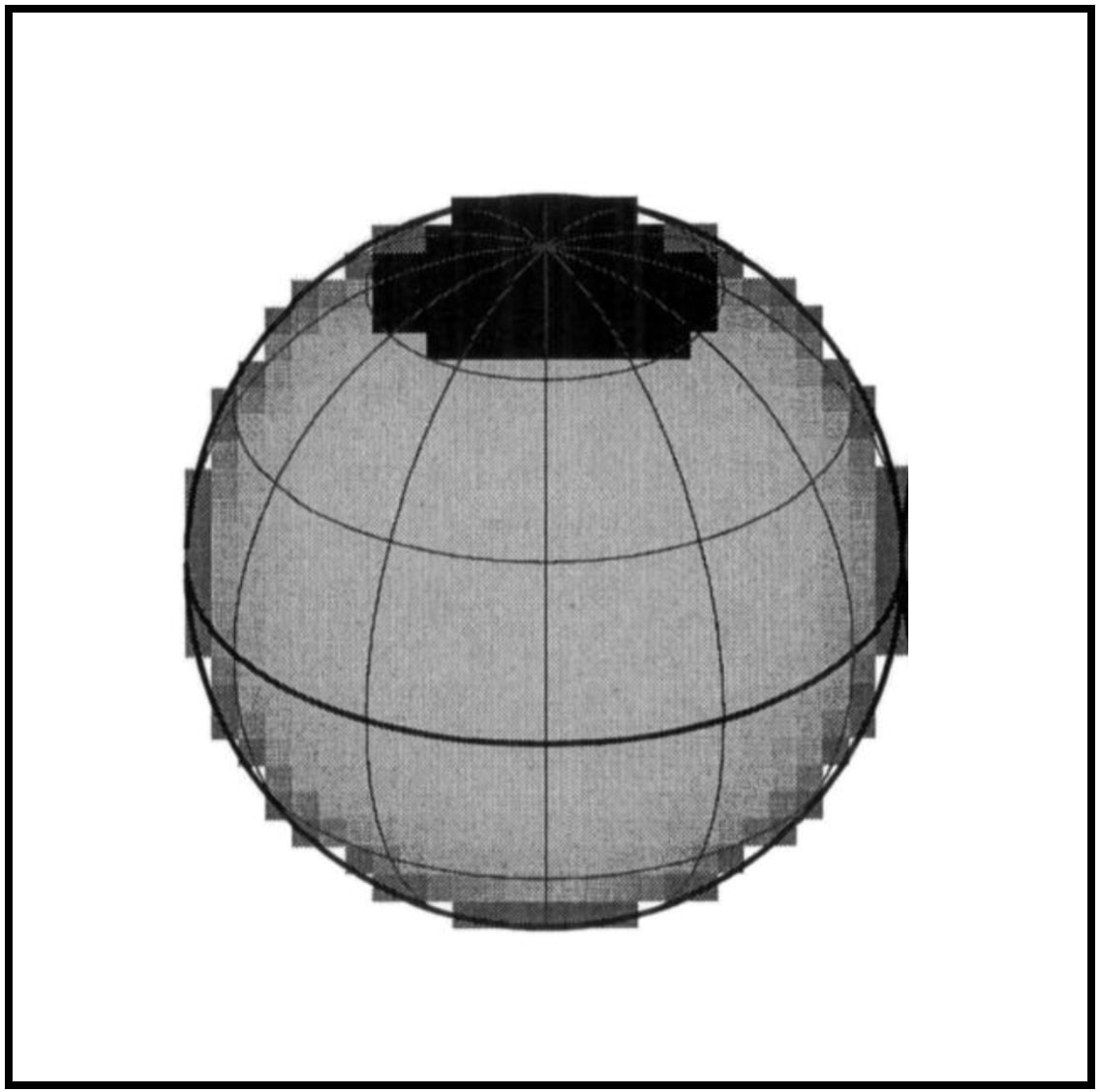}
\caption{A photospheric temperature map of HD~283572 from 
\citet{joncour1994}             
obtained with Doppler imaging and showing a large polar spot. The
maximum temperature difference is 1600~K. 
%[fig:joncour]
\label{fig:joncour}} \end{center}
\end{figure}

\citet{joncour1994} have presented Doppler imaging of the star and
discovered a large polar spot (Figure~\ref{fig:joncour}), which was
confirmed by \citet{strassmeier1998}, who found the polar spot to
be one of the largest and coolest ever observed, with at most few
spots at lower latitudes. 
%   #########################################################################
% They determined an inclination angle of the
% rotational axis of the star to be $\sim$35$^\circ$($\substack{+15 \\ -5}$).

\citet{siwak2011} obtained a detailed light
curve of HD~283572 which shows characteristic low-amplitude
fluctuations ascribed to spots. \citet{davies2014}  suggest a mass of
1.98$\substack{+0.19 \\ -0.25}$~$M_\odot$,
%+0.19-0.25 M$_\odot$, 
a radius of 2.84~R$_\odot$, and an age of 5$\substack{+2.6 \\
-1.2}$~Myr based on evolutionary tracks. As is the case for G2, we
supect that these values are much too high.

When P and \vsini are combined this indicates a \rsini of
2.4~R$_\odot$, which for an inclination of roughly 35$^\circ$
\citep{strassmeier1998} 
suggests a large equatorial radius of
$\sim$4.1~R$_\odot$. If the mass of HD~283572 would be the same
($\sim$0.7~M$_\odot$) as for G2, then it would have about the same oblate
shape as G2, with a flattening (R$_{eq}$/R$_{pol}$ of roughly 1.4. 
%(see Figure~\ref{fig:oblate}).

If HD~283572 is a recent merger, then a major eruption must have taken
place. The star, however, is not directly associated with a dense
cloud, only some wisps of nebulosity, so no surrounding shell has been
formed as in the case of G2. But to the north there is a small cloud,
B214, which has a striking cometary appearance, with a symmetry axis
pointing straight towards HD~283572. No other cometary clouds are seen
in this part of Taurus.  We speculate that the unusual shape of the
B214 cloud is due to an energetic event in HD~283572.

There are several other potential cases, for example HD~30171 
\citep[e.g.,][]{wichman2000}. 
It should be emphasized that it is the combination of the various
peculiarities that indicate a possible merger, not just the G spectral
type. There are many normal young G-stars that are on their way to
become Herbig Ae/Be stars and which have nothing to do with mergers.

The Gaia DR3 proper motions of HD~283572 is almost identical to the
mean of all YSOs closer than 3 arcmin. Its heliocentric radial
velocity is +14.2 \kms \citep{rivera2015}, that is, within
$\sim$5~\kms of the cloud velocity.  Hence HD~283572 is not a walkaway
star.  We have searched the Gaia DR3 catalog for any star moving
directly away from HD~283572, but none were found.

The features of HD~283572 may suggest that it has suffered a past
merging event, when the star was no longer associated with a
substantial molecular cloud core, therefore probably at a slightly
later stage than G2. If so, it may indicate that G2 is not a
pathological case, but represents a not uncommon phenomenon, albeit
infrequent on human time-scales.

% At this late stage it is impossible to determine the properties of the
% pre-merger stars. But we do note that ....

\vspace{0.2cm}

\subsection{When did the merger take place?} \label{subsec:when}

The accurate Gaia proper motions indicate that the triple
disintegration event that produced the walkaway stars G2 and KPNO~15
took place $\sim$5600 yr ago. In principle, the ensuing merger event
could have taken place at any time after that. For lack of better
constraints we speculate that the explosion might have taken place
when G2 was at the center of the bright innermost part of the
reflection nebula, marked by the orange circle in
Figure~\ref{fig:propermotions}.  If so, then it occurred roughly
2000~yr ago.

% {\color{red} Most triple disintegrations take place during the
% embedded protostellar phases \citep[e.g.,][]{reipurth2000}, aided by a
% dissipative environment which is conducive for the inspiraling process
% that can facilitate a merger. In Section~\ref{sec:walkaways}, however,
% it was concluded that there is little evidence for a circumstellar
% disk to G2. We have no information on when the gas disappeared, and
% this is further hampered by the uncertainty about the age of G2. But
% to produce a merger the stars involved must have undergone a number of
% flybys, which would have truncated their disks
% \citep[e.g.,][]{andrews2020}. And the actual merger will cause a
% violent explosion imparting energy into and likely destroying any
% remaining disk material. }

% The structure of the large scale gas environment
% of G2 is further discussed in Section~\ref{sec:mm}. 

\subsection{The Past and Future Evolution of G2} \label{subsec: evolution}

In Section~\ref{subsec:age-massG2} we referred to the derivation of
model isochronal ages and masses for G2 by \citet{rizzuto2020},
\citet{mullan2020} and \citet{garufi2024}, who found ages in the range
3.6-5.0~Myr, significantly higher than the derived age of 2.0$\pm$0.3~Myr for
the neighboring stars in the little T~Tauri group. The derived mass of
$\sim$2~M$_\odot$ is also inconsistent with the mass derived from
analysis of the Gaia proper motions of G2 and KPNO~15.

In light of the merger hypothesis, these discrepancies find a natural
explanation when G2 is seen as the bloated merger of two low-mass
late-type young stars. When the eruption took place G2 reached a much
higher luminosity and an earlier spectral type.  As a result, G2 moved
upwards and to the left in the HR-diagram, to evolutionary tracks for
much more luminous and warm stars with higher masses.

The evolution of mergers can be divided into three phases
with different timescales. The first is the period following the
ejection of the third star during which the newly contracted binary
begins a series of dissipative periastron passages. This can be long
or short depending on the orbital characteristics and the amount of
circumstellar gas. But once the two stars are in contact, the second
phase of actual coalescense and eruption is brief. This is consistent
with observations of V1309~Sco, which erupted and declined to close to
its pre-outburst brightness within about 3~years
\citep{tylenda2011}. However, the resulting star is still out of thermal
equilibrium, so during a third phase it will evolve towards a stable
configuration corresponding to the new higher mass. This will be achieved
on a Kelvin-Helmholtz timescale \\

t$_{KH}$ $\approx \frac{G M^2}{R L} $ \\

where G is the gravitational constant, and  M, R, and L are the mass,
radius, and luminosity of the merger remnant.

For an assumed mass for G2 of $\sim$0.7~M$_{\odot}$ from momentum
conservation with the KPNO~15 walkaway star, and an estimated radius
(assuming an inclination of $\sim$60$^\circ$) of 3.6~R$_\odot$, and an
assumed peak luminosity of 1000~L$_{\odot}$ we find a Kelvin-Helmholtz
time of $\sim$4000~yr. Given the assumptions this number is obviously
uncertain, but it indicates that the star may have observable
peculiarities for several thousand years, consistent with the upper
limit of 5600 yr for the time of the merger. For details on the
post-merger evolution from 3D MHD simulations, see
\citet{schneider2020}.

%\subsection{G2 as a post-FUor?} \label{subsec: fuor}

\begin{figure} 
\begin{center}
\includegraphics[angle=0,width=0.99\columnwidth]{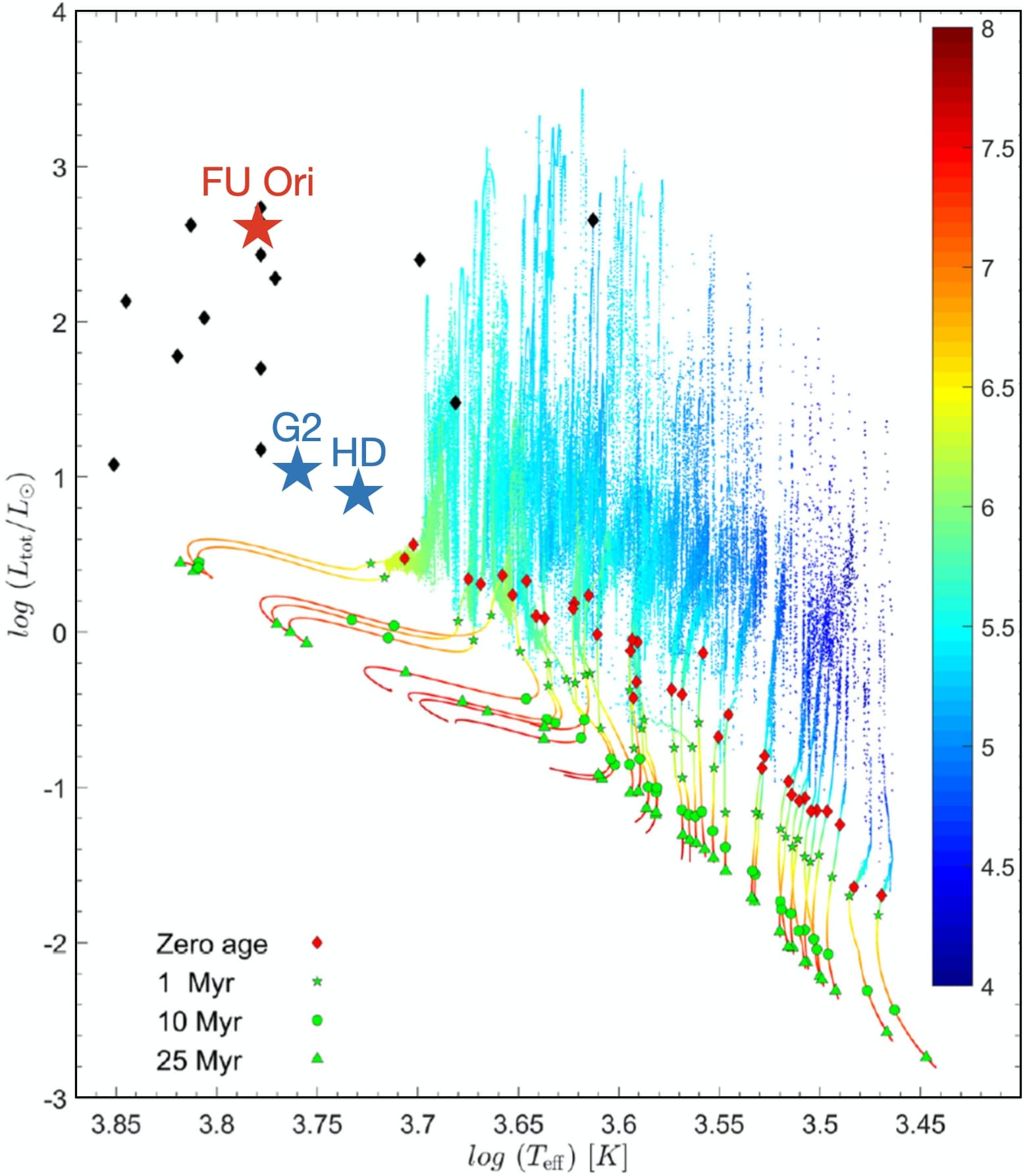}
\caption{The location of a number of FUors (black diamonds) compared 
with evolutionary tracks and models at a range of ages (ages of
models are indicated in the bar) in a figure adapted from Figure~9 of
\citet{elbakyan2019}, see their paper for further details.  The
current location of G2 and HD~283572 are plotted, with temperatures
adopted from their optical spectral types.  Eruptions in low-mass
stars allow them to appear more massive and warmer than their actual
masses would indicate. We argue that G2 and HD~283572 after their
merger events occupied positions further up to the left and as they
are now undergoing re-structuring they gradually move towards the
location of quiescent stars corresponding to their new increased
masses. 
%[fig:elbakyan]
\label{fig:elbakyan}} 
\end{center} 
\end{figure}

The effect of accretion on the luminosity and evolution of low-mass
stars and brown dwarfs has been explored by a number of authors, e.g.,
\citet{vorobyov2017}, \citet{jensen2018}, and \citet{elbakyan2019}. 
In the latter paper, the episodic excursions that low-mass protostars
take in the HR-diagram are compared with observations of FUor
eruptions. Figure~\ref{fig:elbakyan} is from the \citet{elbakyan2019}
paper, and shows the current location of a number of FUor eruptions
(black diamonds) in a theoretical HR diagram.  We have inserted the
current positions of G2 and HD~283572.  They have luminosities that
are roughly similar to what would be expected from young stars with
their optical spectral types, so they have had time to decline from
their peak brightness.

At the time of the mergers, G2 and HD~283572 would have been
significantly more luminous than now, and it is possible that, if they
at the time had been observed, they might have been classified as FUors.  

%+++++++++++++++++++++++++++++++++++++++++++++++++++

In Appendix~F we speculate about the relation between mergers and FUor
eruptions.

%+++++++++++++++++++++++++++++++++++++++++++++++++++

%For a rough estimate of the energy release in merger events of stars
%with different masses see Figure~\ref{fig:merger-diagram}.  Many more
%eruptions have to be studied before observations and models can be
%more reliably tied together.

% \begin{figure} 
% \begin{center}
% \includegraphics[angle=0,width=0.9\columnwidth]{Fig17.jpg}
% \caption{The diagram shows different combinations of masses involved
% in mergers and gives crude order-of-magnitude estimates of the
% gravitational potential energy released in a merger for a wide range
% of masses from brown dwarfs to high mass stars. Labels are
% in ergs and masses in M$_\odot$. See text for further details.  [fig:merger-diagram]
% \label{fig:merger-diagram}} 
% \end{center} 
% \end{figure}

\section{THE REFLECTION NEBULA}\label{sec:reflection}

\begin{figure}
\begin{center}
\includegraphics[angle=0,width=0.80\columnwidth]{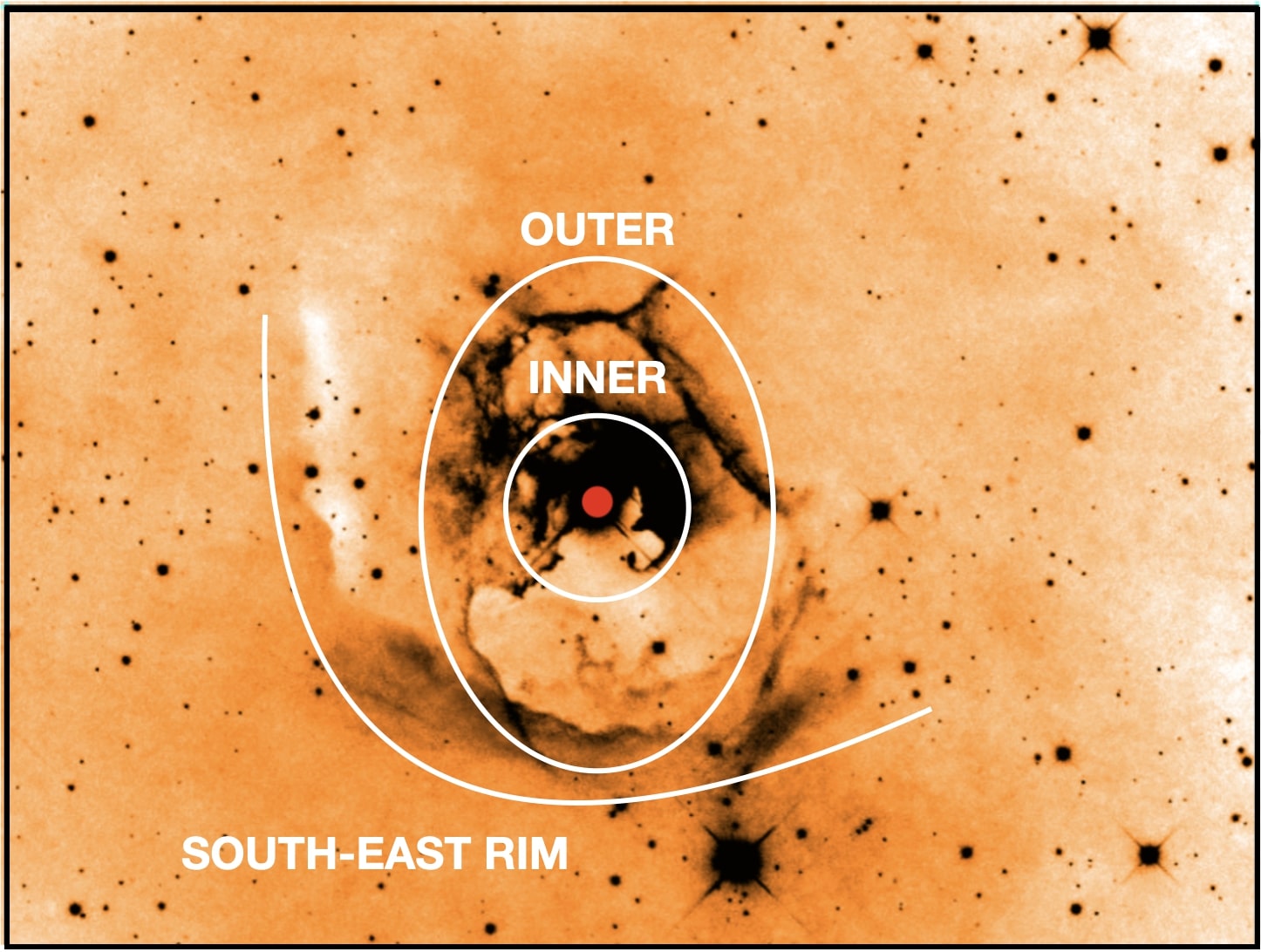} 
\caption{An optical image of the Magakian~77 reflection nebula with the three main rims identified. 
The red dot marks the position of G2. Image courtesy Adam Block.
%[fig:refneb]
\label{fig:refneb}} 
\end{center}
\end{figure}

Our multi-filter APO observations show that the HP~Tau system is
surrounded by a pure reflection nebula, without any shocked emission.
As shown in Figure 1, the reflection nebula consists of a nested set
of partial shells roughly centered on the compact group of stars
consisting of HP Tau, G2, and G3 (Figure~\ref{fig:zaytsev-hanson}).
The brightest emission arises from a corrugated partial arc of
emission which wraps around these three stars and opens-up towards the
south and which we call the inner rim or inner shell (see
Figure~\ref{fig:refneb}).  A protrusion located about 12\arcsec\ north
of G2 exhibits the highest surface brightness.  A sub-cavity with a
radius of about 14\arcsec\ forms an indentation centered on HP Tau.
This inner nebula is surrounded by a dimmer, elongated shell with a
radius of $\sim$100\arcsec\ from HP Tau towards the north and
$\sim$200\arcsec\ towards the south.  This shell has a width ranging
from 90 to 130\arcsec\ in the east-west direction and appears to open,
or simply be much dimmer, towards the south.  In the following, we
refer to this as the outer rim.
%the gourd-shaped shell.  
While the northwest section of the outer rim forms a well defined
straight wall, its northeast portion is filamentary and highly
structured.  Several sub-cavities are seen throughout the nebula.
% surround the star Haro 6-28 and a star 21\arcsec\ to the northwest 
% of Haro 6-28.
The walkaway star G1 is located in the northern part of the inner shell
and is slightly nebulous, indicating its association with the
cloud. The walkaway star KPNO~15 is centered in what appears to be an
evacuated lobe.  Finally, a clam-shell shaped, partial outer rim
envelops the entire group of young stars, here labeled the south-east
rim.

% with a radius of $\sim$160\arcsec\ towards position angle (PA)
% $\sim$135\arcdeg , radius $\sim$260\arcsec\ PA$\sim$45\arcdeg , and
% radius $\sim$100\arcsec\ PA$\sim$125\arcdeg\ surrounds HP Tau, and the
% stars G2 and G3.

The narrow-band \Ha , [\Sii ], and \Htwo\ images failed to reveal any
emission line nebulosity which would have indicated the possible
presence of shocks or fluorescent emission from UV-irradiated cloud
edges.  Thus, the HP Tau system does not currently power Herbig-Haro
objects or molecular hydrogen objects above our
sensitivity limit.

% \begin{minipage}[t]{0.6\textwidth}
% This paragraph of text goes here, occupying about 60\% of the page width.
% You can have multiple lines or even multiple paragraphs here.
% \end{minipage}%
% \hfill
% \begin{minipage}[t]{0.35\textwidth}
%   \centering
%   \includegraphics[width=\linewidth]{2MASS-edgeon-disk.png}
%   \captionof{figure}{A small side figure.}
% \end{minipage}

% THIS IS JOHN'S NEW VERSION NOV 4, 2025

\section{MILLIMETER OBSERVATIONS} \label{sec:mm}

\subsection{Low Velocity \CO\ Streamers from  G2} \label{subsec:outflow}

The HP Tau region has attracted surprisingly few mm studies despite
the concentration of young stars. \citet{moriartyschieven1992} 
made a single $^{12}$CO (J=3-2) pointing towards HP~Tau, and concluded
that there is blue- and red-shifted emission in the region. 

 \begin{figure} \begin{center}
    \includegraphics[angle=0,width=0.99\columnwidth]{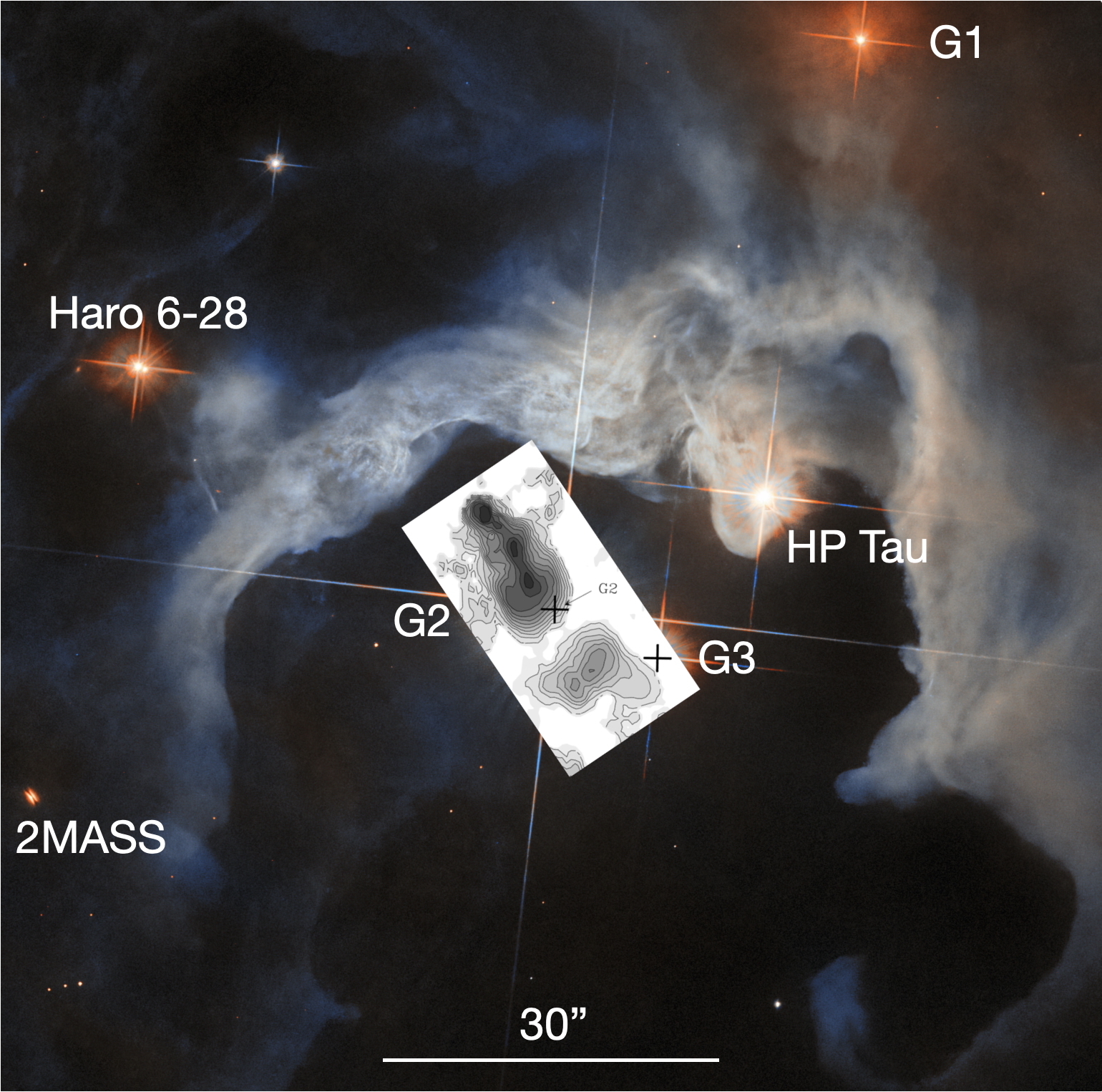}
 \caption{The reflection nebula around G2 as seen with HST 
through several broadband filters 
%[...which filters?...] 
(Credit G. D\^uchene \& G. Kober/ESA). Overlaid is the
 CO(J=2-1) interferometric observations by 
%Duvert et al.(2000) 
\citet{duvert2000} showing the small bipolar outflow emanating from G2
that they discovered with a total extent of about 5,000~AU. The two
crosses represent G2 and G3. The main YSOs are identified. 2MASS
refers to J04355760+2253574, which displays an edge-on disk shadow
(see Figure~\ref{fig:edgeon}).  
%[fig:duchene]
 \label{fig:duchene}} \end{center} \end{figure}
%Figure A5b

\citet{duvert2000} used the IRAM Plateau de Bure interferometer to
observe the $^{12}$CO (J=2-1) line towards G2 with a 3\arcsec\
beam. They found a  compact, very low-velocity entirely redshifted 
flow  with a well defined,  $>$10\arcsec -long ($>$1600 AU)
lobe emanating from G2 towards position angle  
$\sim$25 \arcdeg .   The radial velocity ranges from \Vlsr\ = 8.2 \kms\
near the star to 10.2 \kms\ at the northeast end of the lobe.  In the
counterflow direction, there is  a $\sim$5\arcsec -long ($\sim$800 AU) 
patch which is also redshifted with \Vlsr\ = 10.5 \kms\ 
(Figure~\ref{fig:duchene}).    Thus, this feature only exhibits redshifted 
emission and is not bipolar.

Furthermore,  a bipolar  outflow  is
unexpected for a star apparently with little or no circumstellar
material.   However,   compact bipolar outflows 

are a
characteristic of some recent stellar mergers
\citep{kaminski2015,kaminski2020,schneider2019,schneider2020}.  
It is possible that the low-velocity,   redshifted only  feature may trace 
tidal streamers launched shortly before or during the stellar merger along  
the orbit plane defined by the stars that merged to form G2.

The Taurus outflow survey by 
% Narayanan et al. (2012), 
\citet{narayanan2012} using the FCRAO maps of Taurus \citep{narayanan2008}
%(Narayanan et al. 2008),
did not find a larger-scale outflow. Subsequently, Li et al. (2015)
used the FRCAO data for a deeper survey, and detected blue-shifted
emission in the general region of HP~Tau, which they interpreted as an
outflow (TMO-39).   They likely saw the approaching side
of the expanding shells  (see the following discussion).

Molecular outflows are sometimes associated with optical shocks. 
% Aspin \& Reipurth (2000) 
\citet{aspin2000} obtained [SII] images around HP~Tau, but
found no such shocks. We obtained deeper images with the APO
3.5-m telescope in H$\alpha$ and [SII] and confirm the absence of
Herbig-Haro objects.

\begin{figure*}
 \begin{center}
   \includegraphics[angle=0,width=1.80\columnwidth]{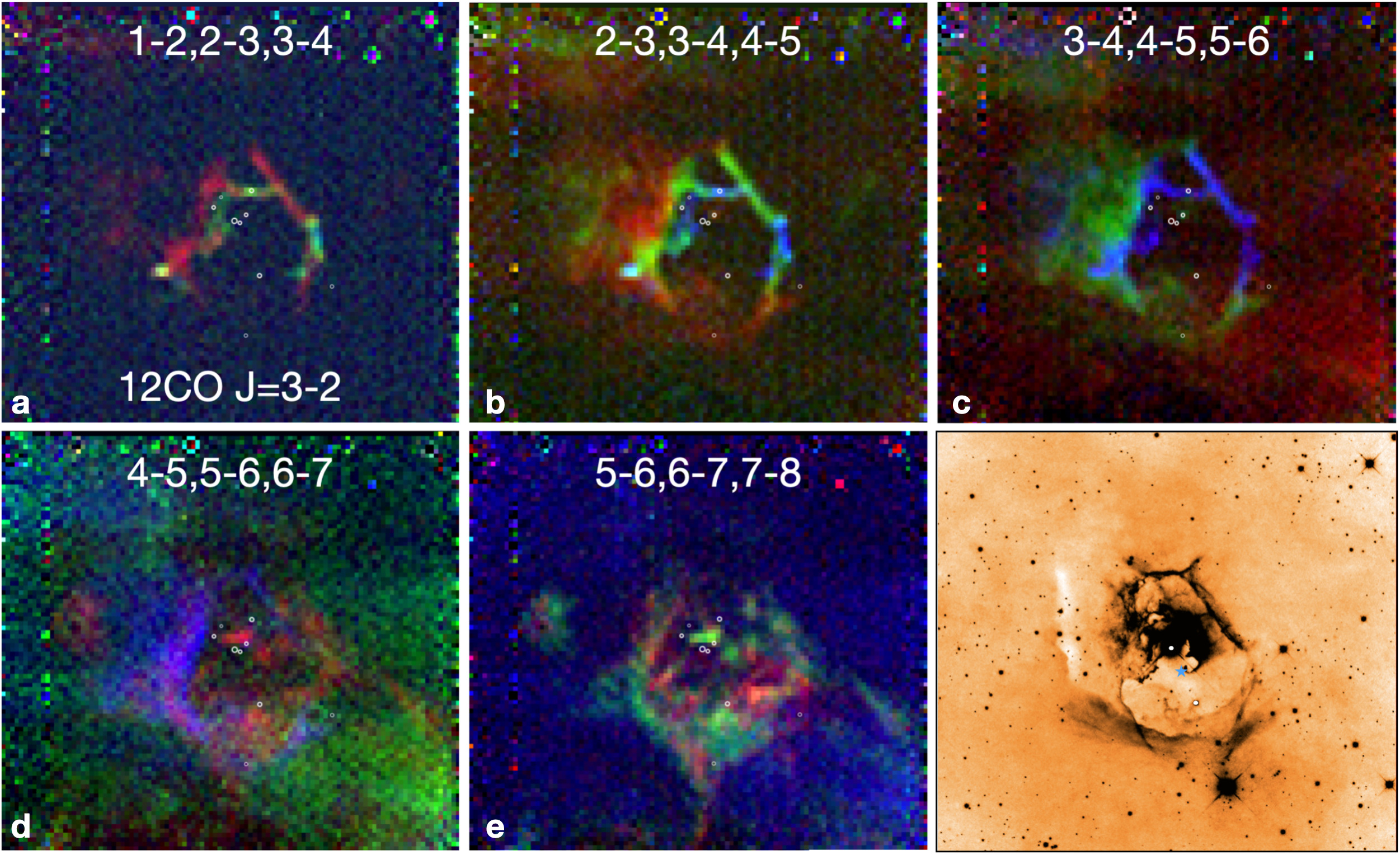}
\caption{$^{12}$CO J=3-2 in 5 channel maps displayed as three-color
images in which three $\sim$1~\kms -wide channels are shown in red,
green, and blue, as indicated. Velocities are in local standard of
rest. The peak of the CO emission is at \Vlsr\ = 5.93 \kms. The
figure shows in finer detail the velocities of the cavity surrounding
G2. The blue asterisk in the last panel shows where the triple system
broke up, and the two white dots represent G2 and KPNO~15. The merger
took place somewhere between the blue asterisk and the current
position of G2. 
\label{fig:johnmosaic}} 
\end{center} 
\end{figure*}

\subsection{New $^{12}$CO and $^{13}$CO Data} \label{subsec:CObubble1}

We  obtained
$^{12}$CO (J=3-2) and $^{13}$CO (J=3-2) maps of the gas around the
area of the Magakian~77 reflection nebula (Figure~\ref{fig:johnmosaic}),
see Section~\ref{sec:obs} for details of the observations.

\begin{figure*}
 \begin{center}
  \includegraphics[angle=0,width=1.90\columnwidth]{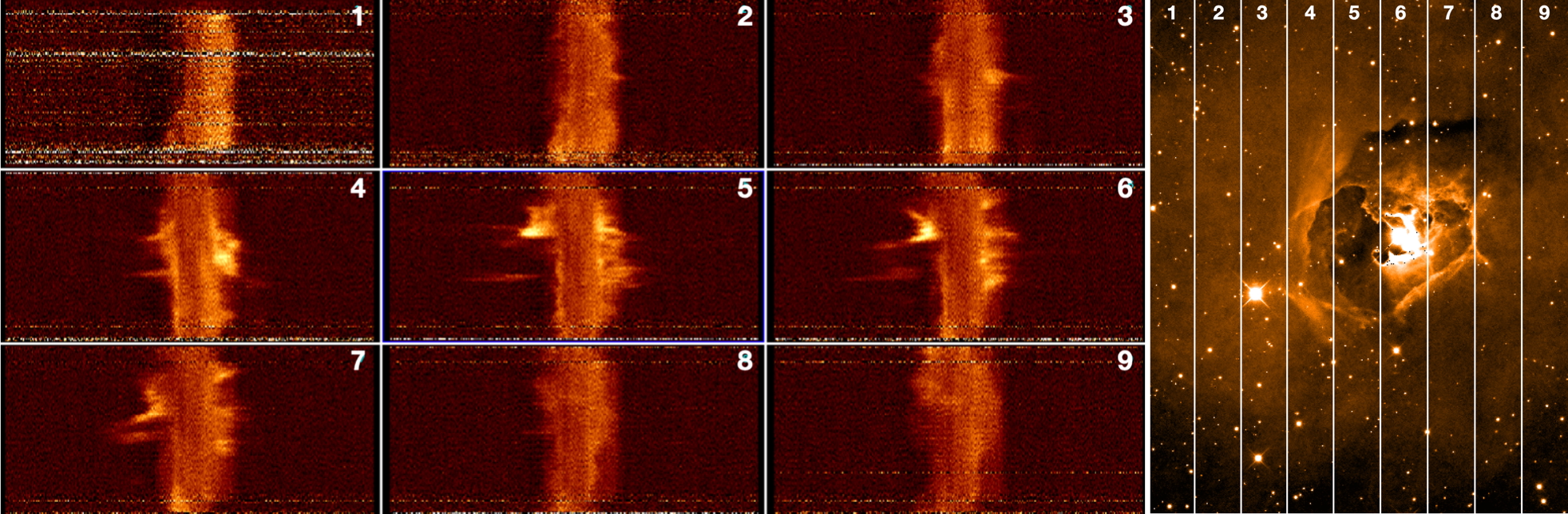}
\caption{{\em left:} Spatial-velocity cuts through the $^{12}$CO data cube, in
which the horizontal axis is the  radial velocity 
ranging  from \Vlsr\ = -1.0 \kms\ on the left to 
\Vlsr\ = +12.0 \kms\ or the right.  The
vertical axis is the right ascension ranging from R.A. = 04:35:22.746
(68.8448) at the bottom of each panel to 04:36:22.790 (69.0950) at the
top.  Each panel is summed over a 75\arcsec\ range in
declination. Panel 1 is the southernmost strip; Panel 9 is
the northernmost strip.  
{\em right:} The locations of the spatial-velocity strips are here shown on a visual-wavelength 
image. North is to the right, and East is up. Image courtesy Adam Block.
%[fig:position-velocity]
\label{fig:position-velocity}} 
\end{center} 
\end{figure*}

\begin{figure}
 \begin{center}
 \includegraphics[angle=0,width=0.60\columnwidth]{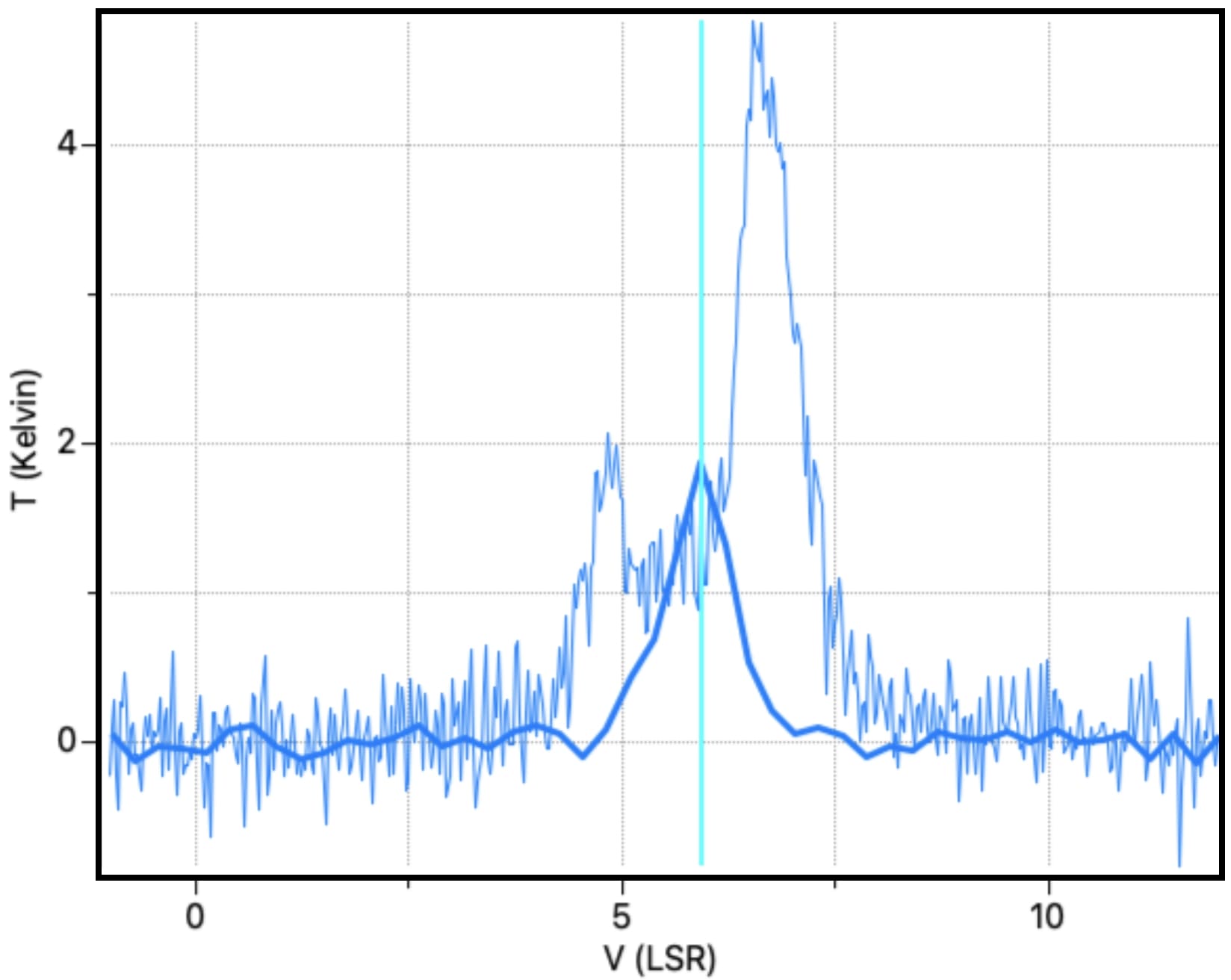}
\caption{Profiles of $^{12}$CO and $^{13}$CO towards a circular area with 
radius 15.5 arcsec and centered at the position 4:35:53.1 +22:54:38. 
The rest velocity of the cloud is V$_{LSR}$ = 5.93 \kms.
%[fig:CO-profiles] 
\label{fig:CO-profiles}} 
\end{center} 
\end{figure}

Figure~\ref{fig:johnmosaic} shows  the kinematics of the
$^{12}$CO emission associated with the HP Tau reflection nebula. 
Figure~\ref{fig:position-velocity} shows spatial velocity cuts
through the \CO\ data cube in which the horizontal axis is the 
radial velocity ranging  from \Vlsr\ = -1.0 \kms\ on the left to 
\Vlsr\ = +12.0 \kms\ on the right.     The vertical axis is the 
right ascension ranging
from R.A. = 04:35:22.746 (68.8448) at the bottom of each panel to
04:36:22.790 (69.0950) at the top.  Each panel is summed over a
75\arcsec\ range in declination.  Panel 1 is the southernmost
strip; Panel 9 is the northernmost strip.
The right panel shows the locations of these
two-dimensional spectra on a visual-wavelength image. Profiles of \CO\
and \tco\ are shown in Figure~\ref{fig:CO-profiles}.

At radial velocities ranging from \Vlsr\
$\sim$5 to $\sim$6.25 \kms , the \CO\ emission is self-absorbed
by  a foreground layer of gas.  
The visibility of the reflection nebula at visual wavelengths
implies that A$_V$  through the foreground sheet 
cannot be  more than a few magnitudes.  
\citet{herczeg2014} measured the extinction
towards five of the young stars, finding a mean A$_V$$\sim$2.65 mag,
implying a column density of gas in front of the stars
of  a few times 10$^{21}$~cm$^{-2}$.    As shown in 
Figure~\ref{fig:johnmosaic}, \CO\ emission at radial velocities which
avoid the absorbing layer trace the visual appearance of the sharp,
inner edges of the reflection
nebula remarkably well.  This indicates that the highest CO 
radial velocities coincide with the bright rims.  

Figure~\ref{fig:position-velocity} shows that Doppler-shifted emission
from the shells protrudes on both the red- and blueshifted sides of
the foreground gas.  This emission can be seen at radial velocities
between \Vlsr\ $\sim$1.0 to 5.3 \kms\ on the blueshifted side and
between 6.7 and 8.0 \kms\ on the redshifted side.  The few \kms\
Doppler shifts are much smaller than what is generally associated with
active outflows powered by forming stars.  Our data does not show
the  outflow seen by \citet{duvert2000} and associated with the star
G2.   Evidently this CO feature  is too beam-diluted in our data to 
be seen by the 15\arcsec\ JCMT beam at our sensitivity.

The sharp, bright rim on the east, north, and west sides of the cavity 
is associated with  \CO\ emission between \Vlsr\ = 2 to 4 \kms .   
At \Vlsr\ = 4 to 5 \kms , \CO\ emission traces emission extending 
$\sim$3\arcmin\ east of star G2 and 2\arcmin\  east of the cavity wall.
The most blueshifted emission (\Vlsr $\sim$1.8 \kms ) 
in the HP Tau shells is located 2.7\arcmin\ east of KPNO~15 at
4:36:03.1, +22:52:48.   A similar blueshifted region occurs on
the west rim at 4:35:43.5, +22:53:47 about 2\arcmin\ northwest
of KPNO15.   These shells disappear behind the foreground  gas at
\Vlsr\ = 5 to 6 \kms .  The east, south, and west rims re-appear at 
\Vlsr\ = 6 to 8 \kms\ on the redshifted side, with  some knots filling-in 
the interior of the cavity  (Figure~\ref{fig:johnmosaic}e).   

The \tco\ data cube probes the gas hidden by foreground \CO\ absorption. 
The upper panel of Figure~\ref{fig:13CO+250} shows a three-color image
of the \tco\ emission from \Vlsr\ = 4.7 to 6.3 \kms.
% of the \tco\ emission from \Vlsr\ = 4.82 to 6.20 \kms
% (actually 4.68 to 6.34 \kms\ because each channel is 0.2767
% \kms\ wide and this figure was made by summing over two channels for
% each of the red, green, and blue frames.)
% (the radial velocities are referenced to the center of each channel).
The brightest \tco\ emission originates from two spots (Peak~1
and 2) close to the central group of stars and associated
with the brightest portion of the HP Tau reflection nebula.   Peak~1 is 
$\sim$25\arcsec\ northwest of  G2 and $\sim$10\arcsec\ north of HP
Tau and has a peak brightness temperature of $\sim$2 K in \tco  .
Peak~2 is  $\sim$40\arcsec\ due west of G3 with a peak brightness 
temperature of also $\sim$2 K.    
At lower intensities between 0.3 to 0.7 K, diffuse \tco\ emission extends 
northwest from Peaks 1  \& 2.  The \tco\ emission peaks at \Vlsr\ = 5.9 \kms 
(corresponding  to V$_{hel}$ = 16.6 \kms ).  

The sharp  inner rims of the reflection traced by \CO\ away from 
the absorbing layer are seen faintly in  \tco\ with peak brightness temperatures
up to $\sim$0.5~K.      The eastern rim exhibits a narrow $\sim$1~\kms -wide
profile centered at \Vlsr $\sim$4.3~\kms\ with a fainter tail extending to about 
6.5~\kms\ at some locations.    The south, southwest, and east  inner
rims shift to $\sim$5.9~\kms\ in \tco .  Toward the south, these rims blend into 
the brighter South-East Rim  near the bottom of  Figure~\ref{fig:zaytsev-hanson}.    
The \tco\  emission from the South-East Rim reaches  $\sim$1.7 K and has  
median intensity of $\sim$1 K.    In \CO , the  South-East Rim is hidden by 
the foreground absorbing layer.   Figure~\ref{fig:13CO+250} shows that this 
clamshell feature is blueshifted with respect to Peak 1 by about 0.5 \kms .
It  has  a radius of about 170\arcsec\ centered near Peaks 1 \&  2 (Fig. 24).      

The \tco\ morphology is similar to  the sub-millimeter wave dust continuum 
traced by the 250 \um\ image (Figure~\ref{fig:13CO+250}). 

%\subsubsection{$^{13}$CO}

% A 100\arcsec\ diameter region centered about 165\arcsec\
% south-southeast of Peak~1 is called Peak 2.  

\subsection{Column Densities and Masses}

We apply the X-factor method \citep{Bolatto2013}
as well as the RADEX code \citep{RADEX2007}
to the estimation of \CO\  column densities.   For the J=3-2 transition,   
\citet{Rigby2025} used 
column densities derived from Herschel Space Telescope sub-mm
dust continuum observations and \CO\ and \tco\ observations
of various regions in the Galaxy
to derive 
X$_{12CO3-2}  = 4.0 \times 10^{20}$  and 
X$_{13CO3-2} = 4.0 \times 10^{21}$ $\rm cm^{-2} (K~km~s^{-1})^{-1}$.   
Thus, N(\Htwo ) = X~I$_{CO}$  ($\rm cm^{-2}$) where I$\rm _{CO}$ is given by
$\int T dV$ (K~\kms )   and T is the brightness temperature 
in Kelvin in each species of CO and the integral is taken over the 
line profile in \kms .  The mass in a region with area A
is then given by $M = 2.7~ m_H~N(H_2) ~A$ where $m_H$ is the
mass of a hydrogen atom.

At radial velocities where most of the \tco\ emission is seen,
\CO\ emission from the HP Tau shells are mostly hidden by the optically 
thick \CO\ foreground layer.    We assume
\tco\ is optically thin, and has an abundance relative to molecular hydrogen,
 [\tco ] / [\Htwo ] $\approx \rm 5 \times 10^{-6}$.  
 
\begin{figure}
\begin{center}
 \includegraphics[angle=0,width=0.77\columnwidth]{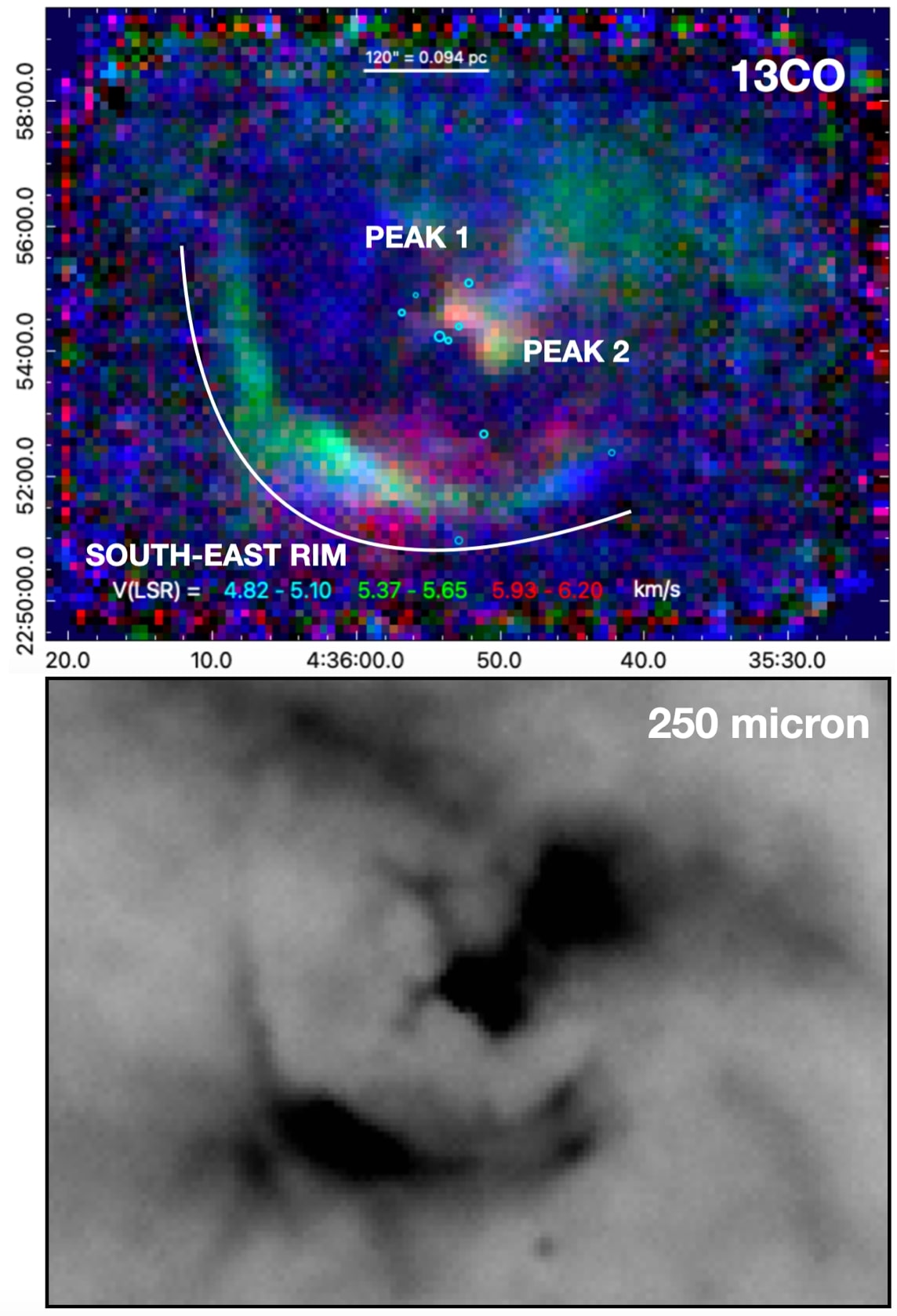} 
\caption{A $^{13}$CO map of the region around the HP~Tau group (upper
panel). Emission from $^{13}$CO is detected only between 4.96 $<$
v$_{lsr}$ $<$ 6.50 {\kms}, and the figure is integrated over this
velocity range. The lower panel is a Herschel 250~$\mu$m image. Both
figures show the same area as the optical Figure~\ref{fig:refneb}.
% [fig:13CO+250]
\label{fig:13CO+250}} 
\end{center}
\end{figure}

We use the X-factor and RADEX  to estimate column densities and 
masses using  \tco\ emission from Peaks 1 and 2 and the 
South-East Rim (the clamshell shaped arc in Figure~\ref{fig:13CO+250}).
Peaks~1 and 2 are close to the stars illuminating the reflection nebula
and heated by starlight.   Acceptable RADEX solutions are
obtained for gas temperatures ranging from 30 to 50 K and \Htwo\
densities ranging from $\rm 10^3$ to $\rm 10^4$~\cms  .   The
\Htwo\ column densities range from $\rm 10^{21}$ to 
$\rm 3 \times 10^{21}$~$\rm cm^{-2}$ and  yield observed 
brightness temperatures of  
$\sim$2 to 3 K  for \tco .   Mean radii of  12\arcsec\ for Peak 1
and 22\arcsec\ for Peak 2 imply  masses of 0.006 to 0.018~\Msol\
for Peak 1 and 0.02 to 0.06~\Msol\ for Peak 2.  The \tco\ X-factor method
gives \Htwo\ column densities of $\rm 4 \times 10^{21}~cm^{-2}$ and
$\rm 6 \times 10^{21}~cm^{-2}$ for Peaks 1 and 2 which implies masses of 
0.024~\Msol\ and 0.12~\Msol , respectively.

The shells seen at visual wavelengths protrude from behind the
foreground veil at red- and blueshifted radial velocities in \CO .  
Some of these features can be faintly seen in \tco\ at  \Vlsr\ 
$\sim$3.6 to 5 \kms on the blueshifted side of the absorbing layer, specifically from 
portions of the east, south, and west rims.  The mean \Htwo\ column densities in
18\arcsec\ by 106\arcsec\ boxes are $\rm 2 \times 10^{20}$, $8 \times 10^{20}$,
and $\rm 6 \times 10^{20}~cm^{-2}$  corresponding to masses of 0.005, 0.02,
and 0.015 \Msol\ in the east, south, and west rims.  Thus, the total mass in
these rims is at least 0.04 \Msol.   This may be a severe lower bound limited
by the faintness of the \tco\ emission. 

The South-East Rim likely has lower gas temperature since it is
located farther from the sources of illumination.  For temperatures
ranging from 10 to 25 K, and \Htwo\ densities ranging from $\rm 10^3$ to
$\rm 10^4$~\cmq ,  \Htwo\ column densities range from
$\rm 3 \times 10^{21}$ to $\rm 10^{22}$ ~\cmq\ which  yield 
 brightness temperatures of 1 to 2 K.   This implies a mass of
 0.55 to 1.8~\Msol\ for an effective area of 0.0084 square parsecs.
 The X-factor method gives a column density of 
N(\Htwo )=$\rm 6 \times 10^{21}~cm^{-2}$ and mass 1.1~\Msol .

We use the X-factor method applied to \CO\ in a small 370\arcsec\ by
370\arcsec\ box 
% (the size of the region mapped) 
centered on the cavity
to estimate column densities and masses in two velocity ranges that
exclude the foreground absorbing layer to estimate upper bounds.
Between \Vlsr = 2.5 and 5 \kms\ we find a mean column density $\rm
N(H_2) = 5 \times 10^{20}~cm^{-2}$ implying a mass of 0.9~\Msol
. Between 6.5 and 8 \kms\ $\rm N(H_2) = 4 \times 10^{20}~cm^{-2}$
implying a mass of 0.7~\Msol .  These estimates include extended
emission away from the sharp rims, and faint emission in the wings of
the foreground layer
\Vlsr\ 4 to 5 \kms\ gas, and thus are severe upper bounds on the shell masses.  
Integrating over the entire line profile and using the X-factor method to a 
slightly smaller  690\arcsec\ by 690\arcsec\  (0.54 by 0.54 pc) region than mapped in 
both \tco\ and \CO, gives a mass of 5.8~\Msol\ derived for \tco\ or 
7.7~\Msol\ derived from \CO .  Most of this mass belongs to the foreground 
layer.   

In summary, the mass in the inner shells, including Peaks 1 and 2, 
is between 0.1 and 0.3 \Msol .  The mass of  the south-east rim is 1 to 2 \Msol .
The remaining mass belongs to the foreground layer and gas beyond the sharp
rims.

\subsection{Shell Morphology}

The Magakian 77 reflection nebula associated with the HP Tau group
consists of sharp, bright rims surrounding  evacuated cavities. 
There are at least three partial shells: The bright
inner half shell located within $\sim$30\arcsec\ of HP Tau, G2, and
G3, the sharp-edged $\sim$130\arcsec\ by 300\arcsec\ diameter outer
rim seen in Figure~\ref{fig:zaytsev-hanson} and best seen in the
blueshifted CO images, and finally the \tco\ $\sim$160\arcsec\ radius
South-East Rim clamshell feature seen in \tco\ and in the sub-mm dust
emission.  

As the stars accreted their masses a few Myr ago, 
they  likely  powered bipolar outflows
which would have created cavities within the parent cloud at early times.  
Over 2 Myr the cavities left behind may have been filled-in and destroyed 
by turbulent motions within the parent cloud.  Turbulent motions with an
amplitude comparable to the sound speed in cold molecular gas
($\sim$0.2 \kms ), or larger,  would have led cavity walls to expand 
and fill-in the cavities on a time-scale shorter than 0.5 Myr.

But  the luminosity and T Tauri stellar winds emitted over 
$\sim$2 Myr by young  stars in the HP Tau compact group
would have exerted sufficient radiation pressure 
to produce the cavities or to keep outflow-generated cavities open. 
While the merger flash is unlikely to have contributed
much momentum, mass loss during a merger and its aftermath 
would likely have injected a strong pulse which could have created
the sharp rims and cavity walls.  

A spherical,  expanding shell such as would be produced by an explosion
is expected to have low radial velocities
with respect to the host cloud at its projected edges located at the 
greatest projected distances from the expansion center.   Radial 
velocity minima and maxima are expected towards the projected center of the shell.   
In the HP Tau region, the highest radial velocity deviations from the host
cloud occur in  or beyond the bright rims seen at visual wavelengths,  
not  at the center of  the cavity.  
This morphology is more consistent with a hole in a sheet of gas where
the rims of the hole have been accelerated orthogonal to the plane of the sheet.

\subsection{Did the Merger Create the Cavity \& Shells?}

The gravitational potential energy released by a merger of a star of
mass $m$ and radius $r$ with a star of mass $M$ and radius $R$ is roughly given by 
$$ E_G \sim f G M m / (R+r) $$
% which for masses of 1~M$_\odot$ and 0.1~M$_odot$ and radii of 
% \approx 2.7 \times 10^{47} ~f M_1 m_{0.1} R^{-1}_{11} ~~~~~~~~~~~~ (erg) 
where $f$ is a parameter of order unity which depends on the final
mass distribution within the merger remnant.  For the merger of a
1~M$_\odot$ and a 0.1~M$_\odot$ star the energy release is of the
order of a few times 10$^{47}$ erg.

% $M_1$ is the mass of the more massive star in units of a Solar mass,
% $m_{0.1}$ is the mass of the lower mass star in units of 0.1~\Msol ,
% and $R_{11}$ is the final effective radius of the merger remnant in
% units of $10^{11}$ cm.

Note that $E_G$ is orders of magnitude larger than the kinetic energy
of motion in the walkaway stars.  If G2 and KPNO15 have masses of
0.7 and 0.45 \Msol ,  and space motions of 5.7 and 9.3 \kms\, (see Table~2), their kinetic
energies are $\rm 2.3 \times 10^{44}$ ergs and  $\rm 3.9 \times 10^{44}$ ergs.

Mergers typically produce a luminous, transient flare lasting anywhere
from weeks to years.  The peak luminosity can range from $10^{4}$ to
over $10^{7}$~\Lsol .    A class of red transients with luminosities between
novae and supernovae  (Luminous Red Novae) are thought to be 
powered by merging stars  \citep{kasliwal2017,karambelkar2023,karambelkar2025}.  
Recent examples include V838 Mon \citep{liimets2023,kaminski2021}, V1309 Sco
\citep{tylenda2016,kaminski2015} (see Appendix~F), and NGC~4490-OT2011
\citep{smith2016}.    The OMC1 cloud core behind the Orion Nebula 
experienced a massive explosion, likely powered by a  protostellar merger 
$\sim$550 years ago \citep{bally2020}.  

Low-luminosity mergers with $M_V$ = -3
or $M_I$ = -4 are relatively common with a Galactic rate of order 0.5
to 0.3 yr$^{-1}$.  Luminous mergers $M_V$ = -7 or $M_I$ = -10 have a
rate of 0.03 yr$^{-1}$; the peak luminosity scales as $M^{2 - 3}$.
For massive stars, the peak luminosity is typically $10^2$ to $10^4$
times the main sequence luminosity of a star having the mass of the
merger remnant \citep{kochanek2014}.

The momentum in the flash is $ P_{rad} = E_G / c$.  For M=1 \Msol  , m=0.1 \Msol,
and f=1 it would be $P_{rad} \approx 9 \times 10^{36}$~ erg~cm~s$^{-1}$ or
0.044 \Msol \kms . 
The maximum speed to which a shell with mass $M_{shell}$  can be accelerated is 
$V_{shell} = P_{rad} /
M_{shell}$, or 
$$
 V_{shell} = f \eta G M m / c R M_{shell}  ~~~~~~~~~~~~~~~~~~~~~~~~~
 $$
where $\eta$ is the fraction of the merger energy that goes into
radiation during the transient.  
The total mass in the shells is estimated above to be $\sim$0.1 to 1~\Msol . 

For m=0.1~\Msol , M=1~\Msol , $\rm R = 10^{11}$~cm,   f=1,  $\eta$=1,
and $\rm V_{shell}$=0.1~\Msol, the maximum speed would be only
0.44~\kms .  
But almost certainly, $\eta <<1$  since most of the
energy of the merger is likely to be advected into the merger remnant 
rather than radiated away in a  prompt flash.    However,  as the merger 
remnant settles toward the main sequence, the merger energy will 
emerge as radiation on a Kelvin-Helmholtz time-scale.
Thus, unless the merger which produced G2 involved a pair stars more massive
than 1 \Msol\ each,  the merger flash is unlikely to be responsible for forming
the HP Tau cavities.     

The total  radiation emitted by the dozen stars
in the HP Tau group over 2 Myr, however could have created the shells.   
The total mass in stars is at least several solar masses.  The total 
radiation emitted by the gravitational contraction and formation of these stars is
of order $\rm E_g \sim G M^2/R >  3 \times 10^{48}$ erg where M is the
total mass of the stars in the HP Tau group.   We assume a severe lower-bound 
on the total mass of M=1~\Msol\ and 
use $\rm R$ from above as an effective radius.
If radiated over 2 Myr, this energy could result in a  mean luminosity of $\sim$10~\Lsol .     
The maximum speed to which a shell with mass $\rm  M_{shell}$ can be accelerated
is $\rm V_{shell} = E_g / c M_{shell}$.  For $\rm  M_{shell}$=0.1~\Msol , 
$\rm V_{shell} \sim$ 4 \kms .    

Young stars produce stellar winds.   The mean  mass-loss rate from young 
T-Tauri star winds is  $\rm \dot M \sim 10^{-9}$ to $10^{-8}$~\Msol ~yr$^{-1}$
with a terminal wind velocity $\rm V_w \sim $ 100 to 200 ~\kms , the momentum 
delivered to the surrounding ISM over 1 Myr would be 0.1 to 2 \Msol~\kms . 
Thus, over 2 Myr, ten stars would deliver 2  to 40 \Msol~\kms .  

Post-merger spectra show that merger remnants typically eject 1 to
10\% of their mass with a speed of a few hundred \kms 
\citep{geballe2025, kochanek2014}.  Since most observations are taken
only a few years to decades after the merger, the ejecta may still
be decelerating as they climb out of the remnant's gravitational potential well.  The
fraction that escapes may contribute momentum to the formation and
acceleration of cavity walls.  

If the G2 merger ejected a wind with a total mass $M_{w}
\sim 0.01$~\Msol\ with a velocity at infinity of $V_w$ = 100 \kms ,
the momentum delivered by such a wind to the surrounding ISM is $P_w
\sim 2 \times 10^{38} M_{w}$~erg~ cm~s$^{-1}$ or 1 \Msol \kms .
For these parameters, and 1\% of the merger remnant's 1 \Msol\ mass
ejected, the wind momentum would be much larger than the momentum in the
merger's transient radiation.  Thus, if mass loss from the G2 merger was similar to 
what is observed in other luminous red novae, the wind momentum would
dominate over radiation.  

The combined effects of protostellar bipolar outflows,  radiation pressure,  and 
T Tauri winds may have created and maintained  the HP Tau cavities.
Radiation pressure and winds from the G2 merger several thousand years 
ago could have created the sharp, arc-second scale
rims surrounding the HP Tau cavities, and led to their acceleration to a speed of 
a few \kms .

%============== END OF JOHN'S NEW WRITEUP OF NOV 4, 2025 ================

\section{SUMMARY AND CONCLUSIONS} \label{sec:conclusions}

% FIGURE OUT HOW TO SLIGHTLY INDENT THE CONCLUSIONS

      {\em 1.} We have carried out a detailed study of the bright
      T~Tauri star HP~Tau/G2, which is part of a little compact
      aggregate of 12 young stars and brown dwarfs associated with
      HP~Tau. HP~Tau/G2 is at the center of a peculiar shell-like
      reflection nebula.

\vspace{0.15cm}

      {\em 2.} Gaia proper motions show that HP~Tau/G2 and KPNO~15,
      when referenced to the rest frame of the cloud, move straight
      away from each other with a tangential velocity difference of
      $\sim$14~\kms. These two walkaway stars were in a common point
      $\sim$5600~yr ago. The ratio of proper motions for the two walkaway
      stars yields their mass ratio. Assuming a mass of 0.45~M$_\odot$
      for the M3 star KPNO~15 this suggests that HP~Tau/G2 has a mass
      of $\sim$0.7~M$_\odot$. This contrasts with the mass of
      $\sim$1.9~M$_\odot$ derived from evolutionary tracks. Similarly
      the age of G2 of 3.6-5.0~Myr derived from evolutionary tracks is
      in conflict with the age of 2.0$\pm$0.3~yr determined for the
      surrounding group of stars.

\vspace{0.15cm}

      {\em 3.} HP~Tau/G2 is an unusual star. In addition to being a
      walkaway star, it is a weak-line T~Tauri star with spectral type
      G2 with low-amplitude variability indicating the presence of spots. 
      It also has an exceptionally large difference of 1400~K between
      temperatures derived in the optical and in the near-infrared. It
      is a fast rotator with a \vsini around 130~\kms\ and a period of
      1.2 days, yielding a \rsini of 3.1~R$_\odot$. For an estimated
      inclination of roughly 60$^\circ$ it follows that HP~Tau/G2 has
      a large radius of 3.6~R$_{\odot}$ and an oblate shape leading to
      significant gravity darkening.  HP~Tau/G2 is also a nonthermal
      radio source, indicating diffusive shock acceleration to mildly
      relativistic velocities, and it is an X-ray source. Finally,
      Gaia DR3 suggests that HP~Tau/G2 may have a faint unresolved
      companion. While all these characteristics in isolation are not
      particularly remarkable, in combination they indicate a peculiar
      object.

\vspace{0.15cm}

      {\em 4.} In order to understand the nature of HP~Tau/G2 we have
      been seeking guidance among late-type contact binaries and
      related stars. While W~UMa stars have no resemblance to
      HP~Tau/G2, we have found an almost perfect match with the
      evolved star FK~Com. The FK~Com stars form a small group of
      stars interpreted as mergers of contact binaries, and thus
      descendants of W~UMa stars. The similarity of HP~Tau/G2 to the
      FK~Com stars, its location at the center of a cavity, the
      discrepant mass from evolutionary tracks vs momentum
      conservation of the HP~Tau/G2 and KPNO~15 walkaway pair,
      together with its unusual properties have led us to propose that
      HP~Tau/G2 is a freshly spun-up merger remnant from the
      coalescence of two low-mass M-dwarfs or, more likely, an M-dwarf
      and a brown dwarf.

%       The paucity of
%       FK~Com stars might suggest that their characteristics are
%       relatively shortlived as they re-organize their interiors until
%       they have the appearance of normal, albeit somewhat more massive
%       stars.

% {\em 5.} We argue that HP~Tau/G2 is a merger of two members of a
%       quadruple system that broke up, sending KPNO~15 in one direction
%       and a compact hierarchical triple system in the other. The
%       break-up caused a contraction of the remaining triple system,
%       which subsequently became hierarchical when one companion
%       (detected only by Gaia) was ejected into a bound orbit and
%       leaving the remaining binary in an even tighter eccentric
%       orbit. This configuration led to a series of increasingly close
%       periastron passages. Due to a dissipative environment the close
%       binary began to spiral in, and ultimately the two components
%       merged, appearing as the peculiar star G2 with a close
%       (unresolved) companion only detected by Gaia. This scenario can
%       explain the available observations.

\vspace{0.15cm}

      {\em 5.} Coalescence of two stars in a binary can occur when the
      Darwin instability has been triggered. Tides will try to spin up
      the primary star at the expense of orbital angular momentum, but
      the spin of the orbit changes faster than the spin of the
      primary, so synchronization is never achieved. As a consequence
      the orbit shrinks and the primary spins up. At some point the
      stars may reach contact and the components will merge on an
      orbital time scale. Binaries with a very small mass ratio are
      much more likely to become Darwin unstable and merge.

\vspace{0.15cm}

      {\em 6.} Such a merger could be the result of the dynamical
       evolution in a higher-order multiple system of low-mass stars. 
      It is impossible thousands of years later to say with any
      certainty how this chaotic process unfolded. During the
      protostellar phase random motions within an initially
      non-hierarchical configuration would after typically
      100~crossing times lead to ejections into both bound orbits and
      escapes, mostly of low-mass objects.  As a result of these
      ejections some of the remaining binaries became highly
      eccentric. In the presence of a dissipative environment the
      eccentric binaries could further tighten, especially during
      periastron passages, and begin to spiral in.  The remaining stars
      would settle into a loosely bound and only marginally stable
      hierarchical architecture.  Simulations show that such a system
      can survive for several million years and eject members,
      sometimes into escapes and sometimes again and again into very
      distant but bound orbits. In the case of HP~Tau/G2 we speculate
      that after $\sim$2~million yr the hierarchical but fragile
      multiple system was perturbed, internally or externally,
% possibly by the passage of the more massive star HP~Tau, 
      leading to the near-simultaneous escape of KPNO~15 and HP~Tau/G1
      about $\sim$5600 and $\sim$4900~yr ago, respectively. As a
      result both the remaining HP~Tau/G2 and HP~Tau/G3 binaries were
      tightened, and in the case of HP~Tau/G2 two of the three already
      tight components merged maybe $\sim$2000~yr ago, while HP~Tau/G3
      ended up as a spectroscopic binary.

%} %end red

\vspace{0.15cm}

      {\em 7.}  It is important to emphasize that, although the merger
      took place only a few thousand years ago, the dynamical
      evolution leading to this event was initiated already in the
      Class~0 or Class~I protostellar phase, at a time when the
      multiple system was embedded and possessed much circumstellar
      and circumbinary material that was continously being
      replenished.

\vspace{0.15cm}
      
%{\color{red} 

      {\em 8.} The actual merger could in principle have taken place
      at any time within the past $\sim$5600~yr. We surmise that it
      occurred roughly 2000~yr ago, when HP~Tau/G2 would have been
      near the center of the inner sharply outlined cavity. This is
      consistent with HP~Tau/G2 cooling on a Kelvin-Helmholtz time
      scale, which is estimated at several thousand years, during
      which the star presumably could have observable peculiarities
      that would account for the current characteristics of the
      star. For an (uncertain) inclination of 60$^\circ$ HP~Tau/G2
      would be rotating at 77\% of its breakup velocity and as a
      result would be very oblate, with a flattening $R_{p}/R_{eq}$ of
      about 1.3, almost as much as the extreme case of the B3V star
      Achernar.

%      We suggest that HP~Tau/G2 could have formed from an inspiraling
%      binary composed of two late-type stars, for example two
%      $\sim$0.35~M$_\odot$ stars. We note that the G2 mass of
%      $\sim$0.7~M$_\odot$ depends on the ejected star KPNO~15 being
%      single, which is the most common situation when a multiple
%      system breaks up.

\vspace{0.15cm}

      {\em 9.} We have searched for additional stars with similar
      unusual properties as HP~Tau/G2 and have found that another
      young star in Taurus, the luminous G5 weakline T~Tauri star
      HD~283572, is a near-twin. This may suggest that HP~Tau/G2 is
      not a pathological case, but exemplifies an event that is not
      uncommon.

\vspace{0.15cm}

      {\em 10.} We speculate that G2 during its outburst might have
      appeared as an FU~Ori event. The star FU~Orionis is a luminous,
      bloated, fast rotating G2 star with a major polar spot. It has a
      massive disturbed disk which accounts for its large infrared
      excess and high luminosity. G2 is older, much less luminous, and
      presumably what might have been left of its disk was mostly
      obliterated in the merger explosion. While the outburst phase is
      long over, it is now in the process of re-organizing its
      interior to its new larger mass.  We suggest that mergers form one
      more of several mechanisms that may cause FU~Ori-like outbursts.

\vspace{0.15cm}

      {\em 11.} The reflection nebula surrounding the HP~Tau group
      shows three partial shells with sharp inner edges. $^{12}$CO and
      $^{13}$CO observations reveal that the gas is currently slowly
      expanding, possibly but not necessarily triggered by an
      explosion. 

\vspace{0.15cm}

      {\em 12.} We cannot {\em prove}, several thousand years later,
      that HP~Tau/G2 suffered a merger, but we believe we have
      presented evidence that makes this {\em a highly probable
      event}. We recognize that there are a number of unresolved
      issues and uncertainties which may be addressed by Gaia DR4 and
      by further observations.

%\clearpage

%      including the following caveats: (a) we do not
%      understand how or from where the third walkaway star G1 was
%      launched; (b) some merger events produce, at least for some
%      period of time, a large cool envelope so the star appears like a
%      red giant, but we do not see that for G2.

%*********** OUTSTANDING ISSUES:

%{\color{red} 
%see Lynne's comments, in particular the G2 time scale of decay is very long 
%Mention FK Com stars and blue stragglers
%}

\begin{center}
ACKNOWLEDGEMENTS
\end{center}

%\begin{acknowledgements}
     
We thank the referee for helpful comments. 
We are also thankful to Christoph Baranec, 
% he tried to check reality of the HST companion of G2
Fernando Comer\'on, % for help with SPHERE data.
Claus Fabricius, % info on RUWE etc
Lynne Hillenbrand, % for discussion about FUors
Ward Howard, % for information on flares
Daniel Huber, % pulsations
Tomasz Kami\'nski, % comments
Nathan Leigh, % mergers
Kevin Luhman, % APOGEE
Andr\'e Maeder, % breakup - shape
Slavek Rucinski, % for discussions and papers
Fabian Schneider, % magnetic fields
Alison Sills, % blue stragglers
Andrei Tokovinin, % for criticism and (unused) speckle observation
Romuald Tylenda, 
Watson Varricatt, % for some observing time, email Nov 2, 2020 
and 
Miguel Vioque % ALMA ACA 1.3mm observation
for help and insightful comments.

We thank Adam Block for the optical image in
Figures~\ref{fig:propermotions}, \ref{fig:refneb},
\ref{fig:johnmosaic}, and \ref{fig:position-velocity},
Vardan Elbakyan for use of Figure~\ref{fig:elbakyan},
 Alexandr Zaytsev and Mark Hanson for the image in
Figure~\ref{fig:zaytsev-hanson}, and C\'edric Mauro and Christophe
Marsaud from Team Astrofleet for Figure~\ref{fig:HD283572}.

% Below is some specifics:

%Christoph Baranec for obtaining an AO image of the G2 region with the Keck telescope,

% Fernando Comer\'on for help with SPHERE data.

%Watson Varricatt for some observing time, see email from him on Nov 2, 2020

This work has made use of data from the European Space Agency (ESA)
mission {\it Gaia} (\url{https://www.cosmos.esa.int/gaia}), processed
by the {\it Gaia} Data Processing and Analysis Consortium (DPAC,
\url{https://www.cosmos.esa.int/web/gaia/dpac/consortium}). Funding for the 
DPAC has been provided by national institutions, in particular the
institutions participating in the {\it Gaia} Multilateral Agreement.
The work presented here is based in part on observations obtained with the
Apache Point Observatory 3.5-meter telescope, which is owned and
operated by the Astrophysical Research Consortium.  We thank the
Apache Point Observatory Observing Specialists for their assistance
during the observations.
The NASA Infrared Telescope Facility is operated by the University of
Hawaii under contract 80HQTR24DA010 with the National Aeronautics and
Space Administration.  MSC is a Visiting Astronomer at the Infrared
Telescope Facility, which is operated by the University of Hawaii
under contract 80HQTR19D0030 with the National Aeronautics and Space
Administration.
Based in part by observations at the Southern Astrophysical Research (SOAR) telescope, which is a joint project of the Ministerio da Ciencia, Tecnologia e Inovacoes do Brasil (MCTI/LNA), the US National Foundation's NOIRLab, the University of North Carolina at Chapel Hill (UNC), and Michigan State University (MSU).
The James Clerk Maxwell Telescope is operated by the East Asian
Observatory on behalf of The National Astronomical Observatory of
Japan; Academia Sinica Institute of Astronomy and Astrophysics; the
Korea Astronomy and Space Science Institute; the Operation,
Maintenance and Upgrading Fund for Astronomical Telescopes and
Facility Instruments, budgeted from the Ministry of Finance (MOF) of
China and administrated by the Chinese Academy of Sciences (CAS), as
well as the National Key R\&D Program of China
(No. 2017YFA0402700). Additional funding support is provided by the
Science and Technology Facilities Council of the United Kingdom and
participating universities in the United Kingdom and Canada. 
Program ID M18BH13A.
This paper includes data collected by the Kepler mission and obtained
from the MAST data archive at the Space Telescope Science Institute
(STScI). Funding for the Kepler mission is provided by the NASA
Science Mission Directorate. STScI is operated by the Association of
Universities for Research in Astronomy, Inc., under NASA contract NAS
5-26555.
This paper includes data collected by the TESS mission, which are publicly available from the Mikulski Archive (MAST).
This paper has used archival data from the Herschel mission. Herschel
is an ESA space observatory with science instruments provided by
European-led Principal Investigator consortia and with important
participation from NASA.
This research has made use of the SIMBAD database, operated at CDS,
Strasbourg, France, and of NASA's Astrophysics Data System
Bibliographic Services.
J.B. acknowledges support by National Science Foundation through grant
No.  AST-1910393.

%\end{acknowledgements}

\vspace{5mm}
\facilities{JCMT(HARP), SOAR (GHTS, TripleSpec),  APO(NICFPS), IRTF (SpeX, iSHELL), Keck(HIRES), ASAS-SN, TESS, HST(WFCM2)}

\software{
CASA  \citep{casa2022}; 
IRAF \citep{tody1986, tody1993};  
PyRAF \citep{pyraf2012}; 
Anaconda Python \citep{anaconda2020}; 
DS9 \citep{joye2003} 
Lightkurve \citep{lightkurve2018}
}

%%%%%%%%%% APPENDIX %%%%%%%%%%%

\clearpage

%\appendix

%\vspace{-0.5cm}

\section{APPENDIX~A:  Members of the HP Tau Multiple System}

The stars HP Tau/G2, KPNO~15, HP~Tau, and HP Tau/G1 are discussed in
Section~{\ref{sec:walkaways}}, and XEST~08-033 in
Section~{\ref{sec:walkaway}}. 
The numbers below refer to Table~1.

{\em 1: FF Tau}: This was first recognized as a T~Tauri star by   
\citet{jones1979}.
\citet{herczeg2014} has determined an optical
spectral type of K8. Identified as a Taurus member by 
\citet{jones1979}
FF~Tau is a binary, detected by Simon et al. 1987 
\citet{simon1987} and
\citet{richichi1994}, with a separation of 36.3 mas and a
$\Delta$M~1.03 mag in a K$_p$ filter 
\citep{kraus2011}.
A dynamical
mass determination by 
\citet{rizzuto2020} yielded a total system
mass of 1.129$\pm$0.027~M$_\odot$.

{\em 2: XEST 08-049}: This is a variable X-ray source  
\citep{guedel2007}, also known as 2MASS J04355286+2250585. It was identified as
a young star by \citet{scelsi2007,scelsi2008}.
%Scelsi et al. (2007, 2008).  
\citet{luhman2017} determined a spectral type of M4.25. 
%{\color{red} *** who found it to be a YSO? Luhman? ***}

{\em 3: HQ Tau}: \citet{herczeg2014} have determined a
spectral type of K2.0. \citet{poully2020} has studied accretion in
the star, and find that a Kepler-K2 light curve exhibits a period of
2.424 d. They also suggest that the star may be a spectroscopic
binary. Long term monitoring with KELT shows a dipper light curve
\citep{rodriguez2017}.
% submm photometry disk Andrews+2013\\
%Radial velocity D.C. Nguyen et al. 2012: 16.65 +/- 0.11 {\kms}\\

%{\em 4: KPNO 15}:\\
%WALKAWAY star Luhman 2018\\

{\em 5: KPNO 9}: Also known as 2MASS J04355143+2249119. This is a brown dwarf with spectral type M8.5V discovered by \citet{briceno2002}.

{\em 6: XEST 08-033}. Also known as 2MASS J04354203+2252226. This is a
medium-strong H$\alpha$ star from Strom et al. (1986) and identified
as an X-ray source by \citet{guedel2007}. \citet{luhman2017} finds a
spectral type of M4.75 and \citet{guo2019} suggest M6. Discrepant
radial velocity measurements from APOGEE suggests that the star is a
spectroscopic binary.

{\em 8:} 
% INCLUDE MENTION OF ANOMALOUS VELOCITY
% although formally not a walkaway star, has an anomalous velocity, and
% is thus included in this section.
{\em HP~Tau} was discovered as a variable star by \citet{badalian1961} and
recognized as an H$\alpha$ emission star, LkH$\alpha~$258, by
\citet{herbig1972}. It has an optical spectral type of K4 according to 
\citet{herczeg2014}, who estimate a mass around 1~M$_\odot$. Our
spectrum, which also indicates a K4 spectral type, shows strong
H$\alpha$ emission with W(H$\alpha$) = 9.1\AA, and it also has
H$\beta$ in emission. \citet{lin2023} present a LAMOST spectrum, which
shows only H$\alpha$ in emission, and they determine a low accretion
rate. Our near-infrared spectrum shows clear Brackett$\gamma$ emission
and strong Paschen$\beta$, $\gamma$, and $\delta$ lines in
emission. It is a close binary found through lunar occultation with
projected separation 17$\pm$1~mas by \citet{richichi1994}, which
corresponds to $\sim$3~AU at the distance of the star.
\citet{cody2022} examined the K2 light curve which shows it is an aperiodic
dipper, with a characteristic timescale of variability of
42.09 days and a normalized flux amplitude of 0.26. \citet{garufi2025}
note that HP~Tau has a very high 2~cm flux that deviates
significantly from a relation between free-free emission and mass
accretion rate, and speculate that it might have undergone a recent
episode of high accretion, or that the radio flux is dominated by
synchroton emission. Our 850~$\mu$m observation reveals a $\sim$100~mJy source, indicating a circumstellar mass of $\sim$0.01~M$_\odot$ of dust and gas.

{\em 9: HP Tau/G3}: 
G3 was identified as a T~Tauri star by \citet{cohen1979}, and 
\citet{herczeg2014} determined a spectral type of M0.6.
 It was found to be a close binary by \citet{richichi1994}, and
\citet{kraus2011} measured a separation of 30.5$\pm$1.5
mas, corresponding to a projected physical separation of about 5~AU, 
and a magnitude difference at K-band of 1.44. 
\citet{rizzuto2020}  determined a total dynamical mass of the system of
1.005$\pm$0.053~M$_\odot$, a period of 27.34~yr and an eccentricity of
0.521. It is an X-ray source \citep{guedel2007}.

{\em 11: Haro 6-28} was identified as an H$\alpha$ emission star by
\citet{haro1953}.  Its optical spectral type is M3.1 as determined
by \citet{herczeg2014}. \citet{leinert1993} used speckle
interferometry to resolve it as a 0.66~arcsec binary.  \citet{cody2022} 
analyzed its Kepler K2 light curve and found a period of
21.56 days with a normalized flux amplitude of 0.23.

\begin{figure} 
\begin{center}
\includegraphics[angle=0,width=0.25\columnwidth]{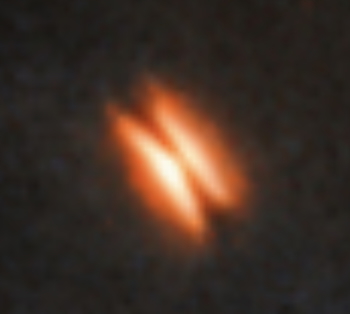}
\caption{
The M5 star 2MASS J04355760+2253574 has a $\sim$1.5$''$ edge-on 
disk shadow bifurcating a compact reflection nebula,  as revealed on the HST image seen in Figure~\ref{fig:duchene}. North is up and east is left.
\label{fig:edgeon}} 
\end{center} 
\end{figure}

{\em 12: 2MASS J04355760+2253574}: This star was identified as a young
very low mass object with spectral type M5 by \citet{guieu2006}. 
It is a periodic dipper with a period of 7.81 day and a normalized flux
amplitude of 0.23 mag \citep{cody2022}. 
On an HST image (Figure~\ref{fig:duchene}) it is seen to display the shadow of an edge-on disk with two illuminated lobes with a
 width of $\sim$240~AU (Figure~\ref{fig:edgeon}).

%=====================================================================

\section{APPENDIX~B: Proper motions of walkaway stars with different group velocities}

When a multiple system breaks up, momentum is conserved. If the
velocities are known, one can determine the mass ratio.  We deduced in
Section~\ref{sec:walkaway} that G2 is $\sim$1.5 times more massive than
KPNO-15. However, this is sensitively dependent on the correction of
the Gaia proper motion by the group velocity. While the relative {\em
angle} of the motion of two stars can be derived merely by subtracting
the two observed Gaia vectors, the {\em ratio} of the velocities
depends on the group velocity. In Figure~B1 we show the vectors of the
four walkaway stars when corrected for different group velocities. In
Panel~A the group velocity of \citet{kerr2021} was determined from
stars in B213 and L1536, Panel~B shows the result of applying the
group velocity of \citet{luhman2018}, Panel~C is our solution based on 7
nearby stars with small proper motions. 
In case~A, the velocity ratio of G2 and KPNO~15 is 0.56, for case~B,
it is 0.53, and for case~C it is 0.66. The mass ratio of G2 to KPNO~15
thus ranges from 1.52 to 1.89.

\begin{figure}
\begin{center}
\includegraphics[angle=0,width=0.5\columnwidth]{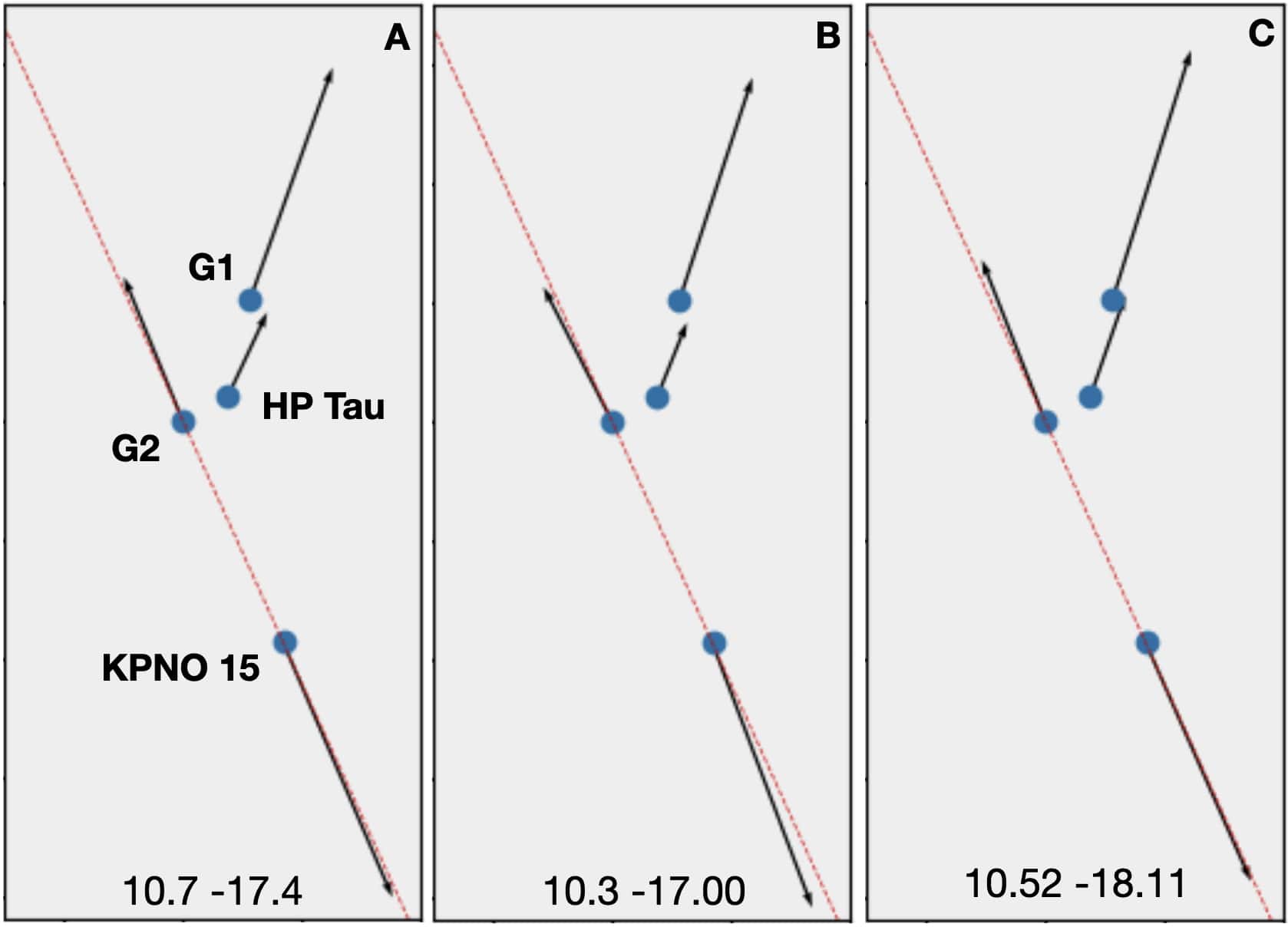} 
\caption{The measured proper motions of the three walkaway stars 
(incl. HP~Tau which, although not formally a walkaway star, has a fairly large proper motion)
relative to
the rest frame of the molecular cloud depend sensitively on what group
velocity is measured for the stars in the group. 
The panels show different determinations, A shows the Gaia
DR2 solution of \citet{kerr2021},
%Kerr et al. (2021), 
B is the Gaia DR3 solution of \citet{luhman2023}, 
%Luhman (2018), 
C is our Gaia DR3 solution based on 7 closeby stars with small proper motions. 
%and D is a fit that arbitrarily shows the best alignment of the
%walkaway vectors of all 4 stars.
The red line is a straight line through G2 and KPNO~15. 
The importance of choosing the correct rest frame is that
the derived mass ratio for a walkaway pair is affected.
The figure is 2.5 $\times$ 5 arcmin.
%[fig:PM-Mosaic]
\label{fig:panel}} 
\end{center}
\end{figure}
%Figure B1

%====================================================================

%{\color{red} 

%\section{APPENDIX~C: A Single Breakup of G2, KPNO~15, and G1?}

% *** Daniel

%Given the presence of three walkaway stars within a very small volume
%we have considered whether these three stars could have originated
%from the single breakup of a small multiple system. However, even with
%a generous interpretation of the Gaia errors, there is no solution
%where the stars were at the same point at the same time. If one
%accepts a 20\% error on the G2 vector, a 10\% error on KPNO~15, and a
%5\% error on G1, there is a solution, but when one applies realistic
%estimates of the masses for the three stars, then momentum is not
%conserved.  It is possible that more accurate data from future Gaia
%releases may provide a fuller understanding of a possible relation
%between these three walkaway stars. It follows that the breakup of G2
%and KPNO~15 occurred separately from the breakup of G1 and G3,
%although separated in time by less than 1000~yr, and hence likely
%triggered by the same perturbation.

%}

%===================================================================

\vspace{0.5cm}

\section{APPENDIX~C: Is G2 Pulsating?}

\begin{figure} 
\begin{center}
\includegraphics[angle=0,width=0.45\columnwidth]{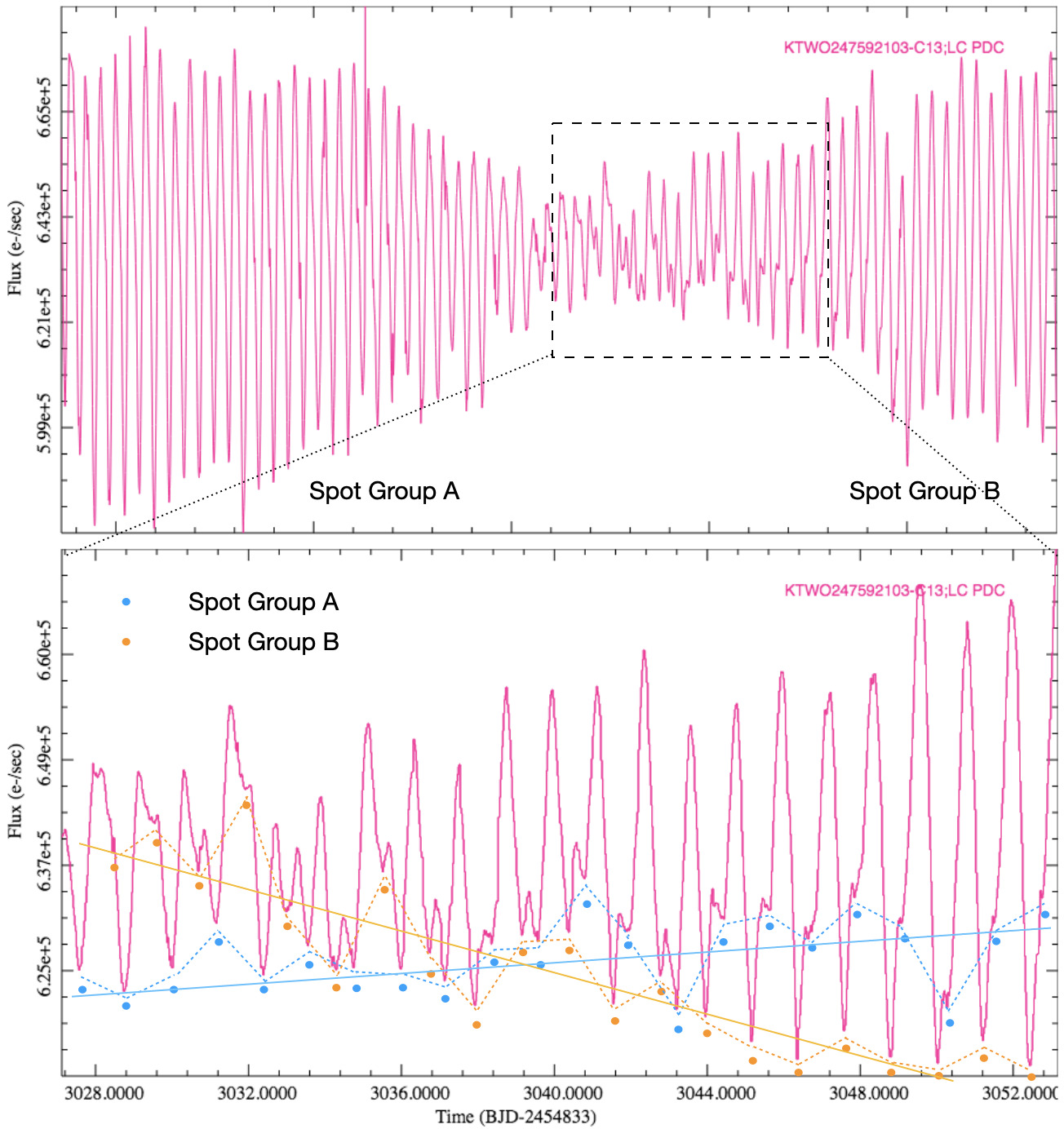}
\caption{The light curve of G2 as observed by the Kepler K2 mission. The upper panel shows the entire light curve observed between March 9 and June 1, 2017, and the lower panel shows 26 days covering the transition phase when spot group A (blue dots, period 1.1978 days) shrinks and the absorption from spot group B (yellow dots, period 1.222 days) becomes dominant. 
%[fig:K2-G2]
\label{fig:K2-G2}} 
\end{center} 
\end{figure}

G2 was observed for 84 days by the K2 mission
(Figure~\ref{fig:K2-G2}-top), much longer than with TESS
(Figure~\ref{fig:TESS-G2}). Based on the K2 data,
\citet{rebull2020} and \citet{cody2022} determined two almost
identical periods, P1=1.1978d and 1.222d. These two distinct periods
can be explained by two separate, large spot groups at slightly
different latitudes in a star with differential rotation. The
photometric variation indicates that at the beginning of the K2 light
curve spot group~A, with P(A) = 1.1978 days, dominates. In the middle
section (dashed box), group~A is gradually fading away, while
simultaneously spot group~B, with P(B) = 1.222 days, grows, and in the final section of the light curve group~B is dominant.

However, the light curve shows a feature that is difficult to explain
by two spot groups with slightly different rotation periods, namely
the bottleneck in the middle of the light curve. The lower envelope
will rise when the spots shrink. But the upper envelope should always
reach maximum light whenever the two spot groups are on the same side
facing away. Only when they are precisely opposite each other would
the upper envelope fade, especially if the spot groups have grown very
large, part of one is always seen. It is also required that the two spot
groups are both changing at the same time to explain the rise of the lower
envelope. Such a situation is possible, but perhaps not common.

On the other hand, such bottle-necks are commonly seen in double-mode
radial pulsations in a star, with two periods creating a beat. And we
note that high-resolution spectra of the H$\alpha$ line of the merger
FK~Com show variability of the profile that has been ascribed to
non-radial oscillations excited by the recent merging of two stars
\citep{welty1994a}.

A simple pulsating model with periods of 1.1978d and 1.222d will
reproduce the large scale behavior seen in
Figure~\ref{fig:TESS-G2}-top while two spot groups with the same
periods will reproduce the detailed structure seen in
Figure~\ref{fig:TESS-G2}-bottom.

Most pre-main sequence stars that are known to pulsate inhabit the
$\delta$~Scuti instability strip \citep[e.g.][]{zwintz2008}, but there
are seven types of pulsations recognized in young stars, either
observationally or theoretically \citep{zwintz2022}. G2 is too cool to
inhabit the instability strips of $\delta$~Scuti and $\gamma$~Doradus
stars, the two most common PMS pulsators. It is, however, close to the
cool edge of the $\gamma$~Doradus stars, and in fact a few pulsators
(not young) are known to inhabit that region in the HR-diagram
\citep{balona2018}. 

The semi-amplitude observed in G2 is $\sim$5$\%$, and that is more
than has been seen in any Gamma Doradus stars \citep{bowman2026}. Also, the 
pulsation-like light curve seen by K2 is not seen in the TESS data
(Figure~\ref{fig:TESS-G2}). Overall, it seems unlikely that G2 is pulsating.

%====================================================================

\section{APPENDIX~D: V838 Mon and V1309 Sco}

The leading interpretation of V838~Mon is that two young ($<$25~Myr)
main sequence binary components of masses $\sim$1.5~M$_\odot$ and
$\sim$0.1-0.5~M$_\odot$ merged \citep{soker2003}.
The star has
an unresolved B3V companion and is also part of a small cluster of young
B-stars (\citet{munari2002}, \citet{afsar2007}.
%Munari et al. 2002, Afsar \& Bond 2007). 
The aftermath of the
spectacular $\sim$10$^6$~L$_\odot$ eruption was followed in detail
spectroscopically and photometrically. Subsequent ALMA observations
revealed a disk-like structure surrounding the star and determined a
projected separation of $\sim$150~AU to the B-companion, indicating
that V838~Mon is the remnant of a binary merger in a triple system
\citep{kaminski2021}.

% Kaminski+2021: V838Mon seen by ALMA
% Smith+2016 - distant extragalactic eruption not relevant 
% Mason+2010 - uninteresting

The eruption of V1309~Sco was extraordinarily well documented, with a
$\sim$10~yr light curve that covers the entire eruption
\citep{tylenda2011}, including several years before the outburst. The progenitor
was a $\sim$1.4~day contact binary with an exponentially decreasing
period prior to the eruption. Analysis of the pre-outburst light curve
indicates that the contact binary had a mass ratio of less than 0.1 
\citep{stepien2011}. 
The luminosity of the star remained $>$10$^4$~L$_\odot$ for a month
during which $\sim$3$\times$10$^{44}$~erg was radiated away
\citep{tylenda2011}.
%(Tylenda et al. 2011). 
At the peak of the eruption V1309~Sco
showed a near-infrared spectrum mimicking an F-giant \citep{rudy2008a},
%(Rudy et al. 2008a), 
which quickly over the next month turned into a late M-giant 
\citep{rudy2008b, mason2010}.
%(Rudy et al. 2008b, Mason et al. 2010). 
This evolution could be a signature of the expulsion of an expanding
envelope from the star. Currently, 17 years after the eruption,
V1309~Sco is a red giant, presumably due to expansion. Analysis of
high resolution spectra show that during the eruption matter was
expelled as dust-laden bipolar ejecta and that the star is surrounded
by a doughnut of gas, which was produced during pre-outburst mass-loss
episodes \citep{mason2022}.
%(Mason \& Shore 2022). 
Far-infrared data from Herschel has confirmed cold dust a few thousand
AU from the star \citep{tylenda2016}.
%(Tylenda \& Kaminski 2016). 
%This dust production has been modeled by \citet{bermudez2024}.
%Berm\'udez-Bustamente et al. (2024).

% Tylenda et al. 2011  propose that the Darwin instability could
% lead to a merger: ``This happens when the spin angular momentum of the
% system is more than a third of the orbital angular momentum. As a
% result, tidal interactions in the system cannot maintain the primary
% component in synchronization anymore. The orbital angular velocity is
% higher than the primary's angular velocity, the tidal forces increase
% their action and rapidly transport angular momentum from the orbital
% motion to the primary's rotation.'' Steinmetz et al. (2024) found that
% an asymmetric outflow is now emerging from the remnant.

\section{APPENDIX~E: FK~COM: H$\alpha$ emission and flares}

The strong H$\alpha$ line in FK Com has 
been studied by many observers,
e.g., \citet{welty1993}, \citet{kjurkchieva2005}, and
\citet{korhonen2009}. It has a complex structure with a broad
absorption $\sim$1000 \kms\ wide overlaid by a double peaked emission
reaching above the continuum with the red and blue components varying
in strength in anti-phase, resulting in an absorption core that shifts
between blue- and red-shift. Given the similarity between G2 and
FK~Com, it is likely that G2's H$\alpha$ profile would be similarly varying
if monitored. 

FK~Com, with a vsini of $\sim$160~\kms , is also the most active of
the little group of FK~Com-type stars. It is a well-known flare star
and the TESS light curve suggests that it has a flare roughly once
every 8 days. Flares have been observed photometrically at optical
wavelengths (...), spectroscopically in the Balmer lines \citep[an
H$\alpha$ superflare was detected by][]{oliveira1999}, in X-rays
\citep{favata1998}, in the ultraviolet \citep{ayres2016}, and a major
flare was detected at millimeter wavelengths by \citet{lovell2024}. 
We note that G2 also shows flares (two flares are seen in the TESS light curve in Figure~\ref{fig:TESS-G2}), and similarly the TESS light curves of HD~283572 show multiple (small) flares.

\begin{figure} \begin{center}
\includegraphics[angle=0,width=0.55\columnwidth]{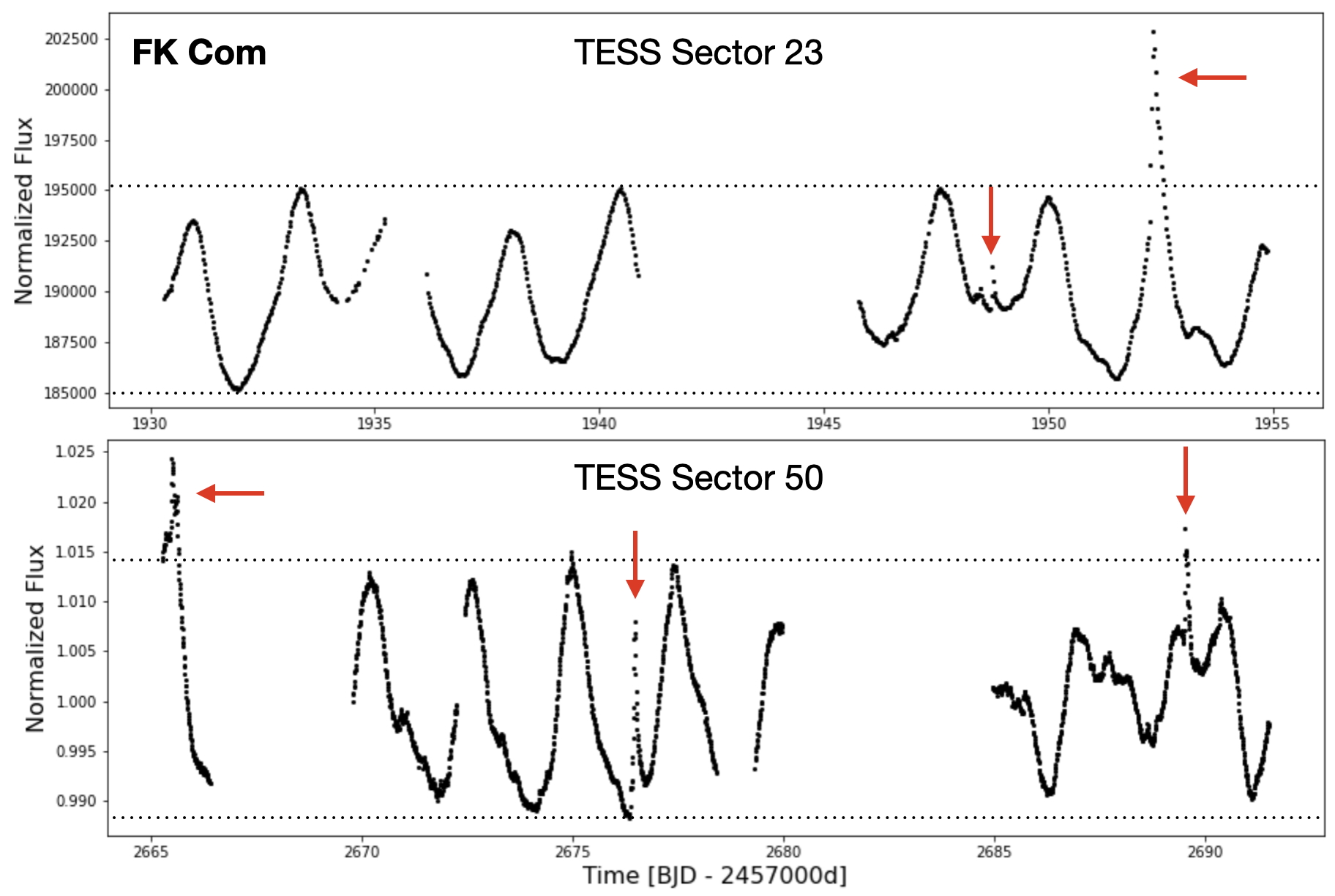}
\caption{The light curve of FK Com as observed by TESS. The
periodocity is well-established, but it is not clear whether the more
detailed structure is due to the presence of a bright very nearby star
or due to technical issues. The more structured variability around
2690 is most likely real. FK~Com is a well-known flare star, and five
flares are seen here, which suggests a rate of roughly one flare per
every 8 days.
\label{fig:TESS-FKCom}} 
\end{center}
\end{figure}

\section{APPENDIX~F: Mergers and FUor Eruptions}

At the time of the eruptions, G2 and HD~283572 would have been much
brighter, and would have been located in the general area of the FUors
in the diagram in Figure~\ref{fig:elbakyan}. We here speculate whether
some FUors could be the result of mergers.

At its most basic level a FUor is a young star
that has undergone a major eruption that led it to be swollen and fast
rotating and with an A-, F-, or G-type optical spectrum displaying prominent
P~Cyg line profiles from a cool wind \citep[e.g.,][]{herbig2003}. 
In the near-infrared they have late-type spectra emitted by an accretion disk. 

We note that when the V838~Mon merger took place, it was initially
suspected of being a FUor based on the high velocity cool winds seen
as P~Cygni profiles at H$\alpha$ and the Sodium doublet, and the
strength of Barium \citep[e.g.,][]{kipper2004}, mimicking a FUor.

\begin{figure} 
\begin{center}
\includegraphics[angle=0,width=0.25\columnwidth]{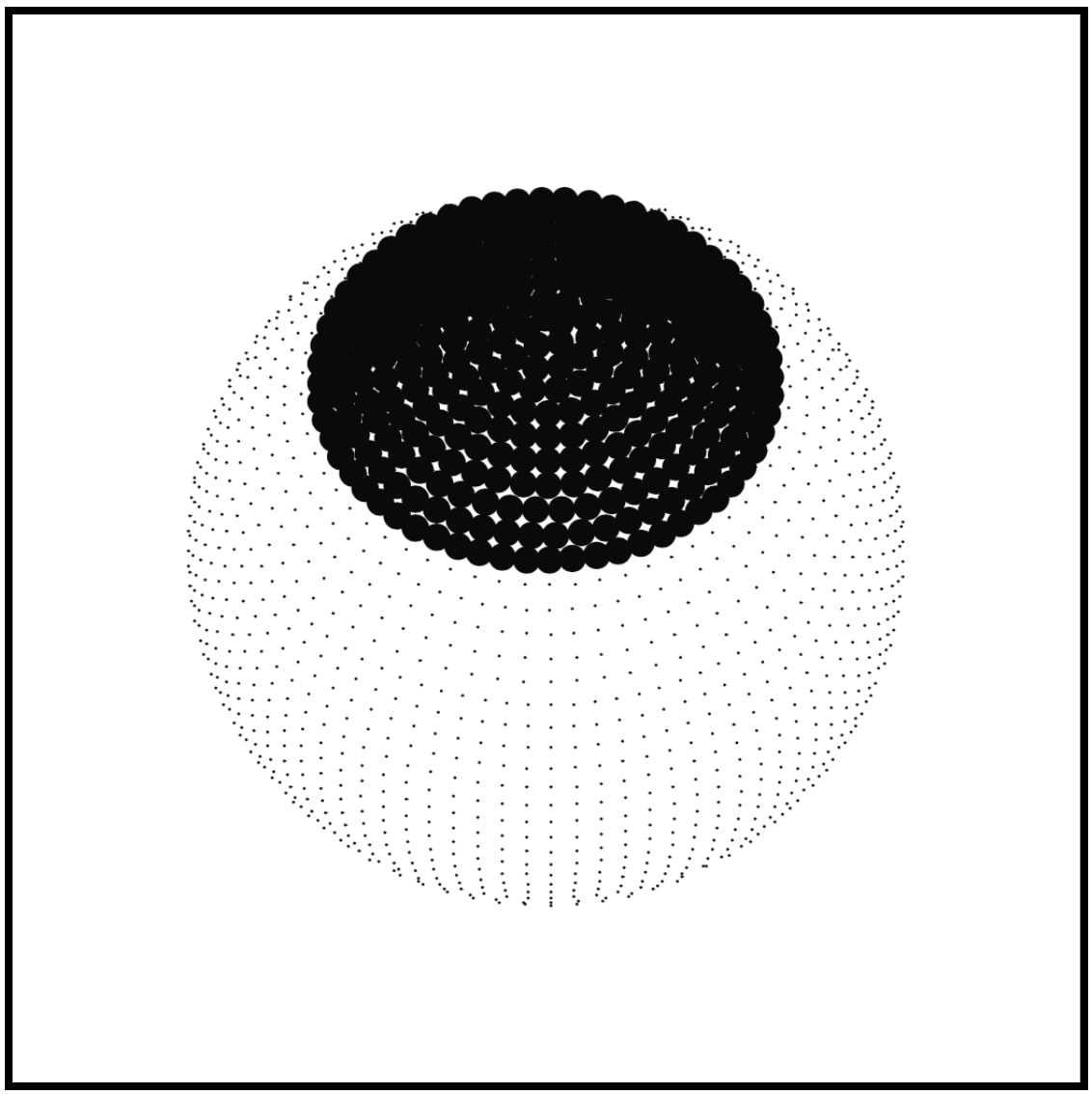}
\caption{A model of the stellar photosphere of FU~Ori from 
\citet{petrov2008} based on stellar line profiles. 
The line profiles reveal a major polar starspot.
%[fig:FUOri-model]
\label{fig:FUOri-model}} 
\end{center} 
\end{figure}

Herbig and Petrov, in a series of important but underappreciated papers,
argue that the metallic line profiles of FU~Orionis in optical high
resolution spectra are well explained as originating in a very
luminous, bloated, and fast rotating star with a major polar star
spot, see Figure~\ref{fig:FUOri-model} and 
\citet{herbig1989}, \citet{petrov1992}, \citet{herbig2003}, 
and \citet{petrov2008}. We are agnostic about whether these optical
lines observed in FU~Orionis originate in the star or from the inner
disk, but merely note that if interpreted as having a photospheric
origin then FU~Orionis would be very similar to what would be expected
from the recent merger of two low-mass stars.

As more and more eruptive events are discovered, the pioneering but
now simple division into FUor and EXor eruptions \citep{herbig1977} is
becoming increasingly inadequate, and many erupting objects are found
with a large range of characteristics \citep[e.g.,][]{hillenbrand2021,
hillenbrand2022, hillenbrand2023}, challenging their classification.

FUors, in the most inclusive sense of the term, can be formed from a
variety of processes \citep[e.g.,][]{vorobyov2021}. We suggest that
stellar mergers can be added to those. Mergers will cause outbursts
that are likely to differ due to individual circumstances, in
particular the range of masses of the merging objects (stars, brown
dwarfs, planets).

It is possible that G2 at the time of its eruption would have appeared
as a FUor eruption, in which case G2 and HD~283572 would be
post-FUors.  If so, then the main difference between G2 as it is seen
today and more recent FUors lies in its evolutionary stage, with
FU~Orionis being much younger and still surrounded by a large cool
disk, while G2 erupted at a later stage and lost what remained of its
circumstellar material in the merger explosion. Hence G2 does not show
the late-type spectrum at near-infrared wavelengths which is
characteristic of most FUors.

It is not known how long FUor eruptions preserve their characteristics.
The longest-lasting FUor known, V883~Ori, erupted
before 1888 \citep{strom1993}, yet still today has a luminosity of
about 220~L$_\odot$ \citep{furlan2016}. FU~Ori erupted in 1938 and
brightened from V$\sim$16.5 mag. to $\sim$9.6~mag. Since then
it has faded only $\sim$0.7~mag. in V. The rate of decline, however,
is variable \citep[e.g.,][]{ibragimov1993}, but if on average the
decline stays constant, then it will take almost 1000~yr for FU~Ori to
fade to its original brightness. At the other extreme, 
V1057~Cyg has since its peak brightness around 1970 already 
faded by almost 4~mag. in V \citep[e.g.,][]{szabo2021}. 

We do not know when the eruption of G2 took place, other
than it was less than 5600 yr, but at least in principle it could be as little
as a few centuries.  

We conjecture that G2 and HD~283572 once appeared as FUor
eruptions and that, as a result of the significant energy release in
the merger, the remaining parts of their accretion disks were
lost. Subsequently the two stars would appear as bloated luminous
stars that began the internal restructuring corresponding to a star
with the combined mass of the precursor binary. This will be completed
on a Kelvin-Helmholtz timescale, although the peculiarities that 
draw our attention today likely would disappear much earlier.

\begin{figure} 
\begin{center}
\includegraphics[angle=0,width=0.35\columnwidth]{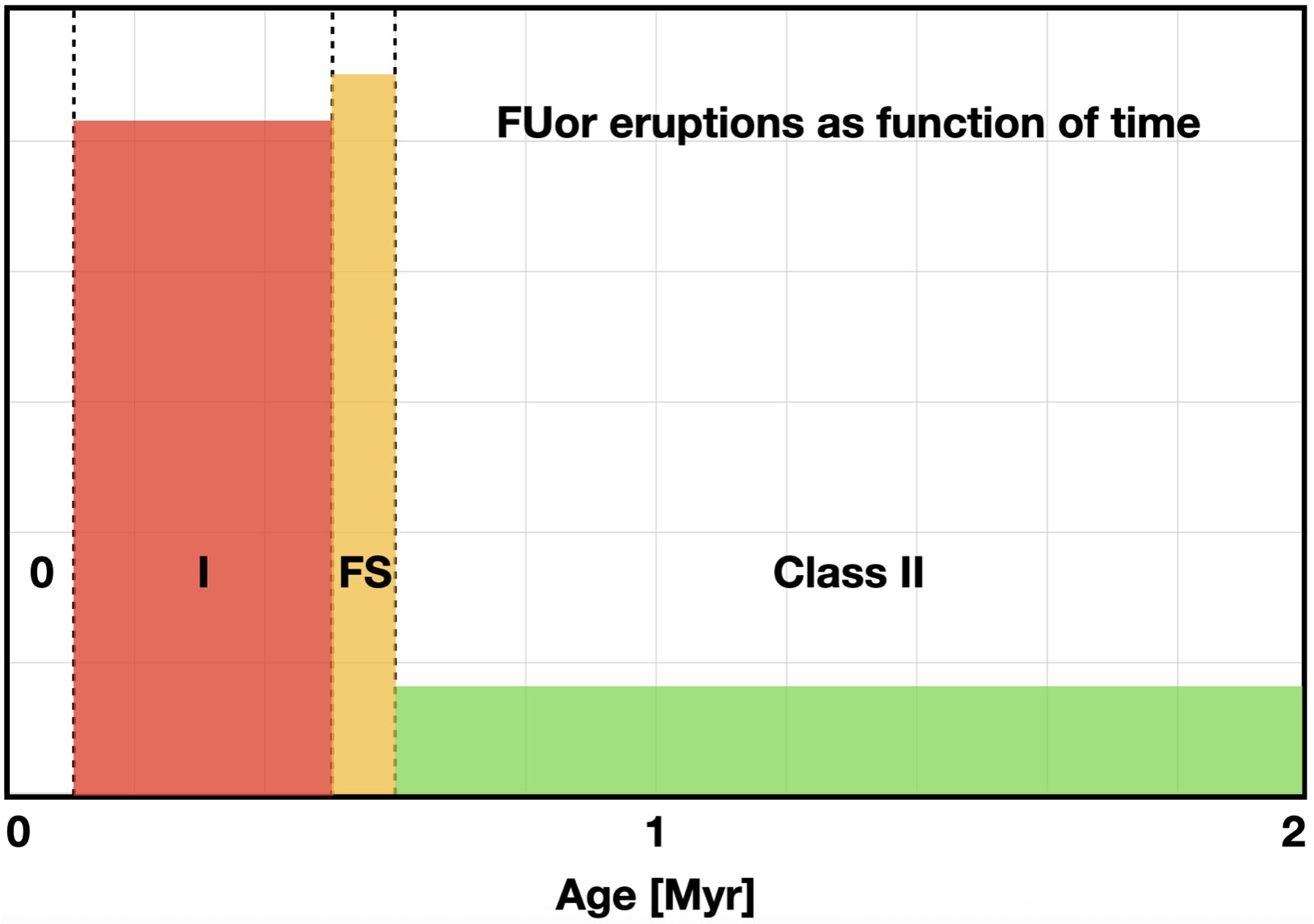}
\caption{ 
The distribution of FUor and FUor-like objects as a function of time. The figure assumes that the Class~0 phase lasts about 0.1~Myr, Class~I about 0.4~Myr, flat-spectrum about 0.1~Myr, and Class~II about 2~Myr. Data from \citet{contreras2025}. 
%[fig:contreras]
\label{fig:contreras}} 
\end{center} 
\end{figure}

Most FUor eruptions occur during the protostellar phase (C. Contreras
et al. 2025, in prep.) and consequently they very often have
substantial accretion disks leading to a strong infrared excess.
Although such outbursts were originally identified among visible stars
\citep{herbig1977}, the most recent work shows that they are far more
common during the protostellar phase. Figure~\ref{fig:contreras}
shows the distribution of FUors and FUor-like objects as a function of
time from the major compilation of young eruptive stars by
\citet{contreras2025}, who tabulated all known major YSO outbursts, 
including 62 (at the time of writing) FUors and FUor-like objects.
This is consistent with the finding through numerical simulations that
triple breakups are much more frequent during the earliest
evolutionary phases.

We wish to stress that we do not suggest that {\em all} FUors are formed in
mergers, just that it is an additional viable explanation for strong
outbursts in young stars.


\begin{thebibliography}{}


\bibitem[M. Afsar \& H.E. Bond (2007)]{afsar2007} Afsar, M. \& Bond, H.E. 2007, AJ, 133, 387
% cluster around V838 Mon


\bibitem[S.M. Andrews (2020)]{andrews2020} Andrews, S.M. 2020, ARAA, 58:483
% disk review


\bibitem[G. Anglada (1996)]{anglada1996} Anglada, G. 1996, ASPC, 93, 3 

\bibitem[Anaconda Software Distribution (2020)]{anaconda2020}
{\em Anaconda Documentation 2020}. Anaconda Inc. Retrieved from https://docs.anaconda.com/

%\bibitem[J.P. Anosova (1986)]
\bibitem[J.P. Anosova (1986)]{anosova1986} Anosova, J.P. 1986, Astrophys. Spa.Sci., 124, 217
% triple evolution


%\bibitem[P. Artymowicz \& S.H. Lubow (1994)]{artymowicz1994} Artymowicz, P. \& Lubow, S.H. 1994, ApJ, 421, 129



%\bibitem[C. Aspin \& B. Reipurth (2000)]
\bibitem[C. Aspin \& B. Reipurth (2000)]{aspin2000} Aspin, C., Reipurth, B. 2000, MNRAS, 311, 522
% search for HH objects around HP Tau

%\bibitem[H. Arce et al. (2011)]
\bibitem[H. Arce et al. (2011)]{arce2011} Arce, H., Borkin, M.A., Goodman, A.A. et al. 2011, ApJ, 742, A105

%\bibitem[T.R. Ayres et al. (2016)]
\bibitem[T.R. Ayres et al. (2016)]{ayres2016} Ayres, T.R., Kashyap, V., Huenemoerder, D. et al. 2016, ApJS, 223:5

%\bibitem[G. Badalian (1961)]
\bibitem[G.S. Badalian (1961)]{badalian1961} Badalian, G.S. 1961, Astron. Tsirk., 224, 22
% discovery of HP Tau as a variable star

%\bibitem[J. Bally et al.(2020)]
\bibitem[J. Bally et al.(2020)]{bally2020} Bally, J., Ginsburg, A., Forbrich, J., et al.\ 2020, \apj, 889, 2, 178. doi:10.3847/1538-4357/ab65f2

\bibitem[L.A. Balona (2018)]{balona2018} Balona, L.A. 2018, Front. Astron. Space Sci. 5:43   doi: 10.3389/fspas.2018.00043


%\bibitem[I. Baraffe et al. (2015)]
\bibitem[I. Baraffe et al. (2015)]{baraffe2015} Baraffe, I., Homeier, D., Allard, F., Chabrier, G. 2015, A\&A, 577, A42

%\bibitem[M.R. Bate et al. (2002)]
\bibitem[M.R. Bate et al. (2002)]{bate2002} Bate, M.R., Bonnell, I.A., Bromm, V. 2002, MNRAS, 336, 705
% Formation of close binaries and orbital decay

%\bibitem[W. Benz \& J.G. Hills (1987)]
\bibitem[W. Benz \& J.G. Hills (1987)]{benz1987} Benz, W. \& Hills, J.G. 1987, ApJ, 323, 614
% 3D hydrodynamic simulations of steller collisions

%\bibitem[M.S. Bessell \& J.M. Brett (1988)]
\bibitem[M.S. Bessell \& J.M. Brett (1988)]{bessell1988} Bessell, M.S. \& Brett, J.M. 1988, PASP, 100, 1134
% Dwarf and giant tracks

%\bibitem[J.H. Bieging et al. (1984)]
\bibitem[J.H. Bieging et al. (1984)]{bieging1984} Bieging, J.H., Cohen, M., Schwartz, P.R. 1984, ApJ, 282, 699
% VLA observations of T Tauri stars. II. A luminosity limited survey ofTaurus-Auriga

%\bibitem[A. Blaauw (1961)]
\bibitem[A. Blaauw (1961)]{blaauw1961} Blaauw, A. 1961, Bull. Astron. Inst. Netherlands, 15, 265
% def of walkaway stars




\bibitem[Bolatto et al.(2013)]{Bolatto2013} Bolatto, A.~D., Wolfire, M., \& Leroy, A.~K.\ 2013, \araa, 51, 1, 207. doi:10.1146/annurev-astro-082812-140944





%\bibitem[I. Bonnell \& P. Bastien (1992)]
\bibitem[I. Bonnell \& P. Bastien (1992)]{bonnell1992} Bonnell, I. \& Bastien, P. 1992, ApJ, 401, L31

%\bibitem[B.W. Bopp \& S.M. Rucinski (1981)]
\bibitem[B.W. Bopp \& S.M. Rucinski (1981)]{bopp-rucinski1981} Bopp, B.W. \& Rucinski, S.M. 1981, in IAU Symp. No. 93, p.177B
% the rapidly rotating giants of the FK Comae-type

%\bibitem[B.W. Bopp \& R.E. Stencel (1981)]
\bibitem[B.W. Bopp \& R.E. Stencel (1981)]{bopp-stencel1981} Bopp, B.W. \& Stencel, R.E. 1981, ApJ, 247, L131
% FK Com stars

%\bibitem[E.M.A. Borchert et al. (2022)]
\bibitem[E. Borchert et al. (2022)]{borchert2022} Borchert, E.M.A., Price, D. pinte, C., Cuello, N. 2022, MNRAS, 517, 4436

%\bibitem[J. Bouvier et al. (1995)]
\bibitem[J. Bouvier et al. (1995)]{bouvier1995} Bouvier, J., Covino, E., Kovo, O. et al. 1995, A\&A, 299, 89
% G2 rotat. period 1.20 days


\bibitem[D. Bowman \& L. Bugnet (2026)]{bowman2026} Bowman, D.M. \& Bugnet, L. 2026, Encyclopedia of Astrophysics, Vol. 2, Elsevier, pp. 133-153


%\bibitem[C. Brice\~no et al.(2002)]
\bibitem[C. Brice\~no et al.(2002)]{briceno2002} Brice\~no, C., Luhman, K.L., Hartmann, L. et al. 2002, ApJ, 580:317

%\bibitem[C. Brice\~no et al.(2019)]
\bibitem[C. Brice\~no et al.(2019)]{briceno2019} Brice\~no, C., Calvet, N., Hernandez, J., et al. 2019, \aj, 157, 85

%\bibitem[C. Brice\~no et al. (2002)]
\bibitem[C. Brice\~no et al. (2002)]{briceno2002} Brice\~no, C., Luhman, K.L., Hartmann, L., Stauffer, J.R., Kirkpatrick, J.D. 2002, ApJ, 580, 317

\bibitem[CASA Team et al. (2022)]{casa2022}  CASA Team, Bean, B., Bhatnagar, S. et al. 2022, in 
PASP, Vol. 134,  issue 1041, id.114501

%\bibitem[C. Cifuentes et al. (2025)]
\bibitem[C. Cifuentes et al. (2025)]{cifuentes2025} Cifuentes, C, Caballero, J.A., Gonz\'alez-Payo, J. et al. 2025, A\&A, 693, A228


\bibitem[C. Claret (2012)]{claret2012} Claret, A. 2012, A\&A, 538, A3 

%\bibitem[J.C. Clemens et al. (2004)]
\bibitem[J. Clemens et al. (2004)]{clemens2004} Clemens, J. C., Crain, J. A., \& Anderson, R. 2004, in Proc. SPIE, Vol. 5492, {\em Ground-based Instrumentation for Astronomy}, ed. A. F. M. Moorwood \& M. Iye, 331-340

%\bibitem[A.M. Cody et al. (2022)]
\bibitem[A.M. Cody et al. (2022)]{cody2022} Cody, A.M., Hillenbrand, L., Rebull, L. 2022, AJ, 163, A212 
% TESS light curves

%\bibitem[M. Cohen \& J.H. Bieging (1986)]
\bibitem[M. Cohen \& J. Bieging (1986)]{cohen1986} Cohen, M. \& Bieging, J.H. 1986, AJ, 92, 1396
% radio variability of G2


%\bibitem[M. Cohen \& L.V. Kuhi (1979)]
\bibitem[M. Cohen \& L. Kuhi (1979)]{cohen1979} Cohen, M. \& Kuhi, L.V. 1979, ApJ Suppl. 41, 743
% Observational studies of pre-main-sequence evolution.


%\bibitem[M.S. Connelley et al. (2007)]
\bibitem[M.S. Connelley et al. (2007)]{connelley2007} Connelley, M.S., Reipurth, B., Tokunaga, A.T. 2007, AJ, 133, 1528

%\bibitem[M.S. Connelley et al. (2008a)]
\bibitem[M.S. Connelley et al. (2008a)]{connelley2008a} Connelley, M.S., Reipurth, B., Tokunaga, A.T. 2008a, AJ, 135, 2496

%\bibitem[M.S. Connelley et al. (2008b)]
\bibitem[M.S. Connelley et al. (2008b)]{connelley2008b} Connelley, M.S., Reipurth, B., Tokunaga, A.T. 2008b, AJ, 135, 2526





\bibitem[C. Contreras Pe\~na et al. (2025)]{contreras2025} Contreras Pe\~na, C., Lee, J.-E., Herczeg, G. et al. 2025, arXiv:2509.24876 






%\bibitem[M.C. Cushing et al.(2004)]
\bibitem[M. Cushing et al.(2004)]{Cushing2004} Cushing, M.~C., Vacca, W.~D., \& Rayner, J.~T.\ 2004, \pasp, 116, 362. doi:10.1086/382907

%\bibitem[S.E. Dahm \& L.A. Hillenbrand (2017)]
\bibitem[S.E.Dahm \& L.A. Hillenbrand (2017)]{dahm2017} Dahm, S.E. \& Hillenbrand, L.A. 2017, AJ, 154, 177
% var. reflec.neb. in NGC 1333
 

%\bibitem[G.H. Darwin (1879)]
\bibitem[G.H. Darwin (1879)]{darwin1879} Darwin, G.H. 1879, Proc. Roy. Soc. London, 29, 168
% Darwin instability


%\bibitem[C.L. Davies et al. (2014)]
\bibitem[C.L. Davies et al. (2014)]{davies2014} Davies, C.L., Gregory, S.G., Greaves, J.S. 2014, MNRAS 444, 1157



\bibitem[S.E. de~Mink et al. (2014)]{demink2014} de Mink, S.E., Sana, H., Langer, N. et al. 2014, ApJ, 782:7
% definition of walkaway star


%\bibitem[A. Domiciano de Souza et al. (2014)]
\bibitem[A. Domiciano de Souza et al. (2014)]{domiciano2014} Domiciano de Souza, A., Kervella, P., Moser Faes, D. et al. (2014), A\&A, 569, A10

%\bibitem[J.-F. Donati et al. (1997)]
\bibitem[J.-F. Donati et al. (1997)]{donati1997} Donati, J.-F., Semel, M., Carter, B.D. et al. 1997, MNRAS, 291, 658

%\bibitem[M.A. Dopita (1978)]
\bibitem[M. Dopita (1978)]{dopita1978} Dopita, M.A. 1978, ApJS, 37, 117

%\bibitem[J.J. Drake et al. (2008)]
\bibitem[J. Drake et al. (2008)]{drake2008} Drake, J.J., Chung, S.M., Kashyap, V., Korhonen, H. et al. 2008, ApJ, 679:1522

%\bibitem[G. Duvert et al. (2000)]
\bibitem[G. Duvert et al. (2000)]{duvert2000} Duvert, G.,  Guilloteau, S., M\'enard, F. et al. 2000, A\&A, 355, 165
% Possible molecular outflow with IRAM Plateau de Bure interferometer


%\bibitem[P.P. Eggleton \& L. Kiseleva-Eggleton (2001)]
\bibitem[P. Eggleton \& L. Kiseleva-Eggleton (2001)]{eggleton2001} Eggleton, P.P. \& Kiseleva-Eggleton, L. 2001, ApJ, 562, 1012


%\bibitem[V. Elbakyan et al. (2019)]
\bibitem[V. Elbakyan et al. (2019)]{elbakyan2019} Elbakyan, V.G., Vorobyov, E.I., Rab, C. et al.  2019, MNRAS, 484, 146

%\bibitem[J.P. Farias et al. (2020)]
\bibitem[J. Farias et al. (2020)]{farias2020} Farias, J.P., Tan, J.C., Eyer, L. 2020, ApJ, 900:14
%  new walkaways from Orion


\bibitem[F. Espinosa-Lara \& M. Rieutord (2011)]{espinosa2011} Espinosa Lara, F. \& Rieutord, M. 2011, A\&A, 533, A43

%\bibitem[F. Favata et al. (1998)]
\bibitem[F. Favata et al. (1998)]{favata1998} Favata, F., Micela, G., Sciortino, S. 1998, A\&A, 337, 413



%\bibitem[E. Feigelson \& J. Kriss (1981)]
\bibitem[E. Feigelson \& J. Kriss (1981)]{feigelson1981} Feigelson, E. \& Kriss, J. 1981, ApJ, 248, L35


%\bibitem[M. Fernandez \& L.F. Miranda (1998)]
\bibitem[M. Fernandez \& L. Miranda (1998)]{fernandez1998} Fernandez, M. \& Miranda, L.F. 1998, A\&A, 332, 629

\bibitem[L. Ferrario et al. (2009)]{ferrario2009} Ferrario, L., Pringle, J.E., Tout, C.A, Wickramasinghe, D.T. 2009, MNRAS, 400, L71


%\bibitem[S. Fitton et al. (2022)]
\bibitem[S. Fitton et al. (2022)]{fitton2022} Fitton, S., Tofflemire, B.M., Kraus, A.L. RNAAS, 6, 18


%\bibitem[C. Flores et al. (2019)]
\bibitem[C. Flores et al. (2019)]{flores2019} Flores, C., Connelley, M.S., Reipurth, B., Boogert, A. 2019, ApJ, 882:75
%Magnetic fields of BP Tau and V347 Aur


%\bibitem[C. Flores et al. (2020)]
\bibitem[C. Flores et al. (2020)]{flores2020} Flores, C., Reipurth, B., Connelley, M.S. 2020, ApJ, 898, 109


%\bibitem[C. Flores et al. (2022)]
\bibitem[C. Flores et al. (2022)]{flores2022} Flores, C., Connelley, M.S., Reipurth, B. 2022, ApJ, 925:21


%\bibitem[E. Furlan et al. (2016)]
\bibitem[E. Furlan et al. (2016)]{furlan2016} Furlan, E., Fischer, W.J., Ali, B. et al. 2016, ApJS, 224, 5



\bibitem[Gaia Collaboration et al. (2016)]{gaia2016} Gaia collaboration,
Prusti, T., de Bruijne, J.H.J. et al. 2016, A\&A, 595, A1

\bibitem[Gaia Collaboration et al. (2021)]{gaia2021} Gaia collaboration, Brown, A.G.A., Vallenari, A. et al. 2021, A\&A, 649, A1


%\bibitem[P.A.B. Galli et al. (2018)]
\bibitem[P. Galli et al. (2018)]{galli2018} Galli, P.A.B., Loinard, L., Ortiz-L\'eon, G.N. et al. 2018, ApJ, 859:33
%distance to G2


%\bibitem[A. Garufi et al. (2024)]
\bibitem[A. Garufi et al. (2024)]{garufi2024} Garufi, A., Ginski, C., van Holstein, R.G., Benisty, M., Manara, C.F. et al. 2024, A\&A, 685, A53
% SPHERE observation of G2

\bibitem[A. Garufi et al. (2025)]{garufi2025} Garufi, A., Carrasco-Gonzalez, C., Macias, E. et al. 2025, A\&A, 694, A290


% apparently not used - nothing in it about G2
%\bibitem[A. Garufi et al. (2025)]
%\bibitem[Garufi et al. (2025)]{garufi2025} Garufi, A., Carrasco-Gonzalez, C., Macias, E. et al. 2025, A\&A, 694, A290

%\bibitem[T.R. Geballe et al.(2025)]
\bibitem[T.R. Geballe et al.(2025)]{geballe2025} Geballe, T.~R., Kaminski, B.~M., Banerjee, D.~P.~K., et al.\ 2025, \mnras, 541, 4, 3331. doi:10.1093/mnras/staf1193

%\bibitem[J.A. Graham \& J.A. Frogel (1985)]
\bibitem[J.A. Graham \& J.A. Frogel (1985)]{graham1985} Graham, J.A. \& Frogel, J.A. 1985, ApJ, 289, 331

%\bibitem[K.N. Grankin et al. (2008)]
\bibitem[K.N. Grankin et al. (2008)]{grankin2008} Grankin, K.N., Bouvier, J., Herbst, W., Melnikov, S.Yu. 2008, A\&A, 479, 827


%\bibitem[M. G\"udel et al. (2007)]
\bibitem[M. G\"udel et al. (2007)]{guedel2007} G\"udel, M., Briggs, K.R., Arzner, K. et al. 2007, A\&A, 468, 353
% XEST survey


%\bibitem[S. Guieu et al. (2006)]
\bibitem[S. Guieu et al. (2006)]{guieu2006} Guieu, S., Dougados, C., Monin, J.-L., Magnier, E., Martin, E.L. 2006, A\&A, 446, 485
% 17 new BDs in Taurus


%\bibitem[E.F. Guinan \& C.R. Robertson (1986)]
\bibitem[E.F. Guinan \& C.R. Robertson (1986)]{guinan1986} Guinan, E.F. \& Robinson, C.R. 1986, AJ, 91, 935


%\bibitem[Y.-X. Guo et al. (2019)]
\bibitem[Y.-X. Guo et al. (2019)]{guo2019} Guo, Y.-X., Luo, A.-L., Zhang, S., Du, B., Wang, Y.-F. et al. 2019, MNRAS, 485, 2167
% spectral classification of XEST 08-033


%\bibitem[G. Haro (1953)]
\bibitem[G. Haro (1953)]{haro1953} Haro, G. 1953, ApJ, 117, 73


%\bibitem[P. Hartigan et al. (1994)]
\bibitem[P. Hartigan et al. (1994)]{hartigan1994} Hartigan, P., Strom, K.M., Strom, S.E. 1994, ApJ, 427,961
% spectral class of G2 


%\bibitem[L. Hartmann et al. (1986)]
\bibitem[L. Hartmann et al. (1986)]{hartmann1986} Hartmann, L., Hewett, R., Stahler, S., and Mathieu, R.D. 1986, ApJ, 309, 275
% Rot. and Rad. vel. of TTS


%\bibitem[L. Hartmann et al. (2016)]
\bibitem[Hartmann et al. (2016)]{hartmann2016} Hartmann, L., Herczeg, G., Calvet, N. 2016, ARAA, 54, 135

%\bibitem[R.M. Heath \& C.J. Nixon (2020)]{heath2020} Heath, R.M. \& Nixon, C.J. 2020, A\&A, 641, A64

\bibitem[J. Henneco et al. (2024)]{henneco2024} Henneco, J., Schneider, F.R.N., Laplace, E. 2024, A\&A, 682, A169


%\bibitem[R.N. Henriksen et al.(1991)]
\bibitem[R.N. Henriksen et al.(1991)]{henriksen1991} Henriksen, R.N., Ptuskin, V.S., Mirabel, I.F. 1991, A\&A, 248, 221


%\bibitem[G.H. Herbig (1977)]
\bibitem[G.H. Herbig (1977)]{herbig1977} Herbig, G.H. 1977, ApJ, 217, 693
% FUors

%\bibitem[G.H. Herbig (1989)]
\bibitem[G.H. Herbig (1989)]{herbig1989} Herbig, G.H. 1989, in {\em Low Mass
Star Formation and Pre-main Sequence Objects}, ESO Conf. Proc. 33,
ed. Bo Reipurth, p.233


%\bibitem[G.H. Herbig \& N.K. Rao (1972)]
\bibitem[G.H. Herbig \& N.K. Rao (1972)]{herbig1972} Herbig, G.H. \& Rao, N.K. 1972, ApJ, 174, 401

%\bibitem[G.H. Herbig et al. (2003)]
\bibitem[G.H. Herbig et al. (2003)]{herbig2003} Herbig, G.H., Petrov, P.P., Duemmler, R. 2003, ApJ, 595, 384
% HIgh-resolution spectra of FU Ori


%\bibitem[W. Herbst (1986)]
\bibitem[W. Herbst (1986)]{herbst1986} Herbst, W. 1986, PASP, 98, 1088
%inclination of G2


%\bibitem[G.J. Herczeg \& L.A. Hillenbrand (2014)]
\bibitem[G.J. Herczeg \& L.A. Hillenbrand (2014)]{herczeg2014} Herczeg, G.J. \& Hillenbrand, L.A. 2014, Ap, 786, A97
% An Optical Spectroscopic Study of T Tauri Stars. I. Photospheric Properties

%\bibitem[L.A. Hillenbrand et al. (2021)]
\bibitem[L.A. Hillenbrand et al. (2021)]{hillenbrand2021} Hillenbrand, L.A., De, K., Hankins, M. et al. 2021, AJ, 161:220

%\bibitem[L.A. Hillenbrand et al. (2022)]
\bibitem[L.A. Hillenbrand et al. (2022)]{hillenbrand2022} Hillenbrand, L.A., Isaacson, H., Rodriguez, A.C. et al. 2022, AJ, 163:115 


%\bibitem[L.A. Hillenbrand et al. (2023)]
\bibitem[L.A. Hillenbrand et al. (2023)]{hillenbrand2023} Hillenbrand, L.A., Carvalho, A., van Roestel, J., De, K. 2023, ApJ, 958:L27


%\bibitem[J.G. Hills \& C.A. Day (1976)]
\bibitem[J.G. Hills \& C.A. Day (1976)]{hills1976} Hills, J.G. \& Day, C.A. 1976, Astrophys. Lett., 17, 87
% collisions in globular clusters





%\bibitem[E.P. Hubble (1916)]
\bibitem[E.P. Hubble (1916)]{hubble1916} Hubble, E.P. 1916, ApJ, 44, 190
% variable nebula NGC 2261

\bibitem[P. Hut (1981)]{hut1981} Hut, P. 1981, A\&A, 99, 126

%\bibitem[D.P. Huenemoerder et al. (1993)]
\bibitem[D. Huenemoerder et al. (1993)]{huenemoerder1993} Huenemoerder, D.P., Ramsey, L.W., Buzasi, D.L., Nations, H.L. 1993, ApJ, 404, 316

%\bibitem[M.A. Ibragimov (1993)]
\bibitem[M.A. Ibragimov (1993)]{ibragimov1993} Ibragimov, M.A. 1993, ARep, 37, 1761

%\bibitem[S.S. Jensen \& T. Haugb\o lle (2018)]
\bibitem[S.S. Jensen \& T. Haugb\o lle (2018)]{jensen2018} MNRAS, 474, 1176

%\bibitem[L. Jetsu et al. (1993)]
\bibitem[L. Jetsu et al. (1993)]{jetsu1993} Jetsu, L., Pelt, J., Tuominen, I. 1993, A\&A, 278, 449

%\bibitem[C.M. Johns-Krull (1996)]
\bibitem[C.M. Johns-Krull (1996)]{johnskrull1996} Johns-Krull C.M. 1996, A\&A, 306, 803

%\bibitem[I. Joncour (1994)]{joncour1994}
\bibitem[I. Joncour (1994)]{joncour1994} Joncour, I., Bertout, C., Bouvier, J. 1994, A\&A, 291, L19
% Doppler imaging of HD 283572

\bibitem[I. Joncour et al. (2018)]{joncour2018} Joncour, I., Duch\^ene, G., Moraux, E. et al. 2018, A\&A, 620, A27



%\bibitem[H. J\"onsson et al. (2020)]{jonsson2020}
\bibitem[H.J. J\"onsson et al. (2020)]{jonsson2020} J\"onsson, H., Holtzman, J.A., Allende Prieto, C. et al. 2020, AJ, 160:120
% APOGEE DR16

%\bibitem[B.F. Jones \& G.H. Herbig (1979)]
\bibitem[B.F. Jones \& G.H. Herbig (1979)]{jones1979} Jones, B.F. \& Herbig, G.H. 1979, AJ, 84, 1872
% proper motions of young stars

%\bibitem[W. Joye \& E. Mandel (2003)]
\bibitem[W. Joye \& E. Mandel (2003)]{joye2003} Joye, W.A. \& Mandel 2003, in {\em Astronomical Data Analysis Software and Systems XII}, ASP Conference Series, Vol. 295,  H. E. Payne, R. I. Jedrzejewski, and R. N. Hook, eds., p.489





%CK Vul
%\bibitem[Kami{\'n}ski et al.(2015a)]{kaminski2015a} Kami{\'n}ski, T., Menten, K.~M., Tylenda, R., et al.\ 2015, Nature, 520, 7547, 322. doi:10.1038/nature14257


%V1309 Sco
\bibitem[T. Kami{\'n}ski et al.(2015)]{kaminski2015} Kami{\'n}ski, T., Mason, E., Tylenda, R., et al.\ 2015, \aap, 580, A34. doi:10.1051/0004-6361/201526212






\bibitem[T. Kami{\'n}ski et al.(2020)]{kaminski2020} Kami{\'n}ski, T., Menten, K.~M., Tylenda, R., et al.\ 2020, \aap, 644, A59. doi:10.1051/0004-6361/202038648


%\bibitem[T. Kami\'nski et al. (2021)]
\bibitem[T. Kami\'nski et al. (2021)]{kaminski2021} Kami\'nski, T., Tylenda, R., Kiljan, A. et al. 2021, A\&A, 655, A32.  doi:10.1051/0004-6361/202141526
% ALMA obs of V838 Mon 



\bibitem[Karambelkar et al.(2023)]{karambelkar2023} Karambelkar, V.~R., Kasliwal, M.~M., Blagorodnova, N., et al.\ 2023, \apj, 948, 2, 137. doi:10.3847/1538-4357/acc2b9

\bibitem[Karambelkar et al.(2025)]{karambelkar2025} Karambelkar, V., Kasliwal, M., Lau, R.~M., et al.\ 2025, , arXiv:2508.03932. doi:10.48550/arXiv.2508.03932


\bibitem[Kasliwal et al.(2017)]{kasliwal2017} Kasliwal, M.~M., Bally, J., Masci, F., et al.\ 2017, \apj, 839, 2, 88. doi:10.3847/1538-4357/aa6978



%\bibitem[S.J. Kenyon \& L. Hartmann (1995)]
\bibitem[S.J. Kenyon \& L. Hartmann (1995)]{kenyon1995} Kenyon, S.J. \& Hartmann, L. 1995, ApJS, 101, 117

%\bibitem[R.M.P. Kerr et al. (2021)]
\bibitem[R. Kerr et al. (2021)]{kerr2021} Kerr, R.M.P., Rizzuto, A.C., Kraus, A.L., Offner, S.S.R. 2021, ApJ, 917, 23
% identification of stellar groups in Taurus


%\bibitem[T. Kipper et al. (2004)]
\bibitem[T. Kipper et al. (2004)]{kipper2004} Kipper, T., Klochkova, V.G., Annuk, K. et al. 2004, A\&A, 416, 1107
% early spectra of V838 Mon


\bibitem[L.G. Kiseleva et al. (1998)]{kiseleva1998} Kiseleva, L.G., Eggleton, P.P., Mikkola, S. 1998, MNRAS, 300, 292



\bibitem[D.P. Kjurkchieva \& D.V. Marchev (2005)]{kjurkchieva2005} Kjurkchieva, D.P. \& Marchev, D.V. 2005, A\&A, 434, 221

\bibitem[H. Knox-Shaw (1916)]{knoxshaw1916} Knox-Shaw, H. 1916, MNRAS, 76, 646
%Variable nebula in CrA


\bibitem[C.S. Kochanek et al. (2014)]{kochanek2014} Kochanek, C.S., Adams, S.M., Belczynski, K. 2014, MNRAS, 443, 1319
% mergers are common

\bibitem[C.S. Kochanek et al. (2017)]{kochanek2017} Kochanek, C.S., Shappee, B.J., Stanek, K.Z. et al. 2017, PASP, 129, 104502  doi:10.1093/mnras/stu1226


\bibitem[Z. Kopal (1959)]{kopal1959}  Kopal, Z. 1959, {\em Close Binary Stars}, Chapman \& Hall, London

\bibitem[H. Korhonen et al. (1999)]{korhonen1999} Korhonen, H., Berdyugina, S.V., Hackman, T. et al. 1999, A\&A, 346, 101
 
\bibitem[H. Korhonen et al. (2000)]{korhonen2000} Korhonen, H., Berdyugina, S.V., Hackman, T. et al. 2000, A\&A, 360, 1067


\bibitem[H. Korhonen et al. (2007)]{korhonen2007} Korhonen, H., Berdyugina, S.V., Hackman, T. et al. 2007, A\&A, 476, 881

\bibitem[H. Korhonen et al. (2009)]{korhonen2009} Korhonen, H., S. Hubrig,  Berdyugina, S.V. et al. 2009, MNRAS, 395, 282


\bibitem[A.L. Kraus et al. (2011)]{kraus2011} Kraus, A.L., Ireland, M.J., Martinache, F. et al. 2011, ApJ, 731, A8
%  Mapping the Shores of the Brown Dwarf Desert. II. Multiple Star Formation in Taurus-Auriga
% identify G3 as visual binary

\bibitem[D.M. Krolikowski et al. (2021)]{krolikowski2021} Krolikowski, D.M., Krays, A.L., Rizzuto, A.C. 2021, AJ, 162:110





\bibitem[R.B. Larson (1980)]{larson1980} Larson, R.B. 1980, MNRAS, 190, 321
% FU Orionis mechanism


% \bibitem[N.W.C. Leigh \& A.M. Geller (2012)]{leigh2012} Leigh, N.W.C. \& Geller, A.M. 2012, MNRAS, 425, 2369

% \bibitem[N.W.C. Leigh \& A.M. Geller (2015)]{leigh2015} Leigh, N.W.C. \& Geller, A.M. 2015, MNRAS, 450, 1724

\bibitem[A.T. Lee et al. (2019)]{lee2019} Lee, A.T., Offner, S.S.R, Kratter, K.M. et al. 2019, ApJ 887, 232


\bibitem[N.W.C. Leigh et al. (2017)]{leigh2017} Leigh, N.W.C., Geller, A.M., Shara, M.M. 2017, MNRAS, 471, 1830

\bibitem[N.W.C. Leigh et al. (2018)]{leigh2018} Leigh, N.W.C., Geller, A.M., Shara, M.M. 2018, MNRAS, 480, 3062
 



\bibitem[E. Leiner et al. (2019)]{leiner2019} Leiner, E., Mathieu, R.D., Vanderburg, A. et al. 2019, ApJ, 881:47


\bibitem[Ch. Leinert et al. (1993)]{leinert1993} Leinert, Ch., Zinnecker, H., Weitzel, N. et al. 1993, A\&A, 278, 129

\bibitem[H. Li et al. (2015)]{li2015} Li, H., Li, D., Xu, D., Goldsmith, P.F. et al. 2015, ApJS, 219:20
% Outflows and Bubbles in Taurus

\bibitem[Lightkurve Collaboration et al. (2018)]{lightkurve2018} Lightkurve Collaboration et al. 2018, Astrophysics Source Code Library, https://ui.adsabs.harvard.edu/abs/2018ascl.soft12013L/abstract



\bibitem[Liimets et al.(2023)]{liimets2023} Liimets, T., Kolka, I., Kraus, M., et al.\ 2023, \aap, 670, A13. doi:10.1051/0004-6361/202142959



\bibitem[C.-L. Lin et al. (2023)]{lin2023} Lin, C.-L. Ip, W.-H., Chang, T.-H., Song, Y., Luo, A.-L. 2023, AJ, 166:82
% Mass accretion, speactral and Photometric properties of T Tauri stars


\bibitem[J.B. Lovell et al. (2024)]{lovell2024} Lovell, J.B., Keating, G.K., Wilner, D.J. et al. 2024, ApJ, 962:L12

\bibitem[Lucy (1967)]{lucy1967} Lucy, L.B. 1967, Z.f.Ap. 65, 89

%\bibitem[Lucy (1968a)]{lucy68a} Lucy, L.B. 1968a, ApJ, 151, 1123

%\bibitem[Lucy (1968b)]{lucy68b} Lucy, L.B. 1968b, ApJ, 153, 877

\bibitem[K.L. Luhman (2018)]{luhman2018} Luhman, K.L. 2018, AJ, 156, A271
% The Stellar Membership of the Taurus Star-forming Region
% discovery of two high proper motion stars near HP Tau

\bibitem[K.L. Luhman (2023)]{luhman2023} Luhman, K.L. 2023, AJ, 165:37 
% identification of YSO's in groups in Taurus

\bibitem[K.L. Luhman (2025)]{luhman2025} Luhman, K.L. 2025, AJ, 170, 19
% spectraltype-mass relation


\bibitem[K.L. Luhman et al. (2003)]{luhman2003} Luhman, K.L., Briceno, C., Stauffer, J.R., Hartmann, L., Barrado y Navascues, D., Caldwell, N. 2003, ApJ, 590:348
%New low-mass members in Taurus  

\bibitem[K.L. Luhman et al. (2009)]{luhman2009} Luhman, K.L., Mamajek, E.E., Allen, P.R., Cruz, K.L. 2009, ApJ, 703, 399

\bibitem[K.L. Luhman et al. (2010)]{luhman2010} Luhman, K.L., Allen, P.R., Espaillat, C. et al. 2010, ApJS, 186, 111

\bibitem[K.L. Luhman et al. (2017)]{luhman2017} Luhman, K.L., Mamajek, E.E., Shukla, S.J., Loutrel, N.P. 2017, AJ, 153, 46


\bibitem[M. MacLeod et al. (2018)]{macleod2018} MacLeod, M., Ostriker, E.C., \& Stone, J.M. 2018, ApJ, 868:136
% bipolar outflow during stellar coalescence 


\bibitem[A. Maeder (2009)]{maeder2009} Maeder, A. 2009, {\em Physics, Formation and Evolution of Rotating Stars (Berlin:Springer}

\bibitem[T.Yu. Magakian (2003)]{magakian2003} Magakian, T. Yu. 2003, A\&A, 399, 141
%Merged catalog


%\bibitem[H.A. McAlister et al. (2005)]{mcalister2005} McAlister, H.A., ten Brummelaar, T.A., Gies, D.R. et al. 
ApJ, 628, 439

\bibitem[L. Malo et al. (2014)]{malo2014} Malo, L., Artigau, E., Doyon, R. et al. 2014, ApJ, 788, A81
% rotation in nearby young groups


\bibitem[R.D. Mathieu et al. (1997)]{mathieu1997} Mathieu, R.D., Stassun, K., Basri, G. et al. 1997, AJ, 113, 1841

\bibitem[R.D.  Mathieu \& O.R. Pols (2025)]{mathieu1997} Mathieu, R.D. \& Pols, O.R. 2025, ARAA, 63-467

\bibitem[E. Mason \& S.N. Shore (2022)]{mason2022} Mason, E., Shore, S.N. 2022, A\&A, 664, A12

\bibitem[E. Mason et al. (2010)]{mason2010} Mason, E., Williams, R.E., Preston, G., Bensby, T. 2010, A\&A, 516, A108

\bibitem[A. McBride \& M. Kounkel (2019)]{mcbride2019} McBride, A. \& Kounkel, M. 2019, ApJ, 884:6
%new walkaways




\bibitem[S. Mohan et al. (2022)]{mohan2022} Mohan, S., Vig, S., Mandal, S. 2022, MNRAS, 514, 3709





\bibitem[G.H. Moriarty-Schieven et al. (1992)]{moriartyschieven1992} Moriarty-Schieven, G.H., Wannier, P.G., Tamura, M., Keene, J. 1992, ApJ, 400, 260
% Molecular outflows in Taurus



\bibitem[D.J. Mullan \& J. MacDonald (2020)]{mullan2020} Mullan, D.J. \& MacDonald, J. 2020, ApJ, 904:108
%Age of G2 from magnetic and non-magnetic models



\bibitem[U. Munari et al. (2002)]{munari2002} Munari, U., Henden, A., Kiyota, S. et al. 2002, A\&A, 389, L51
% eruption of V838 Mon




\bibitem[J.L.A. Nandez et al. (2014)]{nandez2014} Nandez, J.L.A., Ivanova, N., Lombardi, J.C. 2014, ApJ, 786:39
%V1309 Sco

\bibitem[S. Naoz \& D.C. Fabrycky (2014)]{naoz2014} Naoz, S. \& Fabrycky, D.C. 2014, ApJ, 793:137
% Mergers and obliquities in stellar triples


\bibitem[G. Narayanan et al. (2008)]{narayanan2008} Narayanan, G., Heyer, M.H., Brunt, C. et al. 2008, ApJS, 177:341
% FCRAO CO mapping of Taurus



\bibitem[G. Narayanan et al (2012)]{narayanan2012} Narayanan, G., Snell, R., Bemis, A. 2012, MNRAS, 425, 2641
%Molecular outflows in Taurus 


\bibitem[D.C. Nguyen et al. (2012)]{nguyen2012} Nguyen, D.C., Brandeker, A., van Kerkwijk, M.H., Jayawardhana, R. 2012, ApJ, 745, A119
% CLOSE COMPANIONS TO YOUNG STARS. I. A LARGE SPECTROSCOPIC SURVEY
% IN CHAMAELEON I AND TAURUS-AURIGA


\bibitem[S.S.R. Offner et al. (2023)]{offner2023} Offner, S.S.R., Moe, M., Kratter, K.M. et al. 2023, in {\em Protostars and Planets~VII}, S. Inutsuka et al., eds. 

\bibitem[J.M. Oliveira \& B.H. Foing (1999)]{oliveira1999} Oliveira, J.M. \& Foing, B.H. 1999, A\&A, 343, 213

\bibitem[D. O'Neal et al. (1990)]{oneal1990} O'Neal, D., Feigelson, E.D., Mathieu, R.D., Myers, P.C. 1990, AJ, 100, 1610
% Non detection of G2 at 5 GHz


%\bibitem[Bok (1999)]{bok1999} Pakhomov, Y.V., Ryabchikova, T.A., Piskunov, N.E. 2019, ARep, 63, 1010

\bibitem[K. Panov \& D. Dimitrov (2007)]{panov2007} Panov, K. \& Dimitrov, D. 2007, A\&A, 467, 229

\bibitem[R.J. Patterer et al. (1993)]{patterer1993} Patterer, R.J., Ramsey, L., Welty, A.D., Huenemoerder, D.P. 1993, AJ, 105, 1519 


%\bibitem[O. Pejcha et al. (2017)]{pejcha2017} Pejcha, O., Metzger, B.D., Tyler, J.G., Tomida, K. 2017, ApJ, 850:59

\bibitem[H.B. Perets (2026)]{perets2026} Perets, H.B. {\em Evolution of Triple Stars}, in Encyclopedia of Astrophysics Vol.2, ed. I.~Mandel (Editor-in-Chief), Elsevier, p.279-297


\bibitem[H.B. Peters \& D.C. Fabrycky (2009)]{perets2009} Perets, H.B. \& Fabrycky, D.C. 2009, ApJ, 697, 1048
% triple origin of blue stragglers


\bibitem[P.P. Petrov \& G.H. Herbig (1992)]{petrov1992} Petrov, P.P. \& Herbig, G.H. 1992, ApJ, 392, 209
% FU Ori interpretation


\bibitem[P.P. Petrov \& G.H. Herbig (2008)]{petrov2008} Petrov, P.P. \& Herbig, G.H. 2008, AJ, 136, 676
%Line structure in the spectrum of FU Ori




\bibitem[I. Platais et al. (2020)]{platais2020} Platais, I., Robberto, M., Bellini, A. et al. 2020, AJ, 159:272
%ONC walkaways


\bibitem[K. Poully et al. (2020)]{poully2020} Poully, K., Bouvier, J., Alecian, E., Alencar, S.H.P., Cody, A.M. et al. 2020, AA, 642, A99
% Accretion in HQ Tau


\bibitem[S.J.D. Purser et al. (2018)]{purser2018} Purser, S.J.D., Ainsworth, R.E., Ray, T.P. et al. 2018, MNRAS, 481, 5532

\bibitem[A.C. Raga et al. (1990)]{raga1990} Raga, A.C., Canto, J., Binette, L., Calvet, N. et al. 1990, ApJ, 364, 601


\bibitem[L.W. Ramsey et al. (1981)]{ramsey1981} Ramsey, L.W., Nations, H.L., Barden, S.C. 1981, ApJ, 251, L101

\bibitem[K. Rawiraswattana et al. (2012)]{rawiraswattana2012} Rawiraswattana, K., Lomax, O., \& Goodwin, S.P. 2012, MNRAS, 419, 2025


\bibitem[T.P. Ray et al. (1997)]{ray1997} Ray, T.P., Muxlow, T.W.B., Brown, A. et al. 1997, Nature, 385, 415


\bibitem[J. Rayner et al. (2022)]{rayner2022} Rayner, J., Tokunaga, A., Jaffe, D. et al. 2022, PASP, 134, 015002

\bibitem[L.M. Rebull et al. (2002)]{rebull2002} Rebull, L.M., Makidon, R.B., Strom, S.E. et al. AJ, 123, 1528

\bibitem[L.M. Rebull et al. (2004)]{rebull2004} Rebull, L.M., Wolff, S.C., Strom, S.E. 2004, AJ, 127, 1029
% rotation of YSOSs

\bibitem[L.M. Rebull et al. (2020)]{rebull2020} Rebull, L.M., Stauffer, J.R., Cody, A.M., Hillenbrand, L.A., Bouvier, J. et al. 2020, AJ, 159:273
% Rotation periods in Taurus with K2

\bibitem[B. Reipurth (1985)]{reipurth1985} Reipurth, B. 1985, A\&A, 143, 435

\bibitem[B. Reipurth (2000)]{reipurth2000} Reipurth, B. 2000, AJ, 120, 3177 

\bibitem[B. Reipurth \& J. Bally (1986)]{reipurth1986} Reipurth, B. \& Bally, J. 1986, Nature, 320, 336
% Re50

\bibitem[B. Reipurth \& C. Aspin (2004)]{reipurth2004} Reipurth, B. \& Aspin, C. 2004, ApJ, 608, L65

\bibitem[B. Reipurth et al. (2010)]{reipurth2010} Reipurth, B., Mikkola, S., Connelley, M., Valtonen, M. 2010, ApJ, 725, L56
%Orphaned protostars

\bibitem[B. Reipurth \& S. Mikkola (2012)]{reipurth2012} Reipurth, B. \& Mikkola, S. 2012, Nature, 492, 221

\bibitem[B. Reipurth \& S. Mikkola (2015)]{reipurth2015} Reipurth, B. \& Mikkola, S. 2015, AJ, 149, A145


\bibitem[B. Reipurth et al. (2018)]{reipurth2018} Reipurth, B., Herbig, G.H., Bally, J. et al. 2018, AJ, 156, A25

\bibitem[M. Renzo et al. (2019)]{renzo2019} Renzo, M., Zapartas, E., de Mink, S.E. et al. 2019, A\&A, 624, A66

\bibitem[S.P. Reynolds (1986)]{reynolds1986} Reynolds, S.P. 1986, ApJ, 304, 713


\bibitem[A. Richichi et al. (1994)]{richichi1994} Richichi, A., Leinert, Ch., Jameson, R., Zinnecker, H. 1994, A\&A, 287, 145
% New binary stars in Taurus


\bibitem[G.H. Rieke \& Lebofsky (1985)]{rieke1985} Rieke, G.H. \& Lebofsky, M.J. 1985, ApJ, 288, 618
% Reddening vectors





\bibitem[J.L. Rivera et al. (2015)]{rivera2015} Rivera, J.L., Loinard, L., Dzib, S.A. et al. 2015, ApJ, 807, 119


\bibitem[A.C. Rizzuto et al. (2020)]{rizzuto2020} Rizzuto, Aaron C.; Dupuy, Trent J.; Ireland, Michael J.; Kraus, Adam L. 2020, ApJ, 889, A175

%CO 3-2 X-factor
\bibitem[Rigby et al.(2025)]{Rigby2025} Rigby, A.~J., Thompson, M.~A., Eden, D.~J., et al.\ 2025, \mnras, 538, 1, 198. doi:10.1093/mnras/staf278


\bibitem[J.E. Rodriguez et al. (2017)]{rodriguez2017} Rodriguez, J.E., Ansdell, M., Oelkers, R.J., Cargile, P.A., Gaidos, E. et al. 2017, ApJ, 848:97
% HQ Tau dipper light curve 


\bibitem[L.F. Rodriguez et al. (1989)]{rodriguez1989} Rodriguez, L.F., Curiel, S., Moran, J.M. et al. 1989, ApJ, 346, L85


\bibitem[S.M. Rucinski (1990)]{rucinski1990} Rucinski, S.M. 1990, PASP,102, 306

\bibitem[S.M. Rucinski (1991)]{rucinski1991} Rucinski, S.M. 1991, AJ, 101, 2199

\bibitem[S.M. Rucinski et al. (2007)]{rucinski2007} Rucinski, S.M., Pribulla, T., van Kerkwijk, M.H. 2007, AJ, 134, 2353


\bibitem[R.J. Rudy et al. (2008a)]{rudy2008a} Rudy, R.J. Lynch, D.K., Russell, R.W. et al. 2008a, IAUC 8976

\bibitem[R.J. Rudy et al. (2008b)]{rudy2008b} Rudy, R.J. Lynch, D.K., Russell, R.W. et al. 2008b, IAUC 8997



\bibitem[T. Ryu et al. (2025)]{ryu2025} Ryu, T., Sills, A., Pakmor, R. et al. 2025, ApJL, 980:L38


\bibitem[I.S. Savanov et al. (2023)]{savanov2023} Savanov, I.S., Naroenkov, S.A., Nalivkin, M.A., Dmitrienko, E.S. 2023, Astron. Rep. 67, 1394

\bibitem[L. Scelsi et al. (2007)]{scelsi2007} Scelsi, L., Maggio, A., Micela, G. et al. 2007, A\&A, 468, 405
% XEST 08-049 a PMS star

\bibitem[L. Scelsi et al. (2008)]{scelsi2008} Scelsi, L., Sacco, G., Affer, L. et al. 2008, A\&A, 490, 601
% XEST 08-049 a PMS star
 

\bibitem[F.R.N. Schneider (2025)]{schneider2025} Schneider, F.R.N. 2025, arXiv:2509.18421


\bibitem[F.R.N. Schneider et al. (2019)]{schneider2019} Schneider, F.R.N., Ohlmann, S.T., Podsiadlowski, Ph. et al. 2019, Nature, 574, 211
%mergers as the origin of magnetic massive stars


\bibitem[F.R.N. Schneider et al. (2020)]{schneider2020} Schneider, F.R.N., Ohlmann, S.T., Podsiadlowski, Ph. et al. 2020, MNRAS, 495, 2796
%long-term evloution of a magnetic massive merger product


\bibitem[C. Schoettler et al. (2019)]{schoettler2019} Schoettler, C., Parker, R.J., Arnold, B. et al. 2019, MNRAS, 487, 4615
%N-body simulations in SFR




\bibitem[Science Software Branch at STScI (2012)]{pyraf2012} Science Software Branch at STScI 2012,  Astrophysics Source Code Library, record ascl:1207.011


\bibitem[F.G.P. Seidl \& A.G.W. Cameron (1972)]{seidl1972} Seidl, F.G.P. \& Cameron, A.G.W. 1972, Ap\&SS, 15, 44
% head-on collision of two identical stars


\bibitem[J. Serna et al. (2021)]{serna2021} Serna, J., Hernandez, J. Kounkel, M. et al. 2021, ApJ, 923:177 

\bibitem[B. Shappee et al. (2014)]{shappee2014} Shappee, B., Prieto, J.L., Grupe, D. et al. 2014, ApJ, 788, 48


\bibitem[C. Shariat et al. (2023)]{shariat2023} Shariat, C., Naoz, S., Hansen, B.M.S. et al 2023, ApJL, 955:14


\bibitem[C. Shariat et al. (2025)]{shariat2025} Shariat, C., Naoz, S., El-Badry, K. et al. 2025, ApJ, 978:47


\bibitem[M. Simon et al. (1987)]{simon1987} Simon, M., Howell, R.R., Longmore, A.J. et al. 1987, ApJ, 320, 344


\bibitem[M. Simon et al. (1995)]{simon1995} Simon, M., Ghez, A.M., Leinert, Ch. et al. 1995, ApJ, 443,625
%Lunar occultation detection of binaries: HP Tau, FF Tau, CK St 3, HP Tau, Haro 6-28



\bibitem[M. Siwak et al. (2011)]{siwak2011} Siwak, M., Rucinski, S.M., Matthews, J.M. et al. 2011, MNRAS, 415, 1119



\bibitem[Smith et al.(2016)]{smith2016} Smith, N., Andrews, J.~E., Van Dyk, S.~D., et al.\ 2016, MNRAS, 458, 1, 950. doi:10.1093/mnras/stw219



\bibitem[N. Soker \& R. Tylenda (2003)]{soker2003} Soker, N. \& Tylenda, R. 2003, ApJ, 582, L105
% eruption model of V838 MOn


\bibitem[N. Soker \& R. Tylenda (2006)]{soker2006} Soker, N. \& Tylenda, R. 2006, MNRAS, 373, 733
%Violent stellar merger model for transient events



\bibitem[S.W. Stahler (2010)]{stahler2010} Stahler, S.W. 2010, MNRAS 402,1758


\bibitem[K.G. Stassun et al. (1999)]{stassun1999} Stassun, K.G., Mathieu, R.D., Mazeh, T. Vrba, F.J. 1999, AJ, 117:2941


\bibitem[K.G. Stassun et al. (2001)]{stassun2001} Stassun, K.G., Mathieu, R.D., Vrba, F.J. et al. 2001, AJ, 121, 1003

\bibitem[T. Steinmetz et al. (2024)]{steinmetz2024} Steinmetz, T., Kami\'nski, T., Schmidt, M., Kiljan, A. 2024, A\&A, 682, A127

\bibitem[K. Stepie\'n (2011)]{stepien2011} Stepie\'n, K. 2011, A\&A, 531, A18

\bibitem[M.F. Sterzik \& R.H. Durisen (1995)]{sterzik1995} Sterzik, M.F. \& Durisen, R.H. 1995, A\&A, 304, L9
%escape of TTS fom young stellar systems

\bibitem[K.G. Strassmeier (2009)]{strassmeier2009} Strassmeier, K.G. 2009, A\&A Rev. 17:251


\bibitem[K.G. Strassmeier \& Rice (1998)]{strassmeier1998} Strassmeier, K.G. \& Rice, J.B. 1998, A\&A, 339, 497



\bibitem[K.M. Strom \& S.E. Strom (1993)]{strom1993} Strom, K.M. \& Strom, S.E. 1993, ApJ, 412, L63

\bibitem[K.M. Strom \& S.E. Strom (1994)]{strom1994} Strom, K.M. \& Strom, S.E. 1994, ApJ, 424, 237
% SF in L1495E

\bibitem[K.M. Strom et al. (1986)]{strom1986} Strom, K.M. et al. 1986 ApJ Supp. 62, 39

\bibitem[O. Struve \& W.C. Straka (1962)]{struve1962} Struve, O. \& Straka, W.C. 1962, PASP, 74, 474


\bibitem[Zs. M. Szabo et al. (2021)]{szabo2021} Szab\'o, Zs.M., Kospal, A., Abraham, P. et al. 2021, ApJ, 917, 80 


\bibitem[N. Tetzlaff et al. (2011)]{tetzlaff2011} Tetzlaff, N., Neuh\"auser, R., Hohle, M.M. 2011, MNRAS, 410, 190

%\bibitem[J. Tobin \& P. Sheehan (2025)].... Tobin, J.J. \& Sheehan, P.D., ARAA, 62:203


\bibitem[J. Tobin et al. (2022)]{tobin2022} Tobin, J.J., Offner, S.S.R, Kratter, K.M. et al. (2022) ApJ, 925:39

\bibitem[D. Tody (1986)]{tody1986}  Tody, D. 1986, in {\em Instrumentation in astronomy VI}; Proceedings of the Meeting, Tucson, AZ, Mar. 4-8, 1986. Part 2 (A87-36376 15-35). Bellingham, WA, Society of Photo-Optical Instrumentation Engineers, 1986, p. 733.

\bibitem[D. Tody (1993)]{tody1993} Tody, D. 1993, in {\em Astronomical Data Analysis Software and Systems II} (1993), 
A.S.P. Conference Series, Vol. 52, R. J. Hanisch, R. J. V. Brissenden, and Jeannette Barnes, eds., p. 173.

\bibitem[A. Tokovinin (2021)]{tokovinin2021} Tokovinin, A. 2021, Universe, 7, 352

\bibitem[A. Tokovinin \& M. Moe (2020)]{tokovinin2020} Tokovinin, A. \& Moe, M. 2020, MNRAS, 491, 5158

\bibitem[A. Tokovinin et al. (2006)]{tokovinin2006} Tokovinin, A., Thomas, S., Sterzik, M., Udry, S. 2006, A\&A, 450, 681

\bibitem[S. Toonen et al. (2022)]{toonen2022} Toonen, S., Boekholt, T.C.N., Portegies Zwart, S. 2022, A\&A, 661, A61


\bibitem[R.M. Torres et al. (2007)]{torres2007} Torres, R.M., Loinard, L., Mioduszewski, A.J., Rodriguez, L.F. 2007, ApJ, 671:1813

\bibitem[R.M. Torres et al. (2009)]{torres2009} Torres, R.M., Loinard, L., Mioduszewski, A.J., Rodriguez, L.F. 2009, ApJ, 698, 242




\bibitem[R. Tylenda \& T. Kami\'nski (2016)]{tylenda2016} Tylenda, R. \& Kami\'nski, T. 2016, A\&A, 592, A134 doi:10.1051/0004-6361/201527700


\bibitem[R. Tylenda et al. (2011)]{tylenda2011} Tylenda, R. Hajduk, M., Kami\'nski, T. et al. 2011, A\&A, 528, A114
% V1309 Sco eruption and merger

\bibitem[R. Valli et al. (2024)]{valli2024} Valli, R., Tiede, C., Vigna-G\'omez, A. et al. 2024, A\&A, 688, A128


\bibitem[M. Valtonen \& S. Mikkola (1991)]{valtonen1991} Valtonen, M. \& Mikkola, S. 1991, Ann. Rev. Astron. Astrophys., 29, 9
%ARAA article on N-body systems


\bibitem[N. van der Marel (2023)]{vandermarel2023} van der Marel, N. 2023, Eur. Phys. J. Plus, 138, 225
% review of transition disks

\bibitem[van der Tak et al.(2007)]{RADEX2007} van der Tak, F.~F.~S., Black, J.~H., Sch{\"o}ier, F.~L., et al.\ 2007, \aap, 468, 2, 627. doi:10.1051/0004-6361:20066820


\bibitem[V. Vasilyev et al. (2024)]{vasilyev2024} Vasilyev, V., Reinhold, T., Shapiro, A. et al. 2024, Science, 386, 1301
%superflares


\bibitem[K. Vida et al. (2015)]{vida2015} Vida, K., Korhonen, H., Ilyin, I.V. et al. 2015, A\&A, 580, A64

\bibitem[H. von Zeipel (1924)]{vonzeipel1924} von Zeipel, H. 1924, MNRAS, 84, 665
% gravity darkening

\bibitem[E.I. Vorobyov et al. (2017)]{vorobyov2017} Vorobyov, E.I., Elbakyan, V.G., Hosokoawa, T. et al. 2017, A\&A, 605, A77



\bibitem[E.I. Vorobyov et al. (2021)]{vorobyov2021} Vorobyov, E.I., Elbakyan, V.G., Liu, H.B., Takami, M. 2021, A\&A, 647, A44 


\bibitem[F.J. Vrba et al. (1989)]{vrba1989} Vrba, F.J., Rydgren, A.E., Chugainov, P.F., Shakovskaya, N.I., Weaver, W.B. 1989, AJ, 97, 483
% Period of G2 is 1.2 days

\bibitem[P. Vynatheya et al. (2025)]{vynatheya2025} Vynatheya, P., Ryu, T., Wang, C. et al. 2025, arXiv:2510.13736

\bibitem[F.M. Walter \& L.V. Kuhi (1981)]{walter1981} Walter, F.M. \& Kuhi, L.V. 1981, ApJ, 250, 254

\bibitem[F.M. Walter et al. (1987)]{walter1987} Walter, F.M., Brown, A., Linsky, J.L. et al. 1987, ApJ, 314, 297


\bibitem[Wm. Bruce Weaver (1987)]{weaver1987} Weaver, Wm. Bruce 1987, ApJ, 319, L89
% inclinations of TTS


\bibitem[P. Weise et al. (2010)]{weise2010} Weise, P., Launhardt, R., Setiawan, J., Henning, T. 2010, A\&A, 517, A88
% Rotational velocities of nearby young stars



\bibitem[A.D. Welty et al. (1993)]{welty1993} Welty, A.D., Ramsey, L.W., Iyengar, M. et al. 1993, PASP, 105, 1427

\bibitem[A.D. Welty \& L.W. Ramsey (1994a)]{welty1994a} Welty, A.D. \& Ramsey, L.W. 1994a, ApJ, 435, 848
% shape and pulsations of FK Com

\bibitem[A.D. Welty \& L.W. Ramsey (1994b)]{welty1994b} Welty, A.D. \& Ramsey, L.W. 1994b, AJ, 108, 299 
% ROSAT data and X-ray flare

\bibitem[R. Wichmann et al. (2000)]{wichman2000} Wichmann, R., Torres, G.,
Melo, C.H.F. et al. 2000, A\&A, 359, 181


\bibitem[D.T. Wickramasinghe et al. (2014)]{wickramasinghe2014} Wickramasinghe, D.T., Tout, C.A., Ferrario, L. 2014, MNRAS, 437, 675
% proposal that mergers lead to strong magnetic fields


\bibitem[S.J. Wolk \& F.M. Walter (1996)]{wolk1996} Wolk, S.J. \& Walter, F.M. 1996, AJ, 111, 2066

\bibitem[K. Zwintz (2008)]{zwintz2008} Zwintz, K. 2008, ApJ, 673, 1088


\bibitem[K. Zwintz \& T. Steindl (2022)]{zwintz2022} Zwintz, K. \& Steindl, T. 2022,  Front. Astron. Space Sci. 9:914738    doi: 10.3389/fspas.2022.914738

























\end{thebibliography}
\end{document}